\documentclass[aps,floats,floatfix,showpacs,amssymb,prd,twocolumn,superscriptaddress,nofootinbib,nolongbibliography,reprint,preprintnumbers]{revtex4-2}

\usepackage{amssymb,amsmath,verbatim,mathtools,needspace,enumitem,etoolbox,graphicx,physics,microtype,afterpage,xspace,tabularx,lmodern,multirow,bm,booktabs}
\usepackage{textcomp, gensymb}

\usepackage{braket}

\usepackage{relsize}

\usepackage{dcolumn}

\usepackage[normalem]{ulem}
\usepackage{placeins}
\usepackage{comment}
\usepackage[dvipsnames]{xcolor}
\definecolor{linkcolor}{rgb}{0.0,0.3,0.5}
\usepackage[unicode, colorlinks=true, linkcolor=linkcolor, citecolor=linkcolor, filecolor=linkcolor, urlcolor=linkcolor, linktocpage, breaklinks]{hyperref}
\usepackage[all]{hypcap}
\usepackage[utf8]{inputenc}
\usepackage[T1]{fontenc}
\usepackage{mathrsfs}
\hypersetup{colorlinks=true,citecolor=romared,linkcolor=romared,urlcolor=romared}

\definecolor{romared}{RGB}{142,0,28}

\newcommand{\be}{\begin{equation}}
\newcommand{\ee}{\end{equation}}

\def\be{\begin{equation}}
\def\ee{\end{equation}}
\newcommand{\beq}{\begin{eqnarray}}
\newcommand{\eeq}{\end{eqnarray}}

\usepackage{aas_macros}
\usepackage{makecell}
\usepackage{soul}
\usepackage[nolist,nohyperlinks]{acronym}
\usepackage{orcidlink}
\usepackage{lipsum}

\acrodef{LSC}[LSC]{LIGO Scientific Collaboration}
\acrodef{BH}{black hole}
\acrodef{NS}{neutron star}
\acrodef{PN}{Post-Newtonian}
\acrodef{BBH}{binary black-hole}
\acrodef{BNS}{binary neutron-star}
\acrodef{NSBH}{neutron-star black-hole}
\acrodef{NR}{numerical relativity}
\acrodef{GW}{gravitational wave}
\acrodef{PSD}{power spectral density}
\acrodef{aLIGO}{Advanced Laser interferometer Gravitational-Wave Observatory}
\acrodef{AZDHP}{aLIGO zero detuned high power density}
\acrodef{GR}{general relativity}
\acrodef{PE}{parameter estimation}
\acrodef{LAL}{LIGO algorithm library}
\acrodef{TPI}{tensor-product interpolant}
\acrodef{SVD}{singular value decomposition}
\acrodef{SNR}{signal-to-noise ratio}
\acrodef{ODE}{ordinary differential equation}
\acrodef{PDE}{partial differential equation}
\acrodef{ROM}{reduced order model}
\acrodef{QNM}{quasi-normal mode}
\acrodef{IMR}{inspiral-merger-ringdown}
\acrodef{LVK}{LIGO-Virgo-KAGRA}
\acrodef{SXS}{Simulating eXtreme Spacetimes}

\newcommand{\jhu}{\affiliation{William H. Miller III Department of Physics and Astronomy,\\ Johns Hopkins University, 3400 North Charles Street, Baltimore, Maryland, 21218, USA}}
\newcommand{\toulouse}{\affiliation{Université de Toulouse,\\ CNRS/IN2P3, L2IT, Toulouse, France}}

\graphicspath{{../Plots/}}

\newcommand{\ben}{\begin{enumerate}}
\newcommand{\een}{\end{enumerate}}

\def\be{\begin{equation}}
\def\ee{\end{equation}}
\def\beq{\begin{eqnarray}}
\def\eeq{\end{eqnarray}}

\def\apjl{\rm{ApJ}}
\def\apjs{\rm{ApJS}}
\def\aap{\rm{A\&A}}

\graphicspath{{./Plots/}}

\allowdisplaybreaks

\begin{document}

\pagenumbering{arabic}

\title{Systematic biases in parameter estimation on LISA binaries: The effect of \\excluding higher harmonics for non-spinning binaries}

\author{Sophia Yi\texorpdfstring{\,}{ }\orcidlink{0000-0002-9104-1734}}
\email{syi24@jh.edu}
\jhu

\author{Francesco Iacovelli\texorpdfstring{\,}{ }\orcidlink{0000-0002-4875-5862}}
\email{fiacovelli@jhu.edu}
\jhu

\author{Sylvain Marsat\texorpdfstring{\,}{ }\orcidlink{0000-0001-9449-1071}}
\email{sylvain.marsat@l2it.in2p3.fr}
\toulouse

\author{Digvijay Wadekar\texorpdfstring{\,}{ }\orcidlink{0000-0002-2544-7533}}
\email{jayw@jhu.edu}
\jhu

\author{Emanuele Berti\texorpdfstring{\,}{ }\orcidlink{0000-0003-0751-5130}}
\email{berti@jhu.edu}
\jhu

\date{\today}

\begin{abstract}
The remarkable sensitivity achieved by the planned Laser Interferometer Space Antenna (LISA) will allow us to observe gravitational-wave signals from the mergers of massive black hole binaries (MBHBs) with signal-to-noise ratio (SNR) in the hundreds, or even thousands. At such high SNR, our ability to precisely infer the parameters of an
MBHB from the detected signal will be limited by the accuracy of the waveform templates we use. In this paper, we explore the systematic biases that arise in parameter estimation if we use waveform templates that do not model radiation in higher-order multipoles. %
This is an important consideration for the large fraction of high-mass events expected to be observed with LISA. We examine how the biases change for MBHB events with different total masses, mass ratios, and inclination angles. We find that systematic biases due to insufficient mode content are severe for events with total redshifted mass $\gtrsim10^6\,M_\odot$. We then compare several methods of predicting such systematic biases without performing a full Bayesian parameter estimation. In particular, we show that through direct likelihood optimization it is possible to predict systematic biases with remarkable computational efficiency and accuracy. %
Finally, we devise a method to construct approximate waveforms including angular multipoles with $\ell\geq5$ to better understand how many additional modes (beyond the ones available in current approximants) might be required to perform unbiased parameter estimation on the MBHB signals detected by LISA. 

\end{abstract}

\maketitle

\section{\label{sec:intro}Introduction}

As the catalog of gravitational wave (GW) events grows and the detectors improve in sensitivity, our ability to characterize the properties of individual events and of compact binary populations will be increasingly limited by systematic effects. Some of these systematic effects are due to detector noise and astrophysics, but waveform systematics are particularly important, and they have attracted a significant amount of attention in the context of ground-based detectors (see e.g.~\cite{Purrer:2019jcp,Jan:2020bdz,Moore:2021eok,Hu:2022rjq,Hu:2022bji,Read:2023hkv,Owen:2023mid,Puecher:2023rxw,Jan:2023raq,Kapil:2024zdn,Dhani:2024jja,Gupta:2024gun,Pompili:2024yec,Bachhar:2024olc,Chandramouli:2024vhw,Khan:2024whs}).

The Laser Interferometer Space Antenna (LISA) is expected to observe binary black holes (BBHs) with much larger total masses and mass ratios, and with considerably louder signal-to-noise ratios (SNRs), than anything we have seen thus far in ground-based observatories (see e.g. Refs.~\cite{LISA:2017pwj,LISA:2024hlh}). It has long been recognized that systematic biases in waveform modeling will consequently be considerably worse for parameter estimation (PE) with signals detected by LISA~\cite{Berti:2006ew,Cutler:2007mi,Ferguson:2020xnm,LISAConsortiumWaveformWorkingGroup:2023arg}. Although early estimates were already able to demonstrate that systematic errors would dominate over statistical errors for massive LISA sources~\cite{Littenberg:2012uj}, thus far, relatively little detailed work has been done to attempt to understand just how significant these biases will be, and how they may vary across the parameter space.  

One source of systematic bias expected to affect PE with LISA sources is the bias due to neglecting higher harmonics in waveforms. It has been known for some time that more energy is radiated in higher multipoles for systems with larger mass ratios~\cite{Berti:2007fi}, and that the inclusion of higher multipoles will significantly enhance the science return of ringdown observations with LISA~\cite{Baibhav:2017jhs,Baibhav:2018rfk,Baibhav:2020tma}.

Numerous authors have studied the biases introduced by neglecting higher-order modes in PE on BBH signals in current ground-based observatories, generally finding significant systematic errors due to the omission of higher modes for systems with high mass ratio, large total mass, high SNR, and inclination angle close to $\pi/2$ (edge-on systems)~\cite{Varma:2016dnf,CalderonBustillo:2015lrt,Littenberg:2012uj,Varma:2014jxa,Graff:2015bba,Shaik:2019dym,Pang:2018hjb}.
Some of these and other works have additionally found that relying on quadrupole-only waveforms decreases the efficacy of searches for BBH systems in current ground-based detectors, again in particular for heavier, more asymmetric, and more edge-on binaries~\cite{Harry:2017weg,CalderonBustillo:2016rlt,Brown:2012nn}. A recent series of papers found that a combination of downweighting glitches and including higher modes in searches of the data of the third observing run of LVK resulted in 12 new BBH event candidates, in addition to increasing the significance of several candidates previously deemed ``marginal''~\cite{Wadekar:2023gea,Wadekar:2023kym,Wadekar:2024zdq}, as well as increasing the overall sensitive volume of GW searches~\cite{Mehta:2025jiq}. On the other hand, another recent study found that the exclusion of higher harmonics caused relatively minimal biases in parametrized post-Einsteinian (ppE) tests of general relativity, at least compared to the biases induced by other neglected physics (i.e., precession)~\cite{Chandramouli:2024vhw}. However, it is unclear how this result would extend to considerably more massive events. There is clearly motivation to account for higher modes more carefully, both in detection and in PE with GW signals. In this study, we will focus on the importance of subdominant modes in performing PE on BBHs observed by LISA.%

One previous study~\cite{Pitte:2023ltw} demonstrated that, when performing PE on a high-SNR massive black hole binary (MBHB) event of the kind anticipated to be observed with LISA, systematic biases arise due to insufficient mode content in the waveform templates used for signal recovery. This phenomenon was demonstrated explicitly in Ref.~\cite{Pitte:2023ltw} for a single example binary. In this paper, we perform the same analysis of varying the mode content in waveform templates for several example MBHB systems, studying how the aforementioned systematic biases in PE change for events with different total masses, mass ratios, and inclination angles. We then investigate the extent to which these biases can be predicted in a cost-efficient manner, i.e., without having to perform PE for every MBHB event under consideration. We use one such method (direct likelihood optimization; see Sec.~\ref{sec:nelder-mead}), which we find to be remarkably accurate in estimating systematic biases, to set approximate boundaries on the parameter space in which unbiased PE can be performed (i.e., where there is no significant bias due to neglecting a higher-order mode in the waveform template). 

\begin{figure}[t]
\centering
\includegraphics[width=0.48\textwidth]{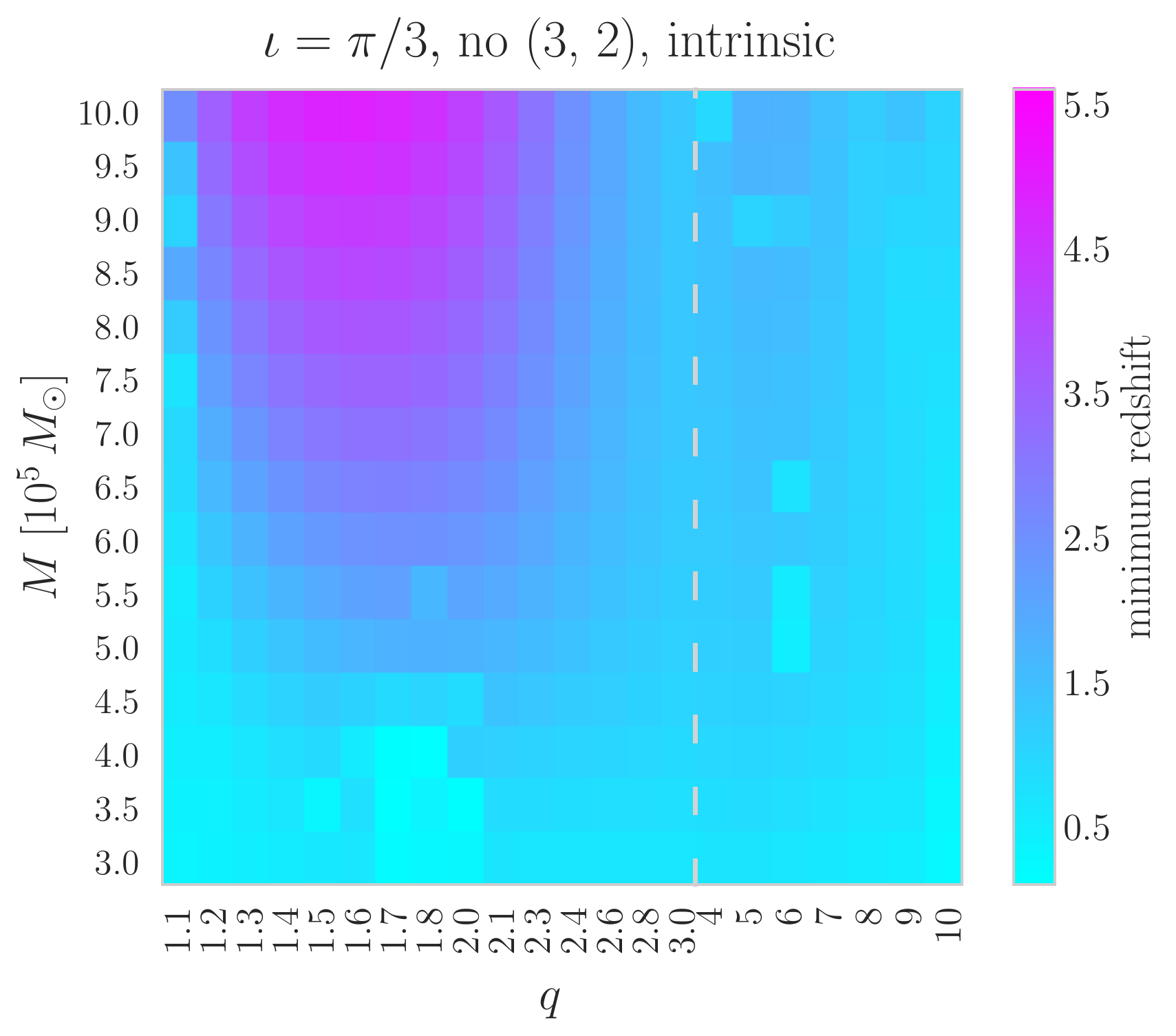}
\caption{%
Minimum redshift at which parameter estimation is unbiased, in the sense that the systematic bias on all four intrinsic parameters $\mathcal{M}_c,\,q,\,\chi_+,$ and $\chi_-$ due to neglecting the $(\ell,|m|)=(3,~2)$ mode in the waveform template is less than the $2\sigma$ statistical error on the parameters. For example, for $M=7\times10^5\, M_\odot$ and $q=2$, the parameter estimation is biased when $z\lesssim3$. The grid is log-spaced between $q\in[1.1,3]$ and spaced linearly between $q\in[3,10]$ (with the transition marked by the dashed gray line), due to the more significant changes in results observed as we move from nearly symmetric to clearly asymmetric binaries. The corresponding plot for extrinsic parameters (sky localization and distance) is given in Fig.~\ref{fig:heatmap-inc3-ext}.}
\label{fig:heatmap-inc3}
\end{figure}

In Fig.~\ref{fig:heatmap-inc3} we show some results of this approximate boundary-setting for biases on the intrinsic parameters 
\begin{eqnarray}\label{eqn:params}    \mathcal{M}_c&=&\frac{\left(m_1m_2\right)^{3/5}}{\left(m_1+m_2\right)^{1/5}}\,,\nonumber\\[.5mm]
    q&=&m_1/m_2>1\,,\\[.5mm]
    \chi_{\pm}&=&\frac{m_1 \chi_1\pm m_2\chi_2}{m_1+m_2}\,,\nonumber
\end{eqnarray}
where $m_1,m_2$ and $\chi_1,\chi_2$ are the individual masses and dimensionless spins of the progenitor BHs, respectively. For each bin in the grid of detector-frame total mass $M/M_\odot\in[3\times10^5,10^6]$ and mass ratio $q\in[1.1,10]$, we show the minimum redshift at which the systematic bias on parameters [$\mathcal{M}_c,q,\chi_+,\chi_-$] 
due to excluding the $(\ell,~m)=(3,~2)$ mode is still less than the $2\sigma$ statistical error on the parameter. Results are shown for systems with inclination angle $\iota=\pi/3$. Moving from left to right in the plot, we see that biases initially become worse as we move toward more massive and more asymmetric binaries, as expected. Continuing further, we have the balancing effect of the chirp mass decreasing going from top left to bottom right, such that mergers occur at higher frequencies. When the merger, where higher-order modes are most important, is pushed toward the higher-frequency region where LISA is less sensitive, the bias due to neglecting these modes becomes less severe again. The reason for the chosen range of total mass ($y$-axis) will be clear in Sec.~\ref{sec:results}, where we find that our PE results on MBHB events with total mass $\lesssim3\times10^5\,M_\odot$ do not exhibit significant biases, whereas the PE for events with total mass $\sim10^6\,M_\odot$ is significantly biased. Details on the construction of Fig.~\ref{fig:heatmap-inc3}, as well as similar plots for extrinsic parameters and for systems with inclination $\iota\approx\pi/2$, are given in Sec.~\ref{sec:heatmaps}.

In the following sections, we lay out the components that go into the construction and interpretation of Fig.~\ref{fig:heatmap-inc3}. In Sec.~\ref{sec:methods}, we outline how we perform both full PE with \texttt{lisabeta}~\cite{Marsat:2020rtl} and several cheaper methods of estimating systematic bias.
In Secs.~\ref{sec:results} and~\ref{sec:results_extrinsic}, we show the results of PE for the intrinsic and extrinsic parameters of about 20 selected binaries, respectively. In Sec.~\ref{sec:CV_results}, we show how our methods of rapidly approximating bias compare to PE and present more plots similar to Fig.~\ref{fig:heatmap-inc3}. Finally, in Sec.~\ref{sec:approx_HM}, we show the results of performing PE with crude waveforms containing more higher-order modes than are currently modeled with the waveform approximants we use for PE (\textsc{IMRPhenomXHM}~\cite{Garcia-Quiros:2020qpx}). Throughout this paper we use geometrical units ($G=c=1$).

\section{Methodology}\label{sec:methods}

\subsection{Parameter estimation with \texttt{lisabeta}}\label{sec:intro_PE}

We examine the extent of systematic biases due to incomplete mode content in the GW templates used for PE on MBHBs of the kind expected to be observed by LISA. We use \texttt{lisabeta}~\cite{Marsat:2020rtl} to perform PE varying the redshifted total mass ($M/M_\odot=10^5,10^6$), mass ratio ($q=1.1, 4, 8$), and inclination angle ($\iota=\pi/12, \pi/3, \pi/2 - \pi/12$). Observing stark differences in the results for events with total mass $10^5\,M_\odot$ vs. $10^6\,M_\odot$, we also run PE on a few additional events at the intermediate mass of $3\times 10^5\,M_\odot$, with the median inclination angle ($\iota=\pi/3$). Altogether, this amounts to a total of 21 PE runs with \texttt{lisabeta}. The rest of the parameters (not modified between runs) are listed in Table~\ref{tbl:grid}, where $\chi_{1,2}$ are the dimensionless progenitor spins aligned with the orbital angular momentum of the binary; $D_L$ is the luminosity distance in Mpc, which we choose to correspond to redshift $z=1$; %
and $\delta t$ is defined through $t_c = t_0 + \delta t$, with $t_c$ the coalescence time and $t_0$ the reference epoch (in years). The angles $\beta_L$ and $\lambda_L$ are the ecliptic latitude and longitude, respectively, in the LISA frame; $\phi$ is the BBH source frame phase; and $\Psi_L$ is the polarization in the LISA frame. Note that while we vary the total mass between runs, we set our sampler to infer the chirp mass, $\mathcal{M}_c$, along with the mass ratio, $q$, both of which are defined in Eq.~\eqref{eqn:params}. Similarly, we sample in spin parameters $\chi_+$ and $\chi_-$, rather than $\chi_1$ and $\chi_2$, although we note that the injected values of these are all zero in the scope of this paper. Altogether, then, the 11 parameters that are simultaneously sampled in each PE run are [$\mathcal{M}_c, q, \chi_+,\chi_-,D_L,\delta t,\beta_L,\lambda_L,\phi,\iota,\Psi_L$].

\begin{table}[t]
\caption{\label{tbl:grid} 
  Parameters of the MBHB sources on which we perform PE with a varying number of modes. The value of $D_L$ is chosen to correspond to $z=1$. For the BBHs with total mass $3\times10^5M_\odot$, we only run PE at the median inclination angle ($\iota=\pi/3$).
}
\begin{tabular}{cccc}
\toprule\midrule
{}& Parameter & Values & \\
\midrule
    & $M$ $(M_\odot)$  & $10^5, 3\times10^5, 10^6$& \\
    & $q$ & $1.1,4,8$ & \\
    & $\chi_{1,2}$ & 0 & \\
    & $D_L$ (Mpc) & 6791.81 & \\ 
    & $\delta t$ (s) & 0.0 & \\ 
    & $\beta_L$ (rad) & $\pi/6$ & \\ 
    & $\lambda_L$ (rad) & 1.8 & \\ 
    & $\phi$ (rad) & 0.2 & \\ 
    & $\Psi_L$ (rad) & 1.2 & \\ 
    & $\iota$ (rad) & $\pi/12,\pi/3,\pi/2-\pi/12$ & \\ 
\midrule\bottomrule
\end{tabular}
\end{table}

For each combination of parameters, we generate the injected waveform for the event with the \textsc{IMRPhenomXHM} waveform model~\cite{Garcia-Quiros:2020qpx,Pratten:2020fqn}, %
including the $(\ell,~m)=$ (2,~2), (2,~1), (3,~3), (3,~2) and (4,~4) modes -- i.e., all the available modes with the \textsc{IMRPhenomXHM} waveform approximant. (From here on, whenever we denote a mode by a comma-delimited pair of numbers in parentheses, we will be referring to the angular harmonic indices of the mode, as done here.) We then perform PE on the signal with \texttt{lisabeta} using first a template with all five modes, then we iterate with one less mode, until we are attempting to recover the signal with just the quadrupole. The authors of Ref.~\cite{Pitte:2023ltw} chose to ``deactivate'' modes in order of SNR contribution, such that a four-mode template excludes only the mode with the lowest SNR, the three-mode template excludes the two ``quietest'' modes, etc. To simplify comparisons between our different events, we always include modes in the following order:
\begin{itemize}
    \item 1 mode: (2,~2) 
    \item 2 modes: (2,~2), (3,~3)
    \item 3 modes: (2,~2), (3,~3), (2,~1)
    \item 4 modes: (2,~2), (3,~3), (2,~1), (4,~4)
    \item 5 modes: (2,~2), (3,~3), (2,~1), (4,~4), (3,~2)
\end{itemize}

\begin{figure*}[ht!]
\centering
\includegraphics[width=0.98\textwidth]{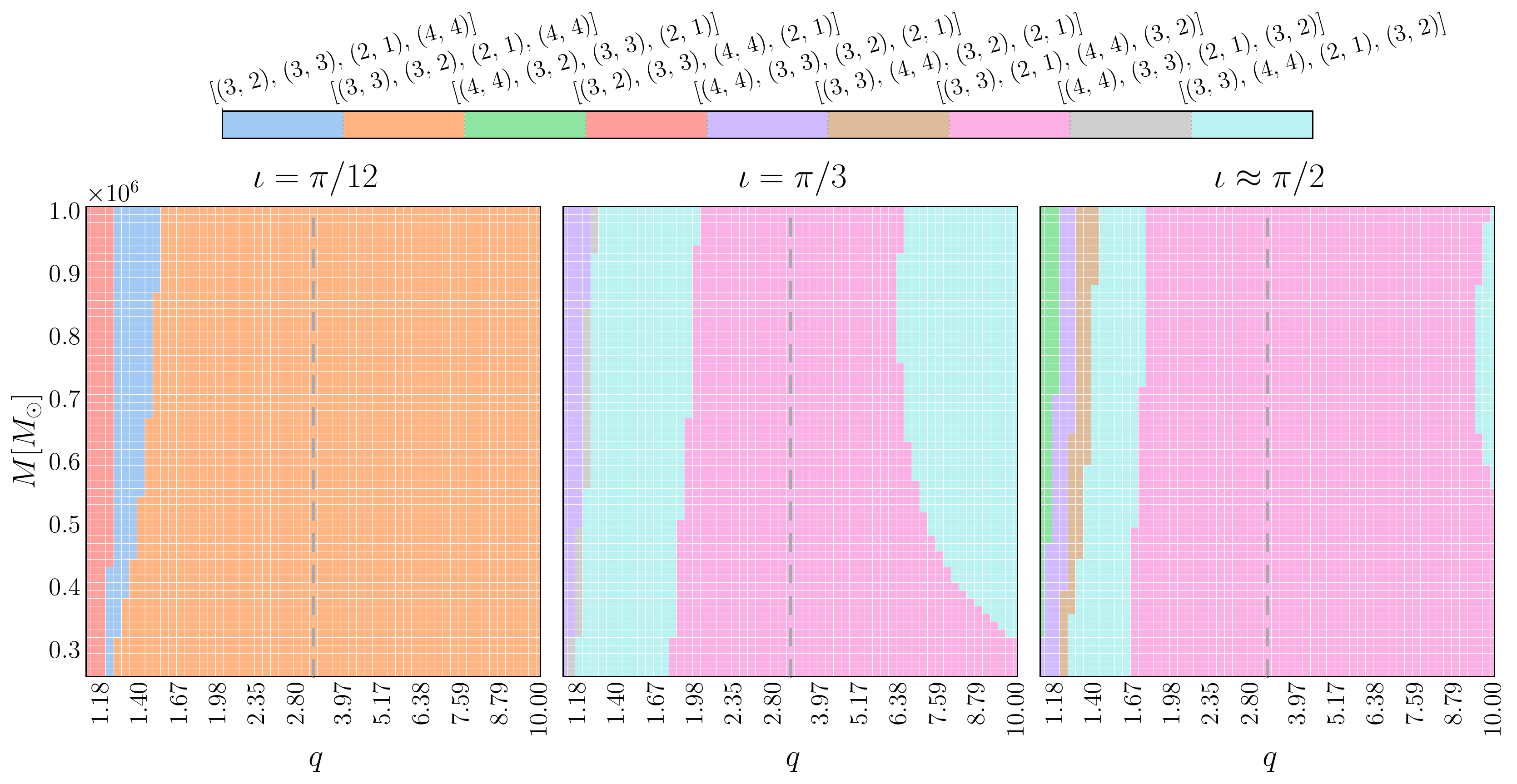}
\caption{Ordering of the subdominant modes by SNR contribution. The (2,~2) mode is always the dominant one. Each color corresponds to the subdominant multipole ordering that we observe within this region of parameter space; in the legend, the multipoles are ranked by SNR. For example, the pink region corresponds to the ordering (2,~2), (3,~3), (2,~1), (4,~4), (3,~2).}
\label{fig:mode-ordering}
\end{figure*}

This ordering was chosen for a number of reasons. First, when examining the mode-by-mode SNR of the 21 events we considered, 9 out of the 21 exhibited the above ordering. In Fig.~\ref{fig:mode-ordering}, we can see the prevalence of this mode ordering within the region of parameter space we consider. For each inclination angle, different colors indicate a different ordering of the subdominant modes (ranked from ``loudest'' to ``quietest''). In all the configurations shown here, the (2,~2) mode dominates, with the (3,~3) mode generally being the second loudest. The ordering of the (2,~1), (4,~4), and (3,~2) modes varies a bit more, with the (4,~4) usually being the quietest for a nearly face-on system, and the (3,~2) generally being the quietest otherwise. %

Second, although the odd-$m$ modes are  suppressed for events with $q\approx1$  (in the case of non-spinning progenitors, which we consider here),
we find that templates with $m\neq2$ nevertheless perform better in breaking degeneracies observed with (2,~2)-only templates. This is due to the fact that modes with the same azimuthal number $m$ (e.g., the (2,~2) and (3,~2)) share the same orbital phasing, whereas modes with different $m$'s exhibit a different phasing behavior that cannot be reproduced by the (2,~2). %
Therefore PE results can be improved with the addition of the (3,~3) or (2,~1) as the second mode after the (2,~2), despite these potentially being more ``quiet'' (i.e., having lower SNR) than the (3,~2) mode.

Lastly, mode mixing makes the differentiability of the \textsc{IMRPhenomXHM} waveforms less clean for the (3,~2) mode. This makes it challenging to perform certain cost-efficient bias estimates which involve derivatives of the waveform. For this reason it is advantageous to include this mode last, in order to minimize unwanted effects on the results of these cheaper methods for events with less than $5$ modes. 

We use the power spectral density (PSD) required by the LISA Science Requirements Document (SciRDv1)~\cite{Babak:2021mhe}. We assume the frequency range of LISA to be $10^{-5}-0.5 \,{\rm Hz}$ and incorporate the full response of LISA~\cite{Marsat:2018oam}. We also account for a background of white dwarf noise as in Sec.~9 of Ref.~\cite{Babak:2021mhe}. 

To sample the GW likelihood, we use the parallel-tempered Markov chain Monte Carlo (MCMC) ensemble sampler \texttt{ptemcee}~\cite{Vousden:2016eeu,Foreman-Mackey:2012any,Goodman:2010dyf}, which is a fast sampler well-adapted to handling complex posteriors. To improve the robustness for degenerate posteriors, the ensemble sampler is enriched with jump proposals designed to explore the possible sky multimodality~\cite{Marsat:2020rtl}. %
For a typical PE run, we use 64 walkers and run \texttt{ptemcee} at 10 temperatures with a max temperature of 100. For each event, we sample in the combinations of mass and spin parameters [$\mathcal{M}_c,\,q,\,\chi_+,$ and $\chi_-$] defined in Eq.~\eqref{eqn:params}. %
At first order in post-Newtonian (PN) theory, the GW frequency evolution is solely dependent on the chirp mass $\mathcal{M}_c$, so this is often the mass parameter that is most easily determined from the data. Similarly, $\chi_+$ (also commonly known as $\chi_{\rm{eff}}$) is generally much easier to determine from the data than the individual spins $\chi_1$ and $\chi_2$. Lastly, $q$ and $\chi_-$ are chosen to complete the set of mass and spin parameters, respectively. %

To perform PE, we first calculate the Fisher covariance matrix and initialize chains from the results of this calculation. We set uniform priors on the chirp mass %
and distance in the range of 0.1 to 10 times their injected values. For the mass ratio, we use a uniform prior over the range $[1,\,10]$, and the spin parameters have a uniform prior over their allowed ranges of $[-1,\,1]$.

\subsection{Predicting biases in the linear signal approximation}\label{sec:CV_intro}

Given the computational cost of full Bayesian PE analyses, it is desirable to have a much faster, cheaper method of predicting systematic biases due to, e.g., neglecting higher-order modes. One technique for accomplishing this is to estimate systematic errors in the linear signal approximation as done by Cutler \& Vallisneri (henceforth, ``CV'') in Ref.~\cite{Cutler:2007mi}. In Ref.~\cite{Antonelli:2021vwg}, the errors calculated in this approximation were found to reliably predict the biases in parameter inference of ET and LISA sources due to overlapping signals, foregrounds created by unfitted sources, and incorrectly removed sources, among other things. The authors of Ref.~\cite{Pitte:2023ltw} examined the usefulness of this technique for the example binary they considered, finding that the errors were not properly reconstructed within the linear signal approximation. Here we expand the study of the usefulness of the CV estimates, covering a much larger region of parameter space than previously covered in Ref.~\cite{Pitte:2023ltw}. In Appendix~\ref{app:CV_formalism}, for completeness, we review the CV formalism. In Sec.~\ref{sec:CV_results} below we compare our PE results with the errors predicted by the CV estimates. %

\subsection{Predicting biases with direct likelihood optimization}\label{sec:nelder-mead}

In Sec.~\ref{sec:CV_results} we will confirm the finding of Ref.~\cite{Pitte:2023ltw}: the biases due to neglecting higher-order modes are not always well-approximated via the CV formalism, in particular for more massive events. Given this result, it is desirable to explore other methods of predicting the biases we see in PE. In this section, we leverage the fact that the systematic bias we consider here is, by definition, the difference between the ``true'' (or injected) parameters and the parameters evaluated at the maximum likelihood given our best-fit waveform. Studying this bias, therefore, in principle reduces to a straightforward optimization problem. Indeed, in theory, performing a rigorous sampling from the posteriors might be considered excessive if one is only interested in studying the distance between the maximum likelihood point and the injection (assuming that the maximum likelihood point can be found without a sampler).

We will see in Sec.~\ref{sec:CV_results} that, just as expected, one can indeed infer the systematic bias with remarkable success when simply optimizing the likelihood directly. In this work, we use the Nelder-Mead algorithm~\cite{Nelder:1965zz} to maximize the likelihood. Essentially, in the Nelder-Mead algorithm, a ``simplex'' with $n+1$ vertices ($n$ being the number of variables) is introduced in the parameter space. The function value is evaluated at each vertex, and then the simplex systematically rearranges (growing, reflecting, contracting, etc.) such that it moves toward and eventually closes in on the location of the minimum. One appealing characteristic of this algorithm is that it is able to minimize a function of $n$ variables without differentiation. 

In this work, we use the implementation of the Nelder-Mead algorithm provided in \texttt{scipy}~\cite{2020NatMe..17..261V}. We initialize our simplex in $(n+1)$-dimensional space by setting each vertex of the initial simplex at the likelihood evaluated with one parameter set to a value $m\sigma$ away from its injected value, where $\sigma$ is calculated from the diagonal elements of the Fisher covariance matrix. We then minimize the function $-\mathrm{ln}(\mathcal{L})$ (i.e., we find the maximum likelihood). To accommodate the large range in parameter values (considering, for example, that we have masses $\mathcal{O}(10^5-10^6)\, M_\odot$ and angles in the range $[-\pi,\pi]$), we rescale parameters by the $\sigma$'s calculated from the Fisher analysis to perform this minimization. During the maximization procedure, we enforce the same prior boundaries on the parameters used in PE.

After using the Nelder-Mead algorithm to compute the location of the maximum likelihood a single time, we then take the location of this calculated maximum as an initial guess for the subsequent iteration of the algorithm, allowing the algorithm to determine the initial simplex based on this initial guess. 
We repeat this process until we find convergence. We find some dependence on the ``size'' of the initial simplex ($m$), as we will discuss in Sec.~\ref{sec:CV_results}. Generally, we find good performance with a rather broad initial simplex, i.e., setting something like $m=20$. This suggests that, given the high dimensionality of the problem and the potential presence of multiple local minima throughout the parameter space, the algorithm performs best when the initial simplex is large enough to enclose the global minimum, which apparently does not necessarily coincide with the injected value, especially for the angles. 

As we will see in Sec.~\ref{sec:CV_results}, we find that the parameters evaluated at the maximum likelihood found by the Nelder-Mead algorithm generally show good agreement with the median values of the posterior distributions recovered by a full Bayesian PE. In the few cases where the agreement is not very good, we attempted to see if the Nelder-Mead algorithm mistakenly finds incorrect (local) minima. To check this, we utilize the ``Basin-hopping'' technique as implemented in \texttt{scipy}, which attempts to find the global minimum by allowing the algorithm to ``hop'' between different regions of parameter space and explore the various local minima it encounters. We find that Basin-hopping generally does not improve the results of the optimization, suggesting that the Nelder-Mead algorithm is generally successful in finding the global minimum.  %

Importantly, we find that the biases can be estimated by directly maximizing the likelihood in this manner for all 5 mode configurations for a given event in as little as 9 seconds to $\sim1$ minute in a \texttt{Jupyter Notebook} running on a single core of an 8-core CPU (8GB memory). %
This is dramatically faster than the PE runs, which take $\sim3-15$ minutes \emph{per mode configuration} when parallelized across 48 cores (3.9GB RAM per CPU).

In Sec.~\ref{sec:CV_results}, we show how directly optimizing the likelihood with the Nelder-Mead algorithm can result in much better estimates of systematic biases than we are able to obtain with the CV formalism. 

\section{Results: Parameter Estimation with \texttt{lisabeta}}
\label{sec:results}

We now present PE results across our grid of parameter space, examining how biases arise due to an insufficient number of modes. In this section, we focus on examining the biases in intrinsic parameter recovery. Corresponding discussions for extrinsic parameters are given in Sec.~\ref{sec:results_extrinsic}.  We remind the reader that while we divide our discussion of intrinsic and extrinsic parameters for a clearer presentation of results, all 11 parameters (intrinsic and extrinsic) were simultaneously sampled in each of our PE runs. For the present, we restrict our analysis to injected signals with non-spinning progenitors ($\chi_1=\chi_2=0$). %
We place our events at a distance corresponding to $z=1$. For the rest of the extrinsic parameters (source localization, etc.), which we expect to have a minimal impact on our analysis, we simply take the arbitrary values listed in Table~\ref{tbl:grid}.

\subsection{Lowest mass events (\texorpdfstring{$M=10^5 M_\odot$}{M=1e5~Msun})}\label{sec:M1e5}

We begin with results for the lowest-mass MBHBs, with total redshifted mass equal to $10^5\, M_\odot$. The total SNRs of the events investigated in this section range from 188.6 to 692.7. 

\subsubsection{Events with \texorpdfstring{$q=1.1$}{q=1.1}}\label{sec:M1e5q1}

\begin{figure}[tbp]
\centering
\includegraphics[width=0.42\textwidth]{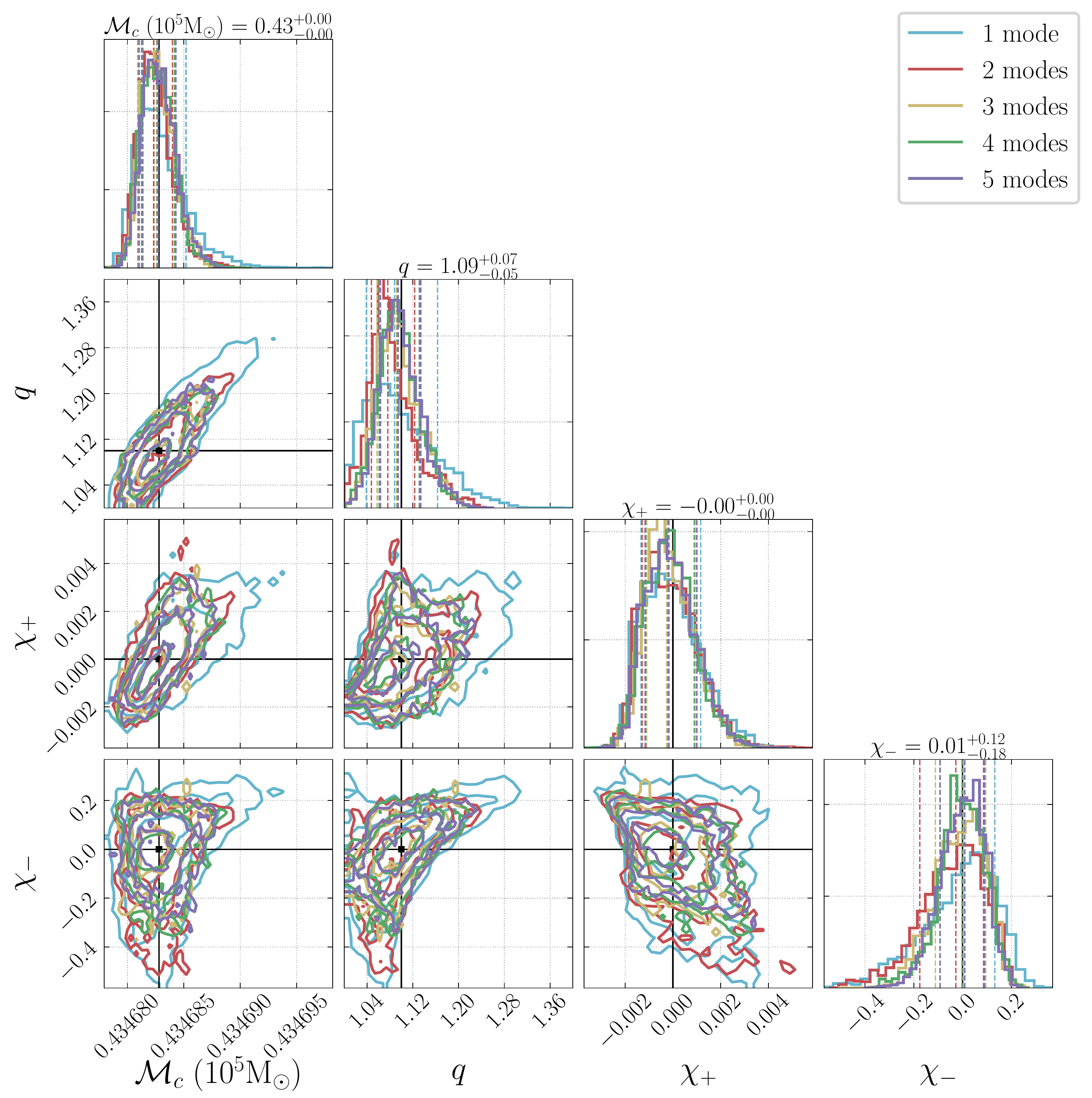} 
\includegraphics[width=0.42\textwidth]{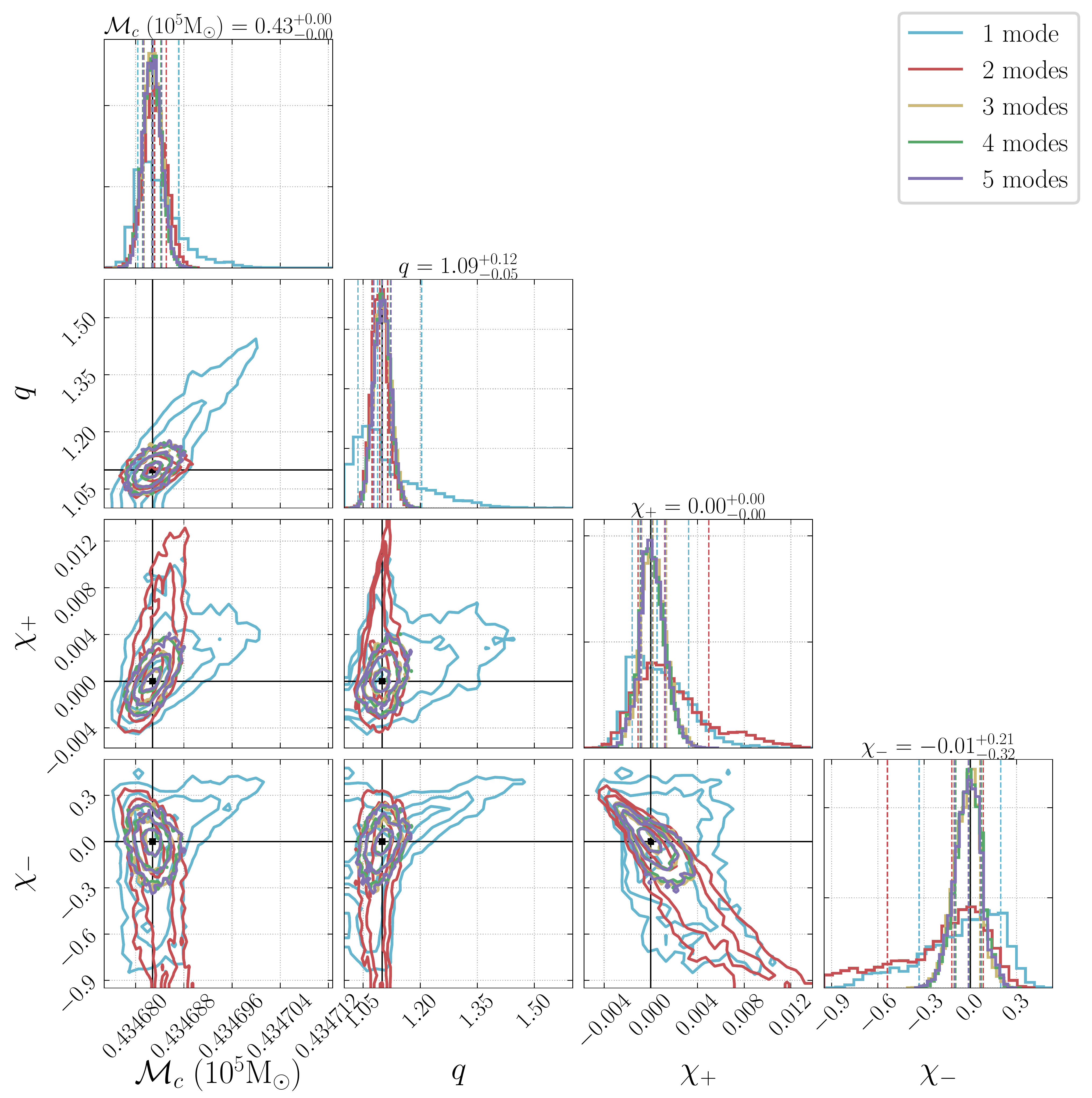} 
\includegraphics[width=0.42\textwidth]{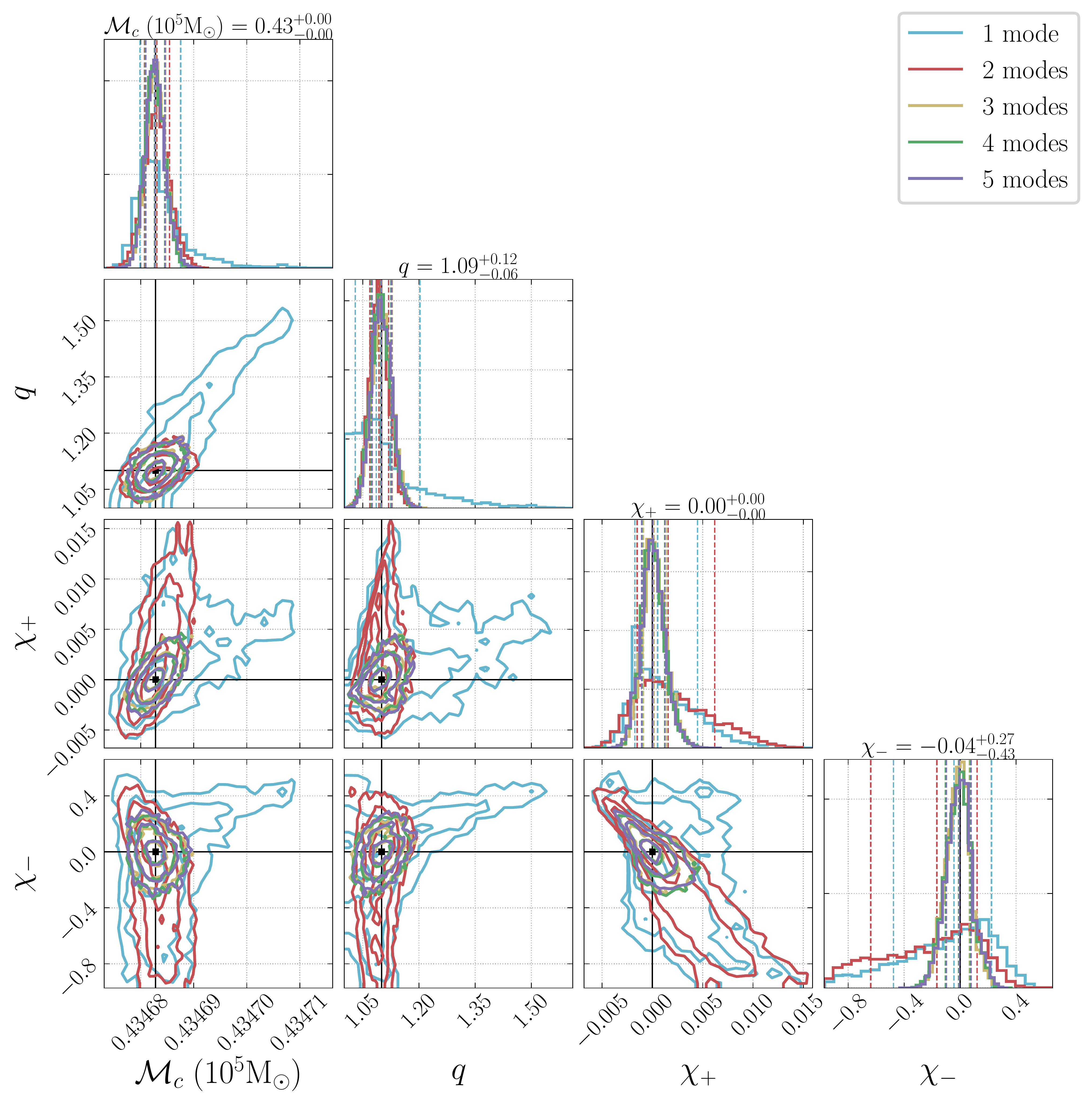} 
\caption{$M=10^5\, M_\odot, q=1.1$. Top, center, bottom: inclination angle and total SNR vary from $\iota=[\pi/12, \pi/3, \pi/2-\pi/12]$ and [692.7, 404.1, 298.4].}
\label{fig:pM5s0q1}
\end{figure}

The first set of events with $M=10^5\, M_\odot$ that we examine have mass ratio $q=1.1$ and differ in inclination angle $\iota=[\pi/12, \pi/3, \pi/2 - \pi/12]$. The posteriors on the intrinsic parameters $\mathcal{M}_c, q, \chi_+,$ and $\chi_-$ generated for these events are presented in Fig.~\ref{fig:pM5s0q1}. The colors correspond to posteriors obtained with a different number of modes in the template used for recovery, as indicated in the legends. The top labels of the 1D histograms display the median values recovered with the \emph{quadrupole-only} templates, which are generally the most biased values. 

In general, we see that the posteriors close in more tightly around the injected values (marked by the black lines) when more modes are included in the template. For the nearly edge-on system ($\iota \approx \pi/2$) and the system with median inclination ($\iota=\pi/3$), the posteriors on the spin parameters $\chi_+$ and $\chi_-$ are rather broad when inferred with only 1- or 2-mode templates. Nevertheless, the posteriors still peak around the correct (injected) value, and there are in general no strong biases.%

\subsubsection{Events with \texorpdfstring{$q=4$}{q=4}}\label{sec:M1e5q4}

\begin{figure}[tbp]
\centering
\includegraphics[width=0.42\textwidth]{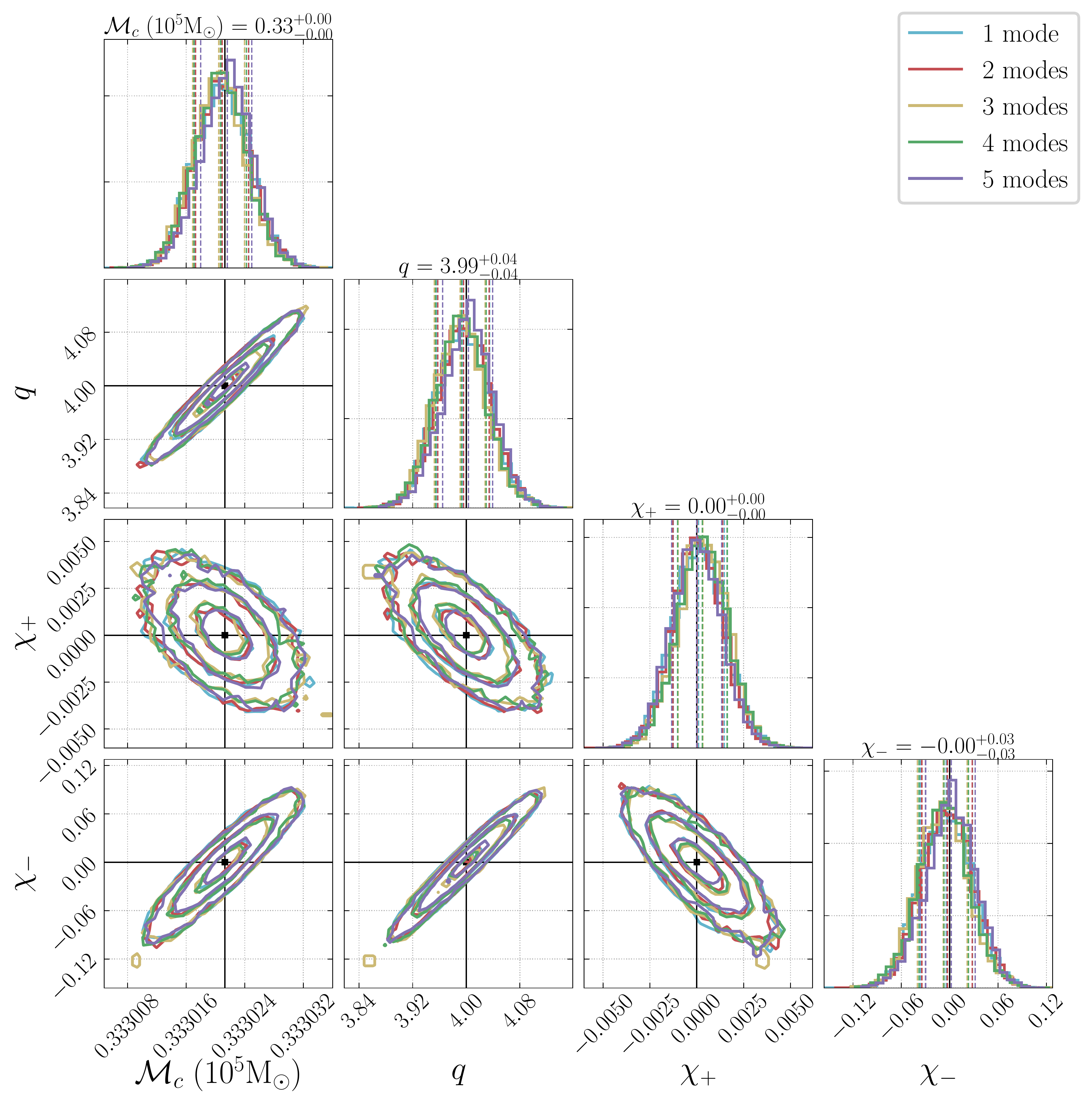} 
\includegraphics[width=0.42\textwidth]{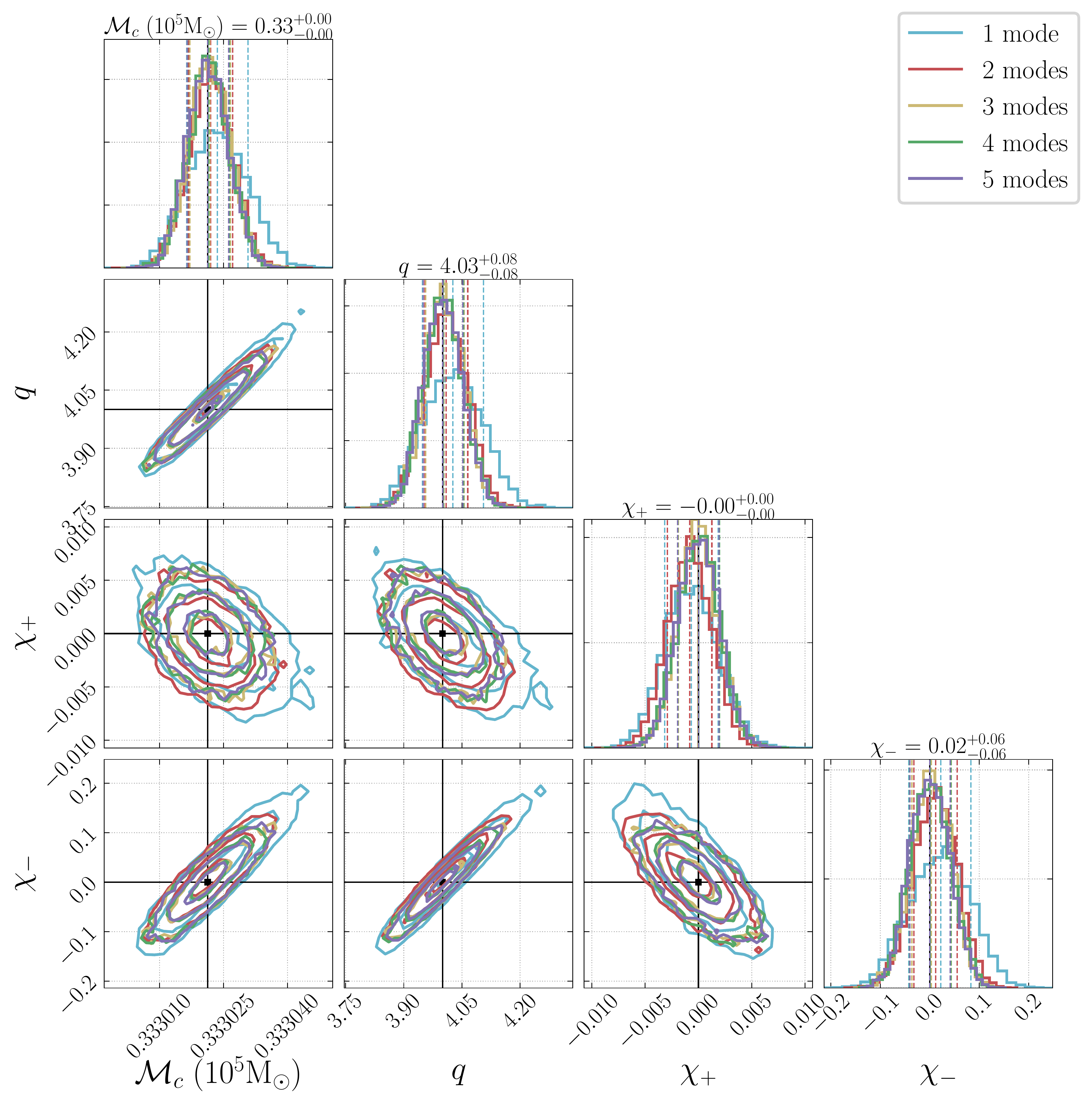} 
\includegraphics[width=0.42\textwidth]{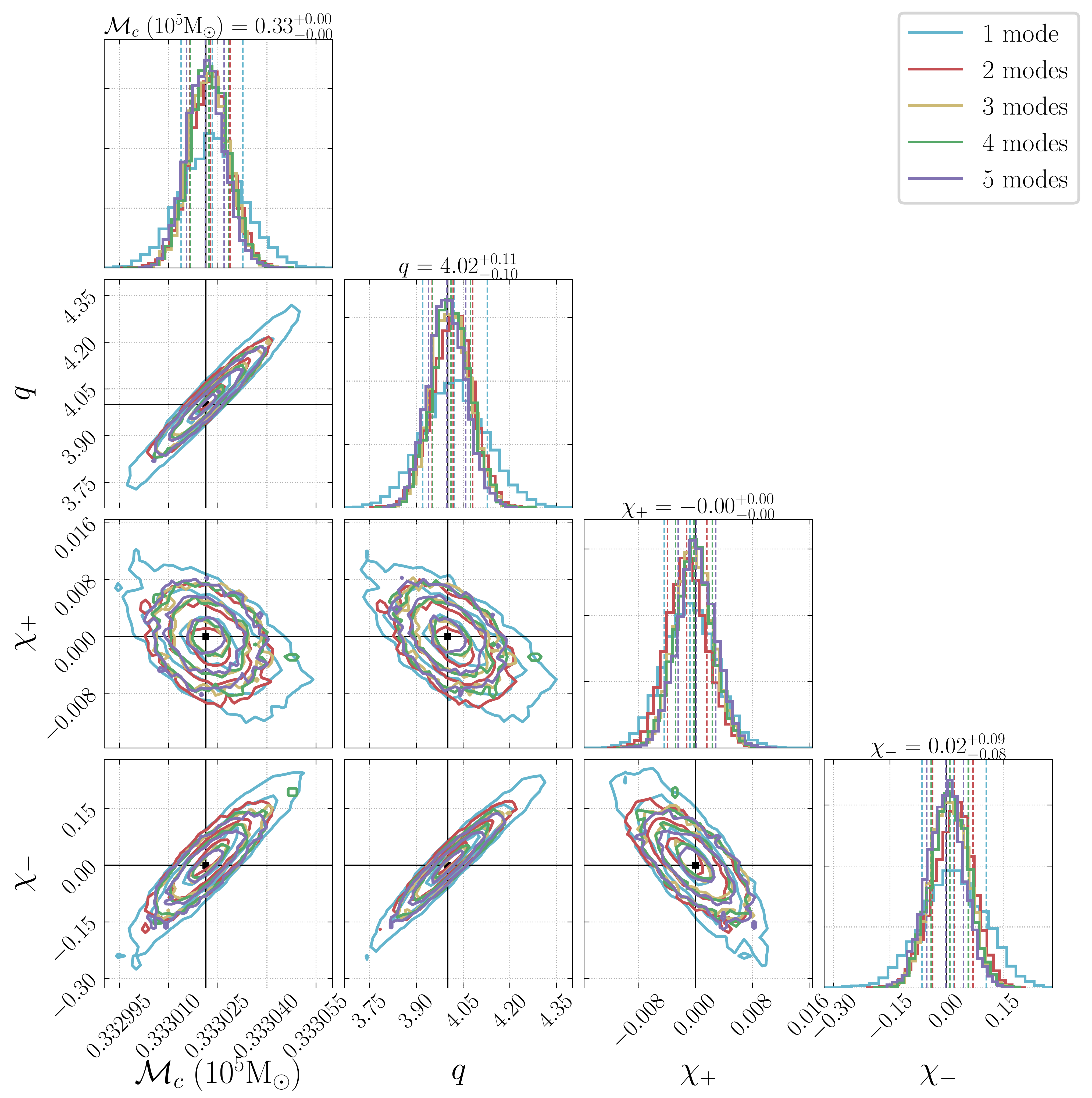} 
\caption{$M=10^5\, M_\odot, q=4.$ Top, center, bottom: inclination angle and total SNR vary in the range $\iota=[\pi/12, \pi/3, \pi/2-\pi/12]$ and [551.8, 323.4, 238.9].}
\label{fig:pM5s0q4}
\end{figure}

Next, we consider $M=10^5\, M_\odot$ events with $q=4$, again varying the inclination angles. The results are presented in Fig.~\ref{fig:pM5s0q4}. 

Compared with the approximately equal-mass events in Sec.~\ref{sec:M1e5q1}, the posteriors are considerably more tightly centered around the true values. Moreover, there do not seem to be significant biases even for the spins. In fact, the posteriors in Fig.~\ref{fig:pM5s0q4} look about the same across all 5 mode configurations, with the exception that using the 1-mode template can result in slightly broader posteriors.

This result may reflect the fact that, despite higher-order modes being more significant for more asymmetric binaries (larger $q$), with a larger $q$ there is also comparatively more SNR in the inspiral compared to the merger/postmerger, where higher-order modes are most important. This could be a significant balancing effect for this set of lowest-mass events, for which higher modes are pushed to the high-frequency region of lower sensitivity of LISA. Moreover, note that the signal amplitude for a fixed total mass, and thus the SNR, decreases with increasing $q$.

\subsubsection{Events with \texorpdfstring{$q=8$}{q=8}}\label{sec:M1e5q8}

\begin{figure}[tbp]
\centering
\includegraphics[width=0.42\textwidth]{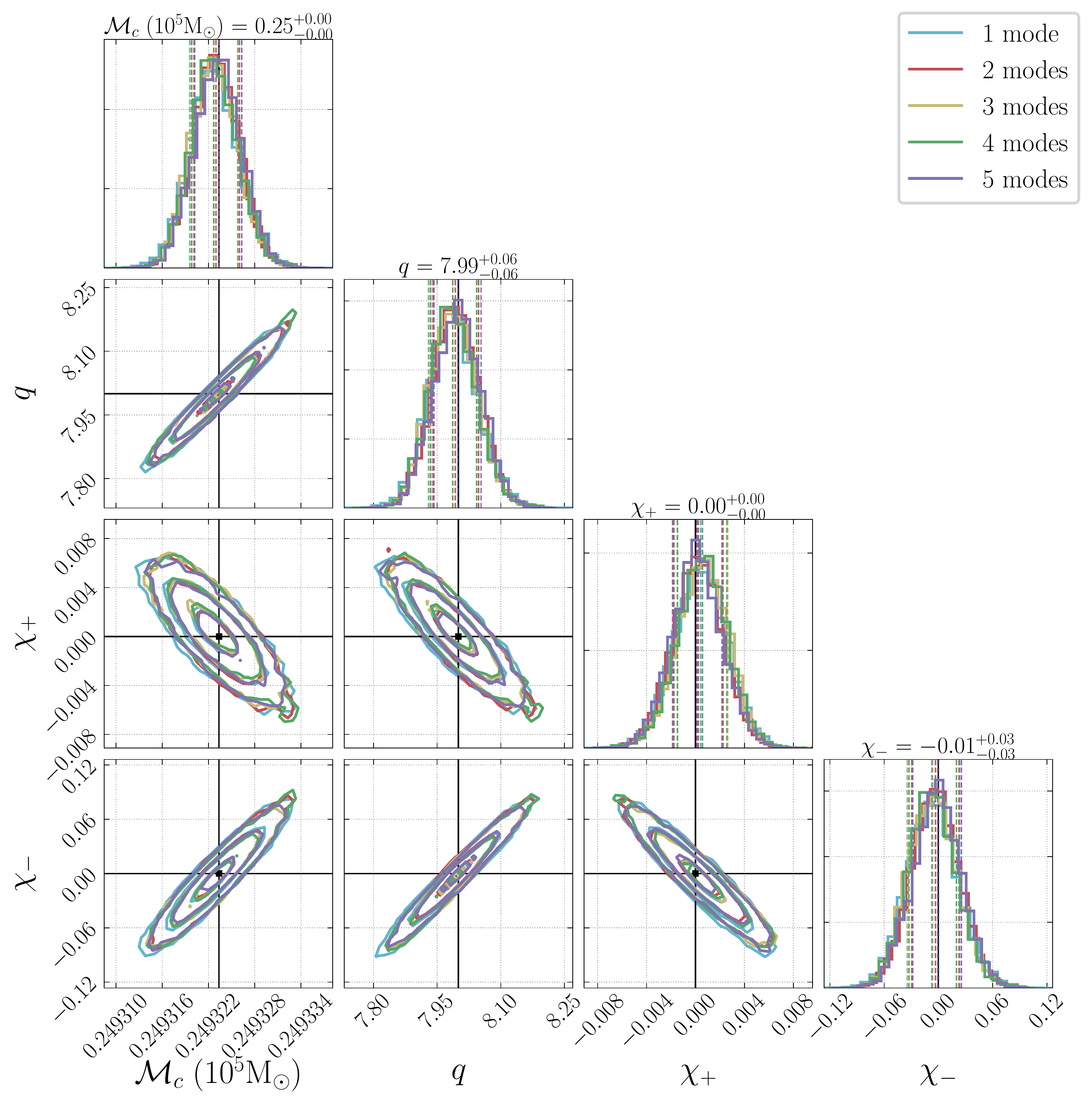} 
\includegraphics[width=0.42\textwidth]{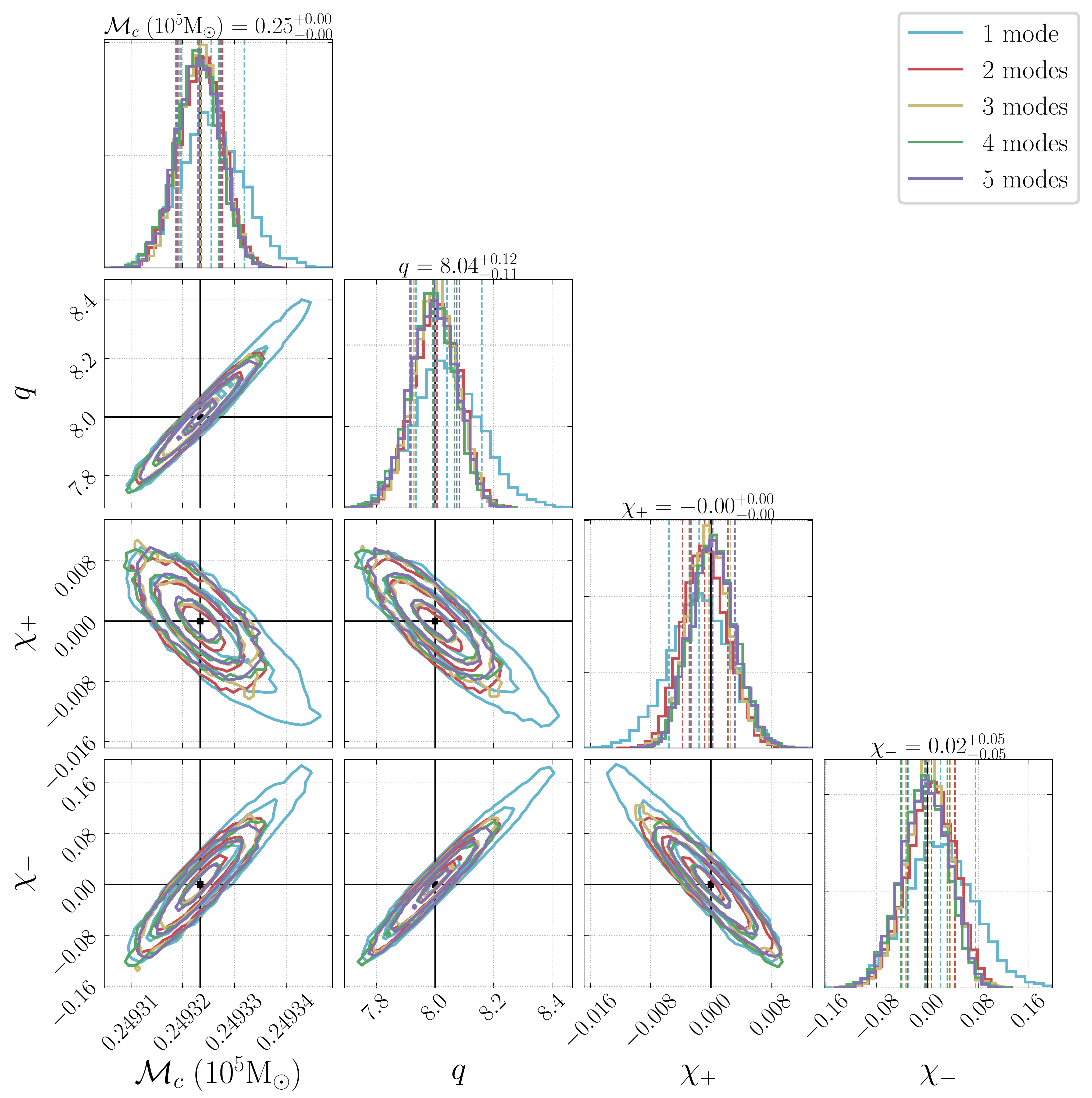} 
\includegraphics[width=0.42\textwidth]{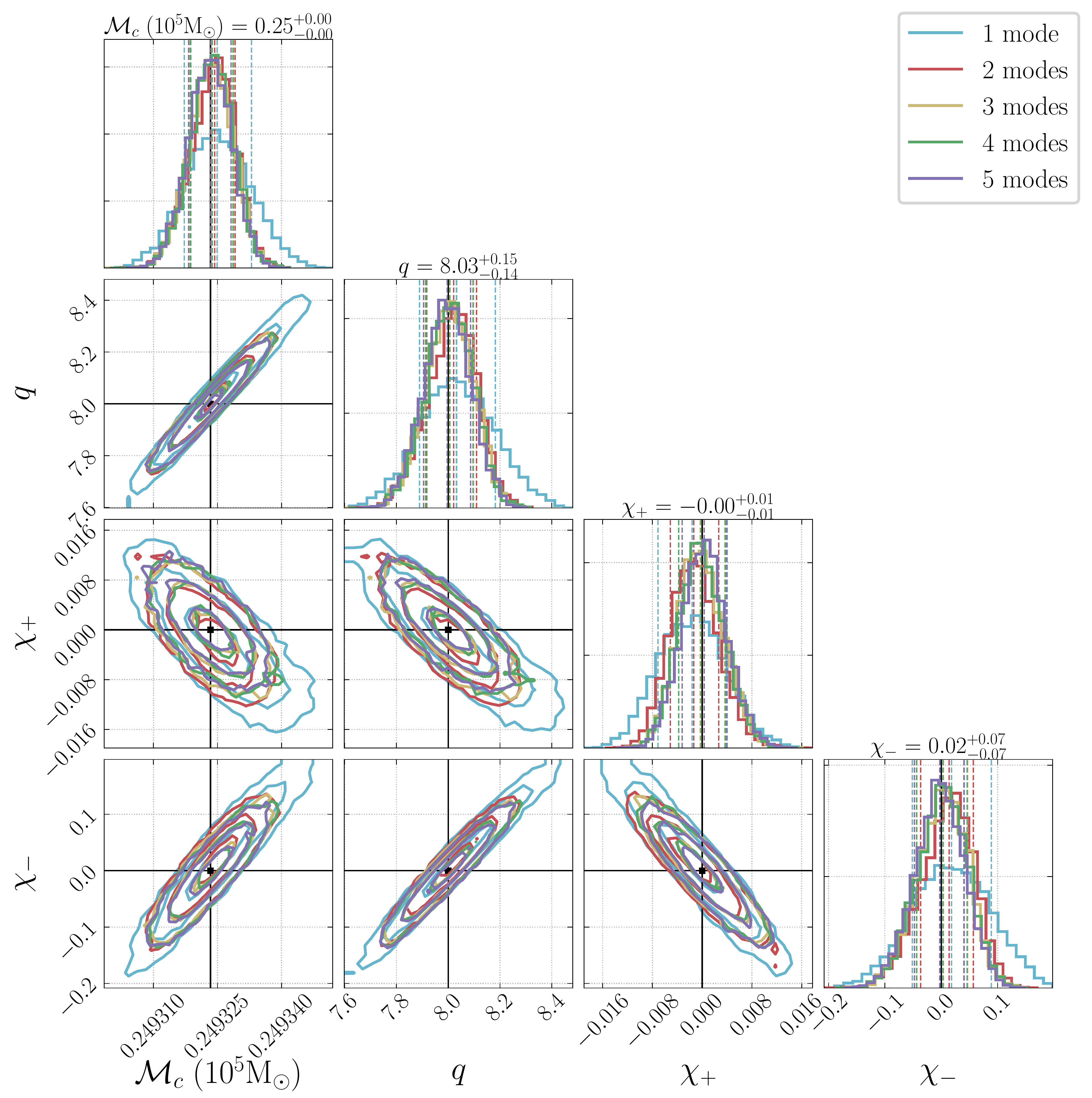} 
\caption{$M=10^5\, M_\odot, q=8$. Top, center, bottom: inclination and total SNR vary in the range $\iota=[\pi/12, \pi/3, \pi/2-\pi/12]$ and [433.7, 255.1, 188.6].}
\label{fig:pM5s0q8}
\end{figure}

The last set of $M=10^5\, M_\odot$ events we consider have mass ratio $q=8$ and posterior distributions shown in Fig.~\ref{fig:pM5s0q8}. The results are quite similar to those shown in Fig.~\ref{fig:pM5s0q4}, i.e., there are no significant biases and somewhat tighter constraints on parameters than was observed in Fig.~\ref{fig:pM5s0q1}. 

To summarize Secs.~\ref{sec:M1e5q1}--\ref{sec:M1e5q8}, we see that the biases on the intrinsic parameters for this set of lowest-mass binaries are fairly negligible, and are certainly much less severe than the biases observed for the considerably heavier MBHB system examined in Ref.~\cite{Pitte:2023ltw}. 

\subsection{MBHBs with \texorpdfstring{$M=10^6\, M_\odot$}{M=1e6~Msun}}\label{sec:M1e6}

We now consider events with $M=10^6\, M_\odot$. We will see that, as expected, the biases from insufficient mode content in the templates are more significant for these more massive events, which have total SNR in the range of $873.1-4105.2$, compared to the much lower SNRs observed in Sec.~\ref{sec:M1e5}. %

\subsubsection{Events with \texorpdfstring{$q=1.1$}{q=1.1}}

\begin{figure}[tbp]
\centering
\includegraphics[width=0.42\textwidth]{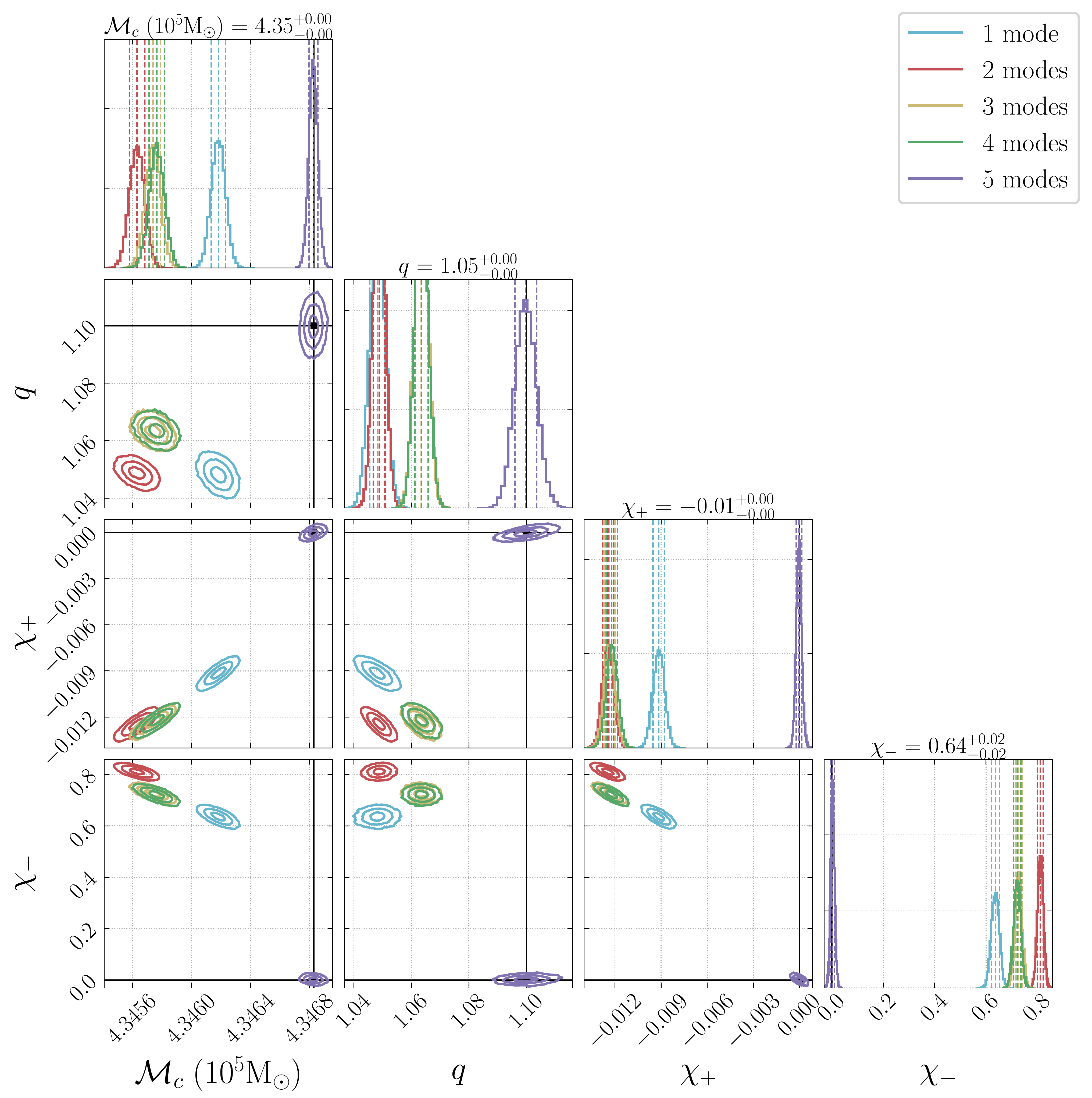} 
\includegraphics[width=0.42\textwidth]{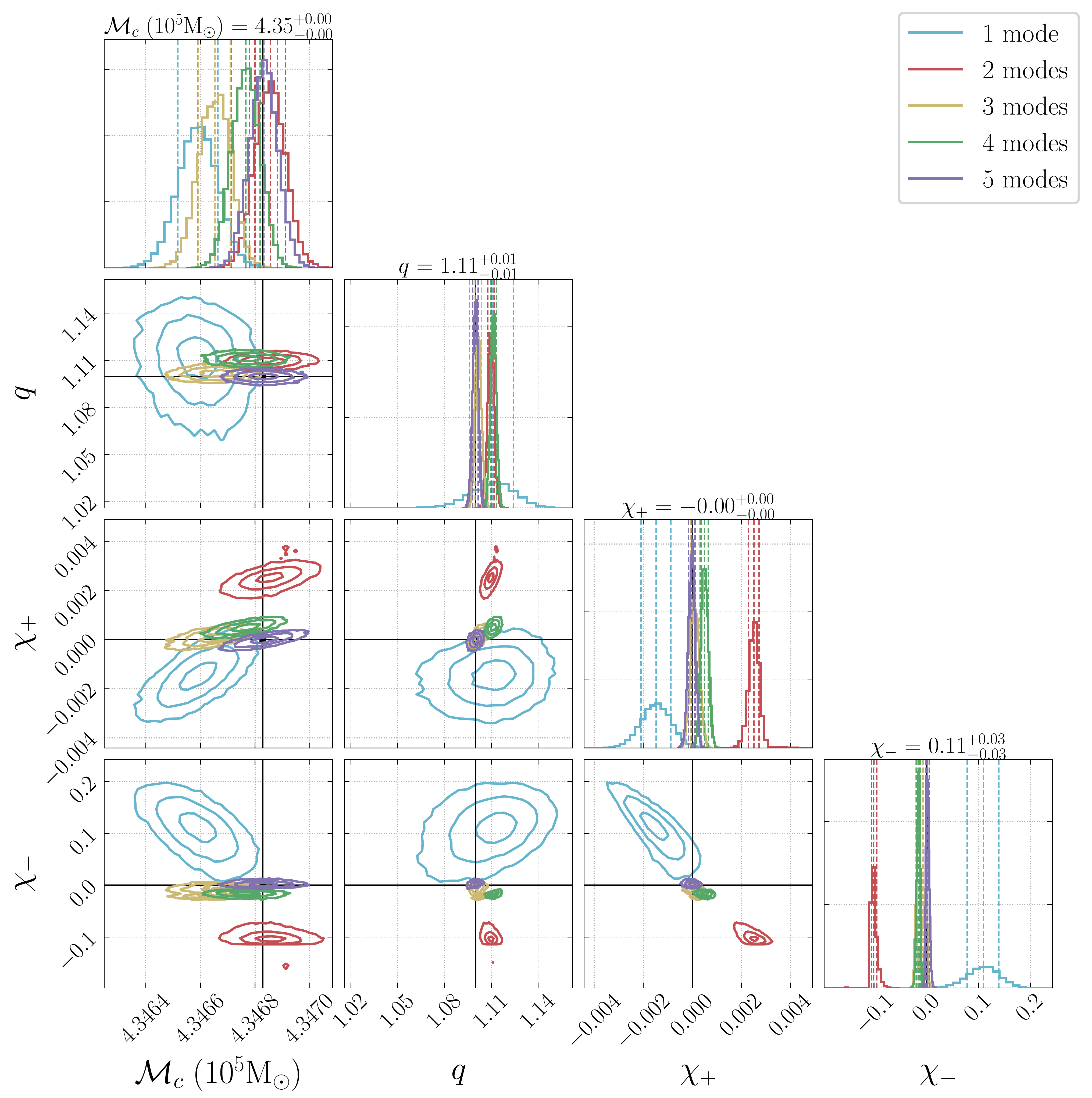}
\includegraphics[width=0.42\textwidth]{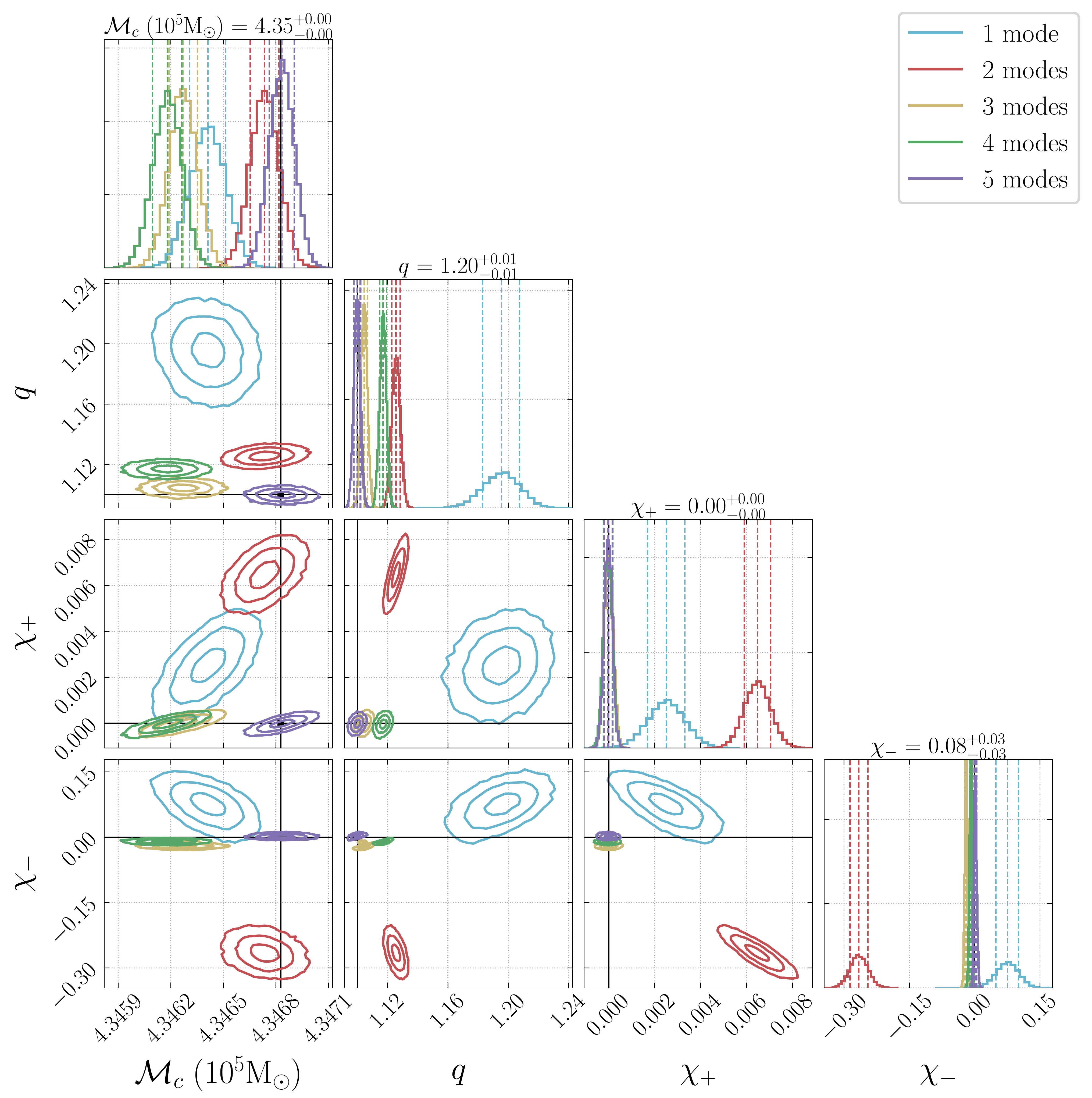} 
\caption{$M=10^6\, M_\odot, q=1.1, \iota=[\pi/12,\pi/3, \pi/2-\pi/12]$, total SNR = [4105.2, 2317.3, 1678.3]. %
}
\label{fig:pM6s0q1}
\end{figure}

Beginning with the nearly equal-mass events (see Fig.~\ref{fig:pM6s0q1}), we observe significant biases in parameter recovery with less than five modes in the template. We note that for the nearly face-on event, the posteriors on the chirp mass and spin parameters recovered by the (2,~2)-only template are all closer to the injected value than posteriors with 2, 3, and 4 modes. %
This is a feature we also see in the nearly face-on events with other mass ratios (see top panels of Figs.~\ref{fig:pM6s0q4} and~\ref{fig:pM6s0q8}). We recall that at inclinations close to zero, the radiation is overwhelmingly in the quadrupole, with additional modes (particularly the $m\neq2$ modes) being very subdominant. 

For the events with other inclination angles, parameter recovery with just the quadrupole is again sometimes apparently less biased than recovery with at least two modes; however, the posteriors are also considerably broader with the 1-mode template. %

\subsubsection{Events with \texorpdfstring{$q=4$}{q=4}}
\begin{figure}[tbp]
\centering
\includegraphics[width=0.42\textwidth]{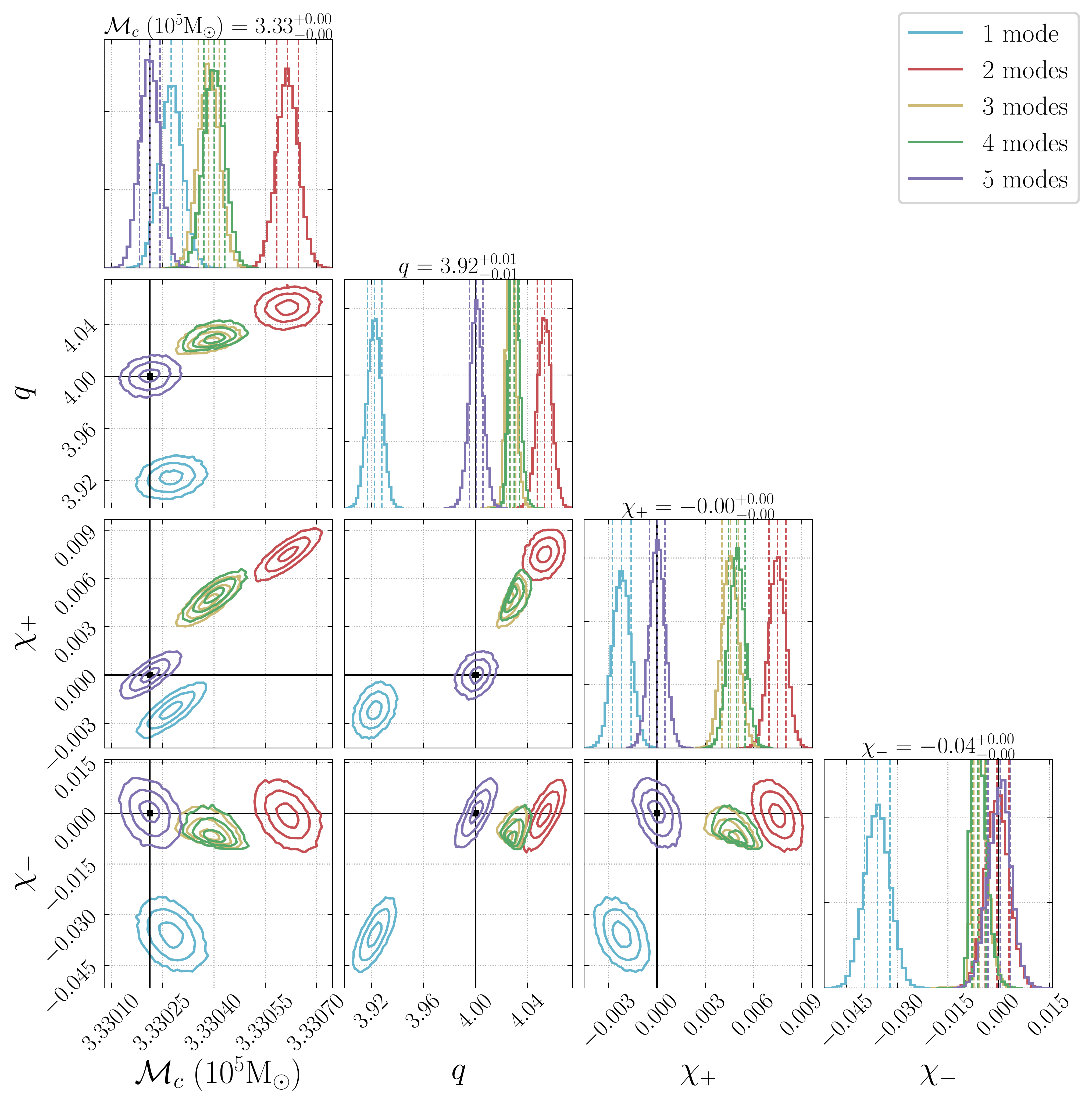} 
\includegraphics[width=0.42\textwidth]{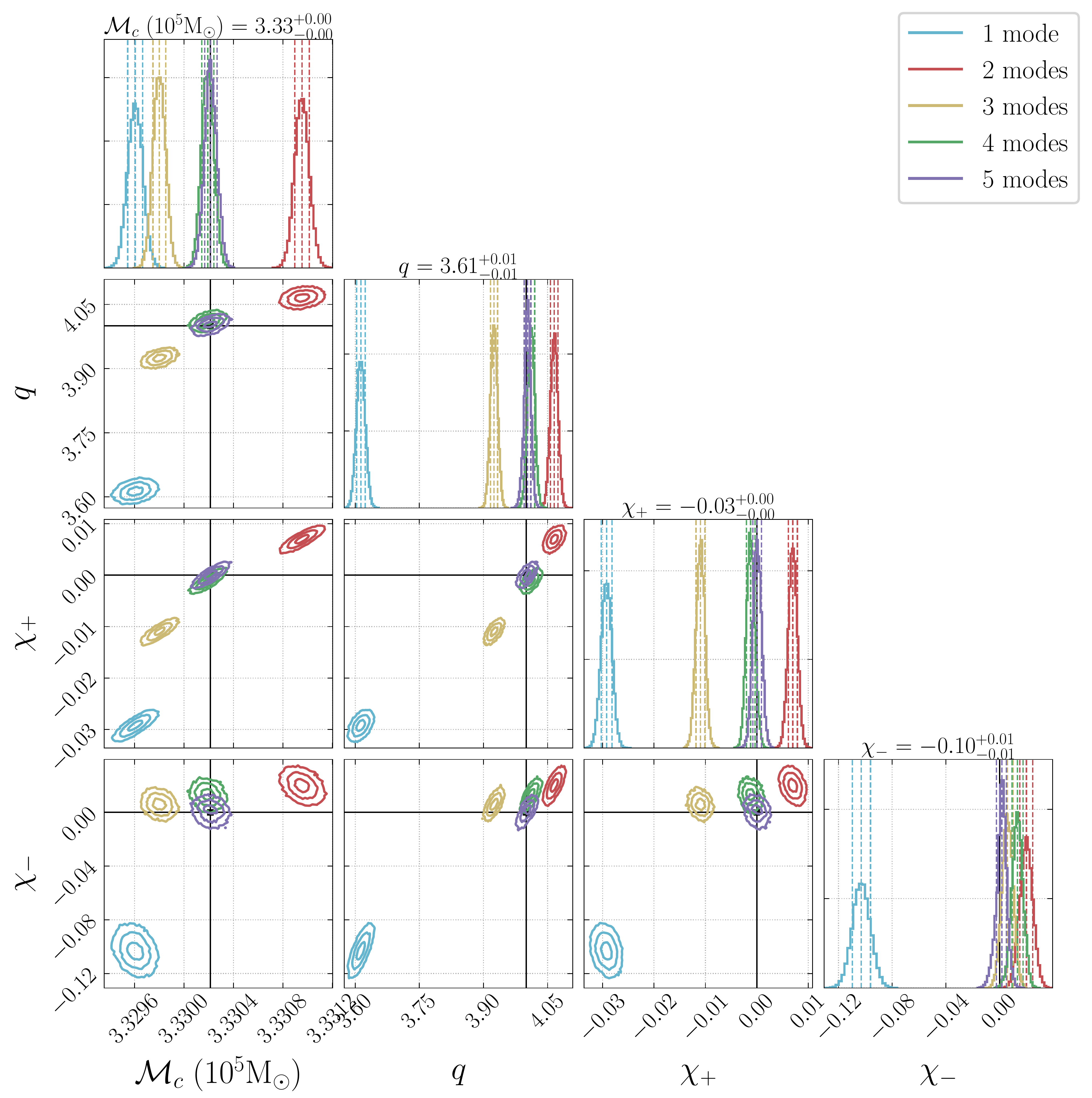} 
\includegraphics[width=0.42\textwidth]{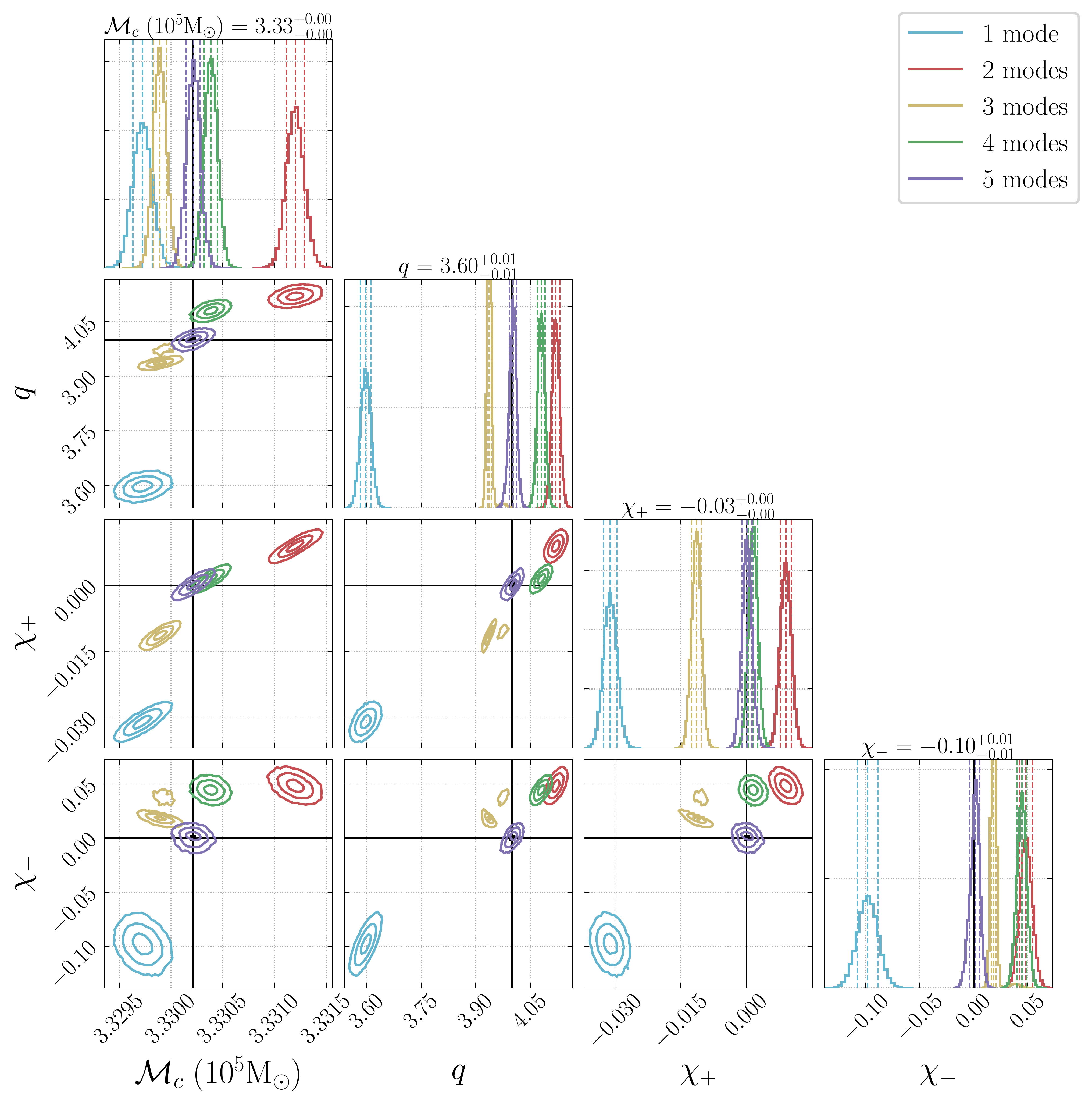} 
\caption{$M=10^6\, M_\odot, q=4, \iota=[\pi/12,\pi/3, \pi/2-\pi/12]$, total SNR = [2777.1, 1658.0, 1228.5].}
\label{fig:pM6s0q4}
\end{figure}
Compared to the $M=10^6\, M_\odot, q=1.1$ events, the width of the posteriors for events with $q=4$ shown in Fig.~\ref{fig:pM6s0q4} is generally more consistent across different mode configurations. Here, we see the expected trend a bit more clearly for the $\iota=\pi/3$ and $\iota\approx\pi/2$ events, in the sense that the 1-mode template generally results in more biased posteriors than templates with more modes. For the $\iota=\pi/12$ event, we again see that the quadrupole-only template can perform better than templates with 2, 3, and 4 modes, particularly in recovering the injected values for $\mathcal{M}_c$ and $q$. 
Again, this could be a consequence of the fact that the (2,~2) mode is considerably more dominant for face-on events.

\subsubsection{Events with \texorpdfstring{$q=8$}{q=8}}

\begin{figure}[tbp]
\centering
\includegraphics[width=0.42\textwidth]{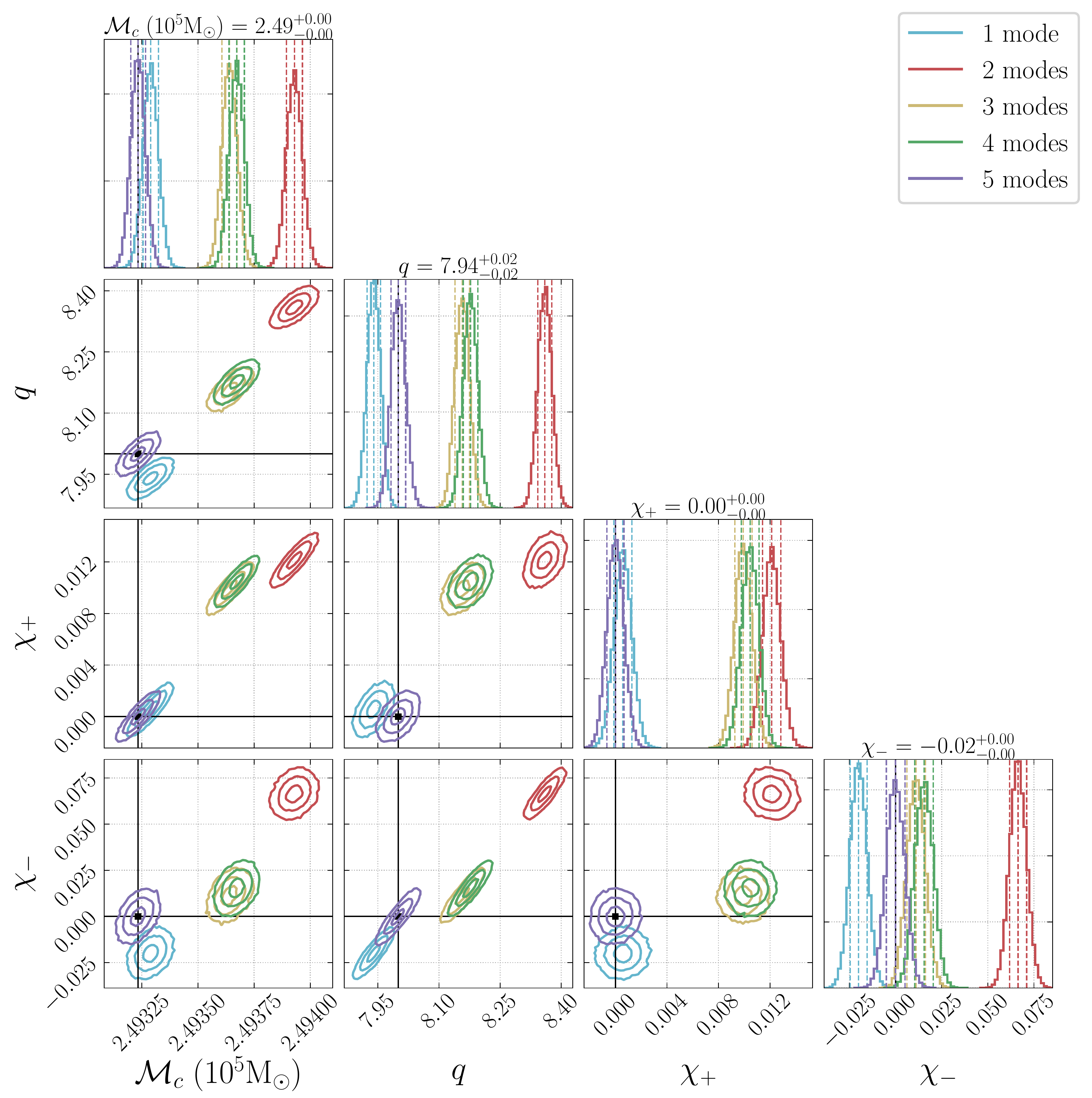} 
\includegraphics[width=0.42\textwidth]{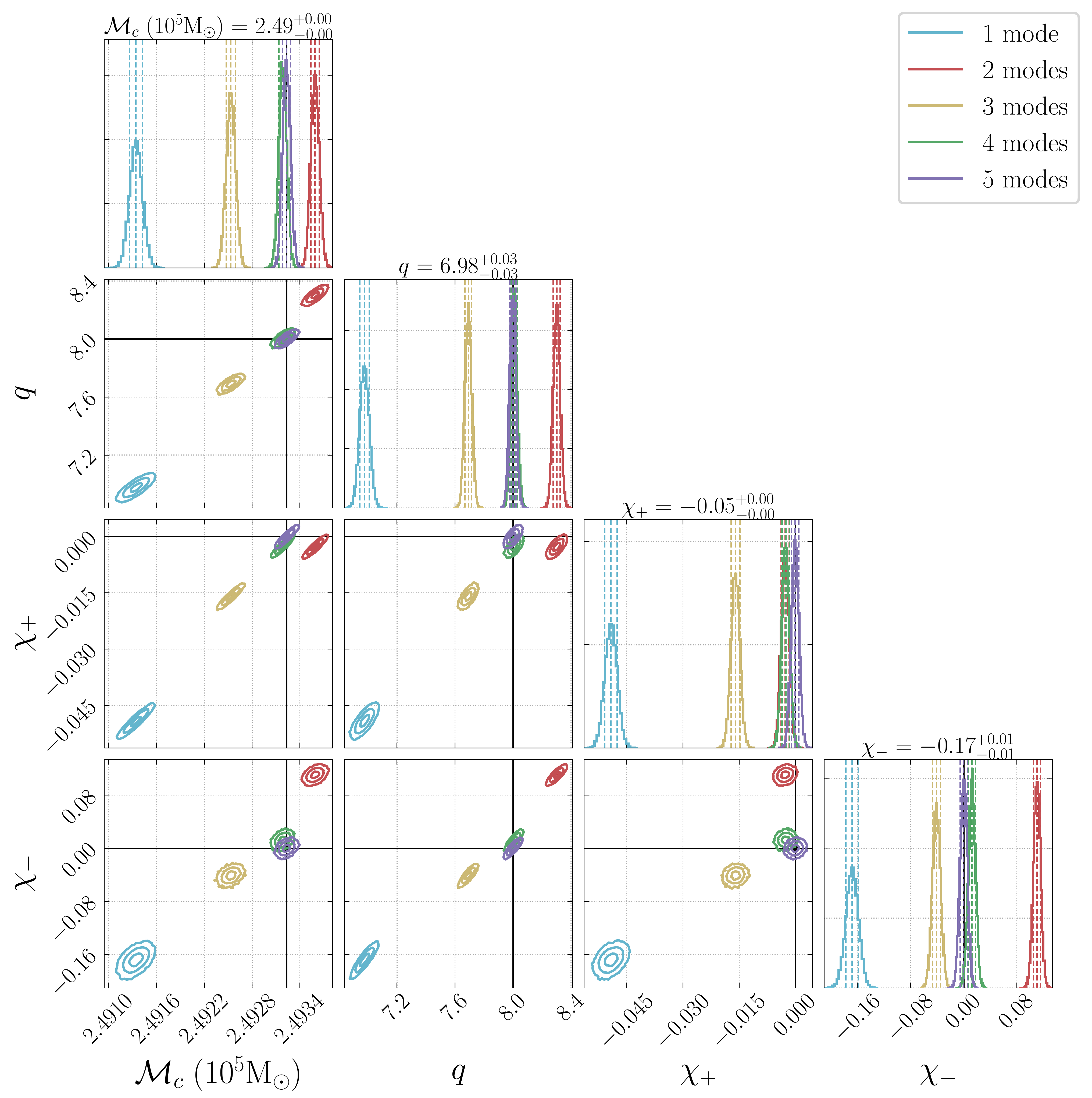} 
\includegraphics[width=0.42\textwidth]{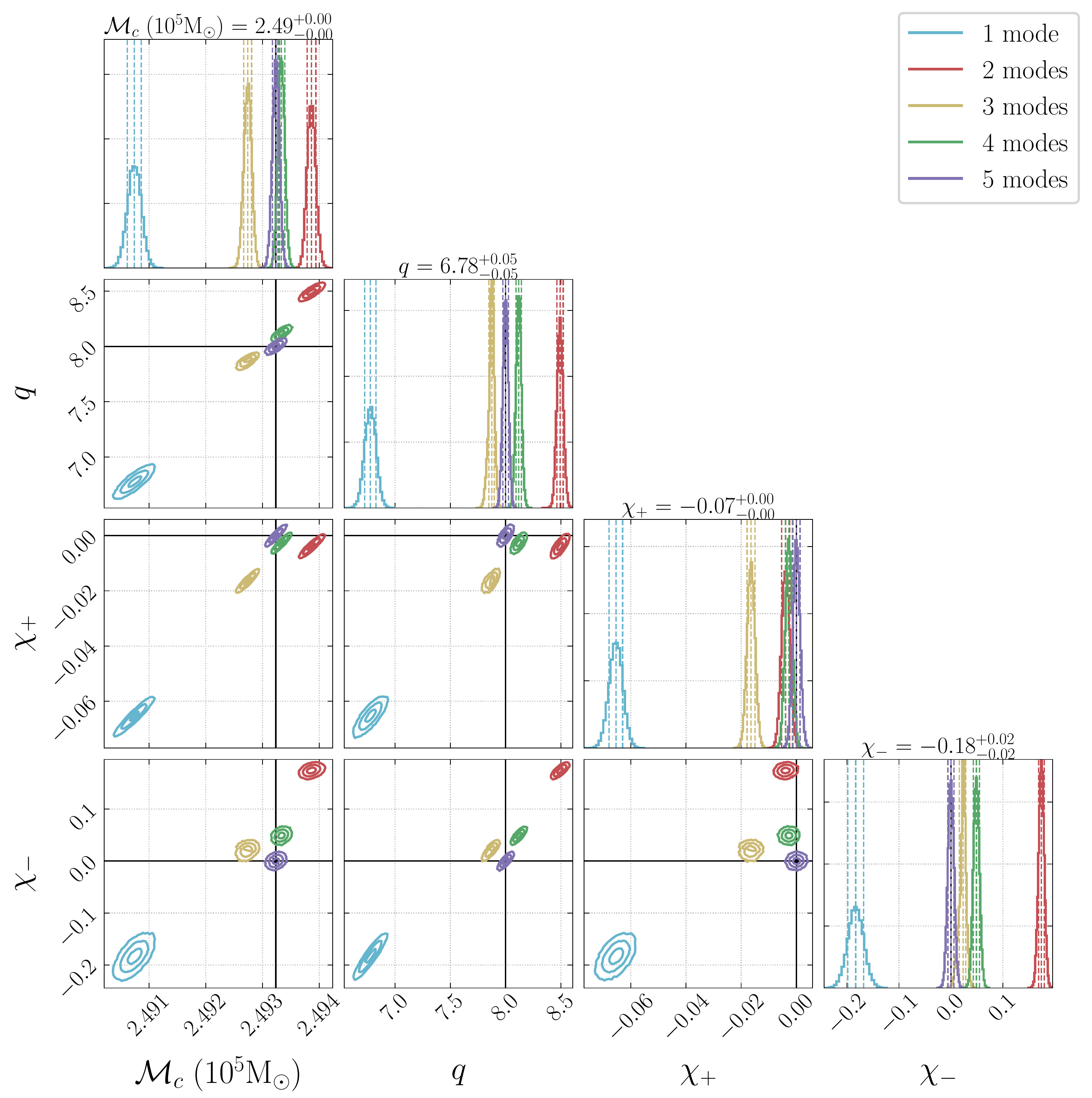} 
\caption{$M=10^6\, M_\odot, q=8, \iota=[\pi/12,\pi/3, \pi/2-\pi/12]$, total SNR = [1876.8, 1161.1, 873.1].}
\label{fig:pM6s0q8}
\end{figure}
The trends for the $M=10^6\, M_\odot,\, q=8$ events are largely the same as the trends for $q=4$ and the same total mass, only the biases are noticeably more pronounced, especially at larger inclination angles. This is consistent with what we expect: the higher mode content is generally more significant for more asymmetric binaries. 

As with the $q=4$ and $q=1.1$ events, we see that the 1-mode template performs better than the 2, 3, and 4 mode templates in the $\iota=\pi/12$ case.

\subsection{MBHBs with \texorpdfstring{$M=3\times10^5\, M_\odot$}{M=3e5~Msun}}\label{sec:M3e5}
\begin{figure}[tbp]
\centering
\includegraphics[width=0.42\textwidth]{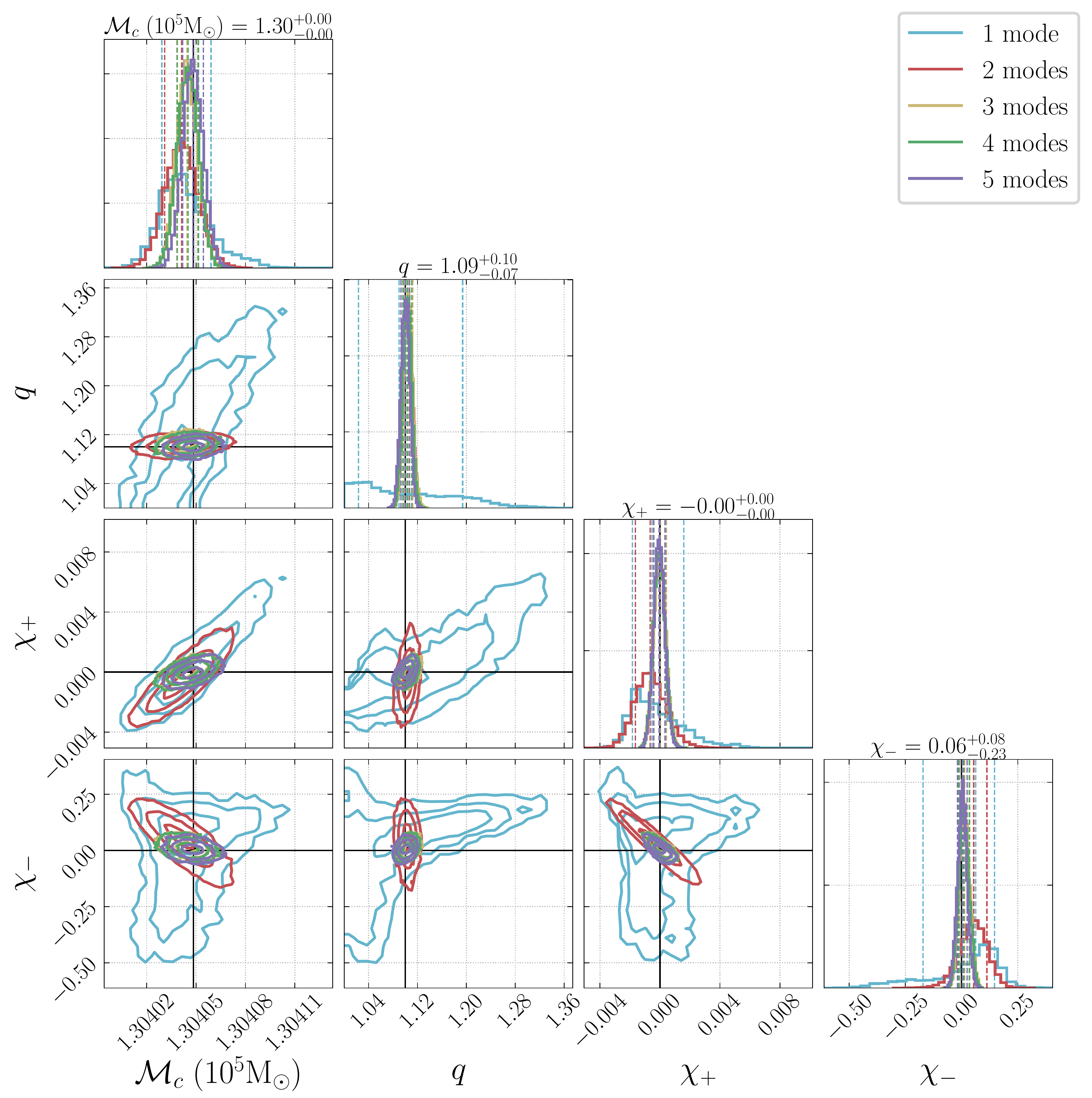} 
\includegraphics[width=0.42\textwidth]{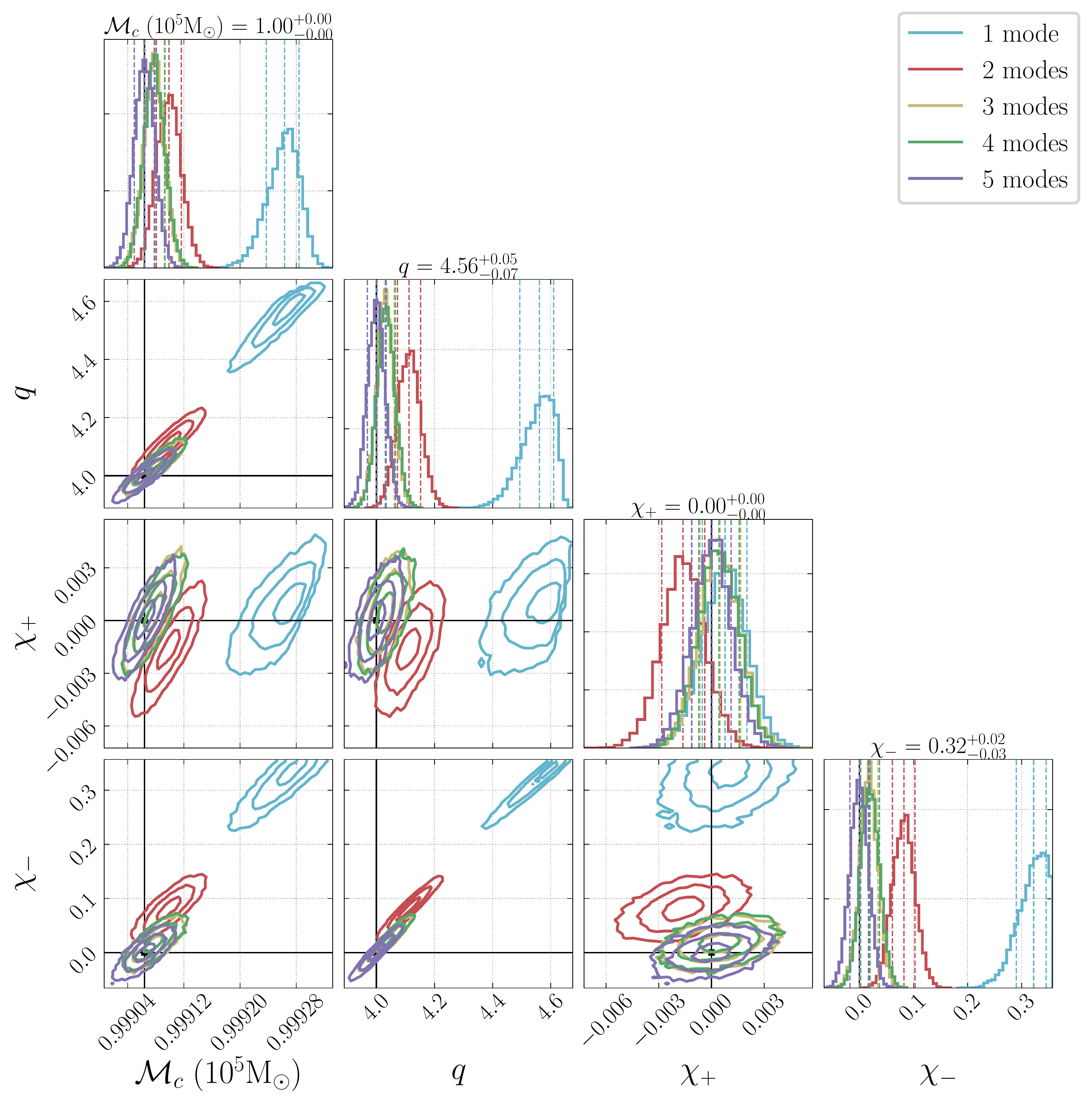} 
\includegraphics[width=0.42\textwidth]{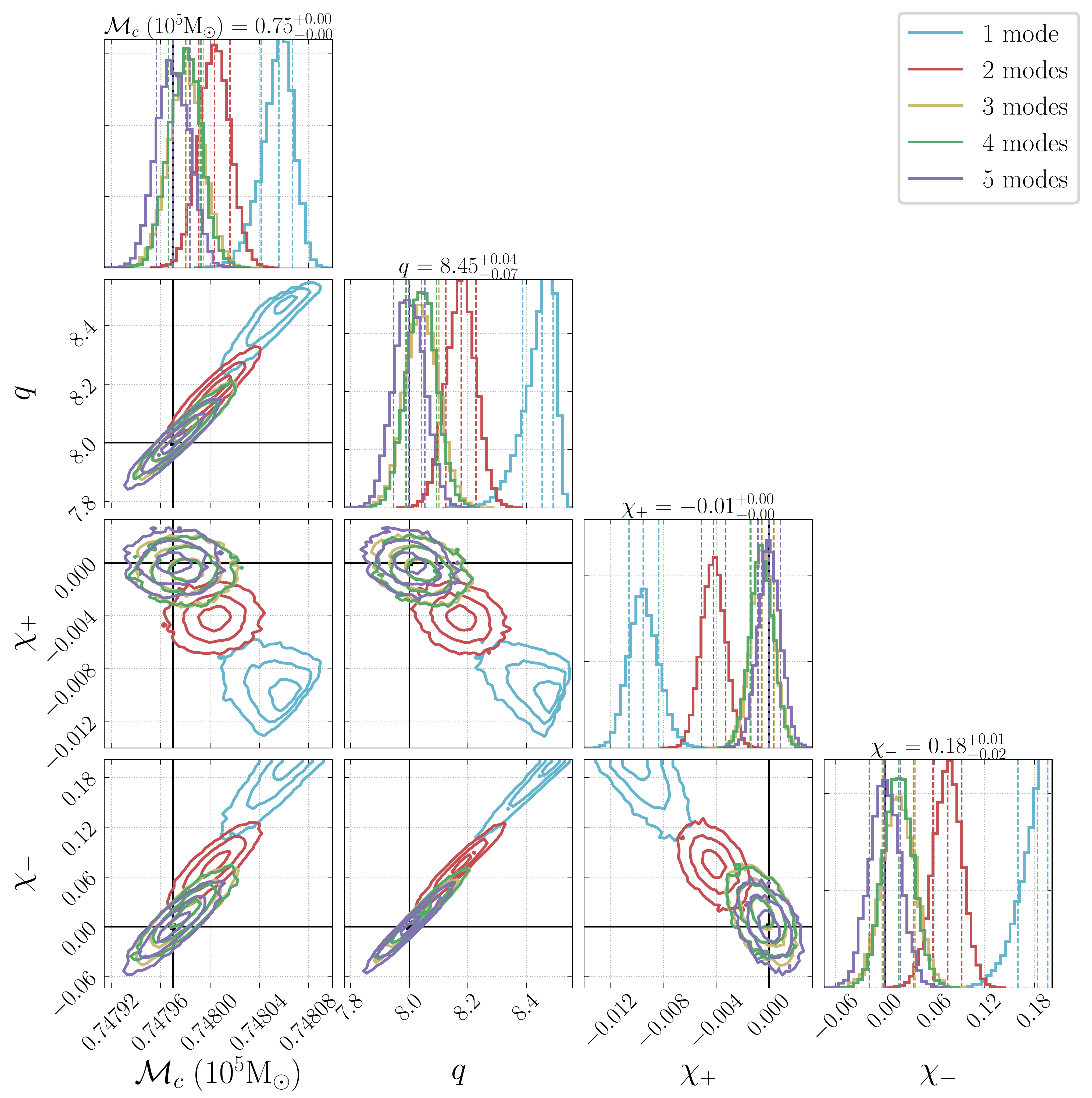} 
\caption{$M=3\times10^5\, M_\odot, \iota=\pi/3, q=[1.1,4,8]$, total SNR = [932.4, 729.5, 563.2].}
\label{fig:pM3e5}
\end{figure}

Noting the stark difference between biases for the $10^5\, M_\odot$ and $10^6\, M_\odot$ total mass binaries, we performed PE on a few events with intermediate masses to see if we could determine when biases begin to appear. The PE results for events with total mass $M=3\times10^5\,M_\odot$ are shown in Fig.~\ref{fig:pM3e5}. We performed all of the intermediate-mass PE runs at $\iota=\pi/3$. 

We first note that the general behavior is as we expect: the posteriors are tighter (looser) and biases are worse (better) than they were with the $10^5\, M_\odot$ ($10^6\, M_\odot$) events. Moreover, the performance with different mode configurations is quite consistent with what we expect: the 1-mode template performs worst, then the 2-mode template, then 3-mode, then 4-mode, and then (as always) we recover the injected signal with all 5 modes. We note that the posteriors with one mode are quite broad in the $q=1.1$ case, similar to what we found for the nearly symmetric binaries with $M=10^5\,M_\odot$. The performance for the $q=4$ and $q=8$ events are quite similar.

\section{Biases introduced in extrinsic parameter recovery}\label{sec:results_extrinsic}

We now observe how changing the mode content of the templates affects the recovered posteriors on the extrinsic parameters. The sky localization of events detected by LISA will be crucial for performing follow-up electromagnetic observations of MBHB systems. Previous studies have shown that LISA's inference of a system's sky position can be multimodal, and is strongly informed by higher harmonics, particularly at high masses~\cite{Marsat:2020rtl}. As a consequence, mismodeling the higher-modes content of the signal can be expected to lead to biases. This contrasts with LVK observations, where most of the information about the sky position comes from triangulating the time of arrival in different detectors, which is mostly independent of the physical content of the waveforms. 
We attempt to understand how biases can change with differing waveform mode content. 

A few representative corner plots illustrating our inference of extrinsic parameters are given in Fig.~\ref{fig:extrinsic}. Below, we highlight a few patterns that we observe: 

\begin{figure*}[t]
\centering
\includegraphics[width=0.48\textwidth]{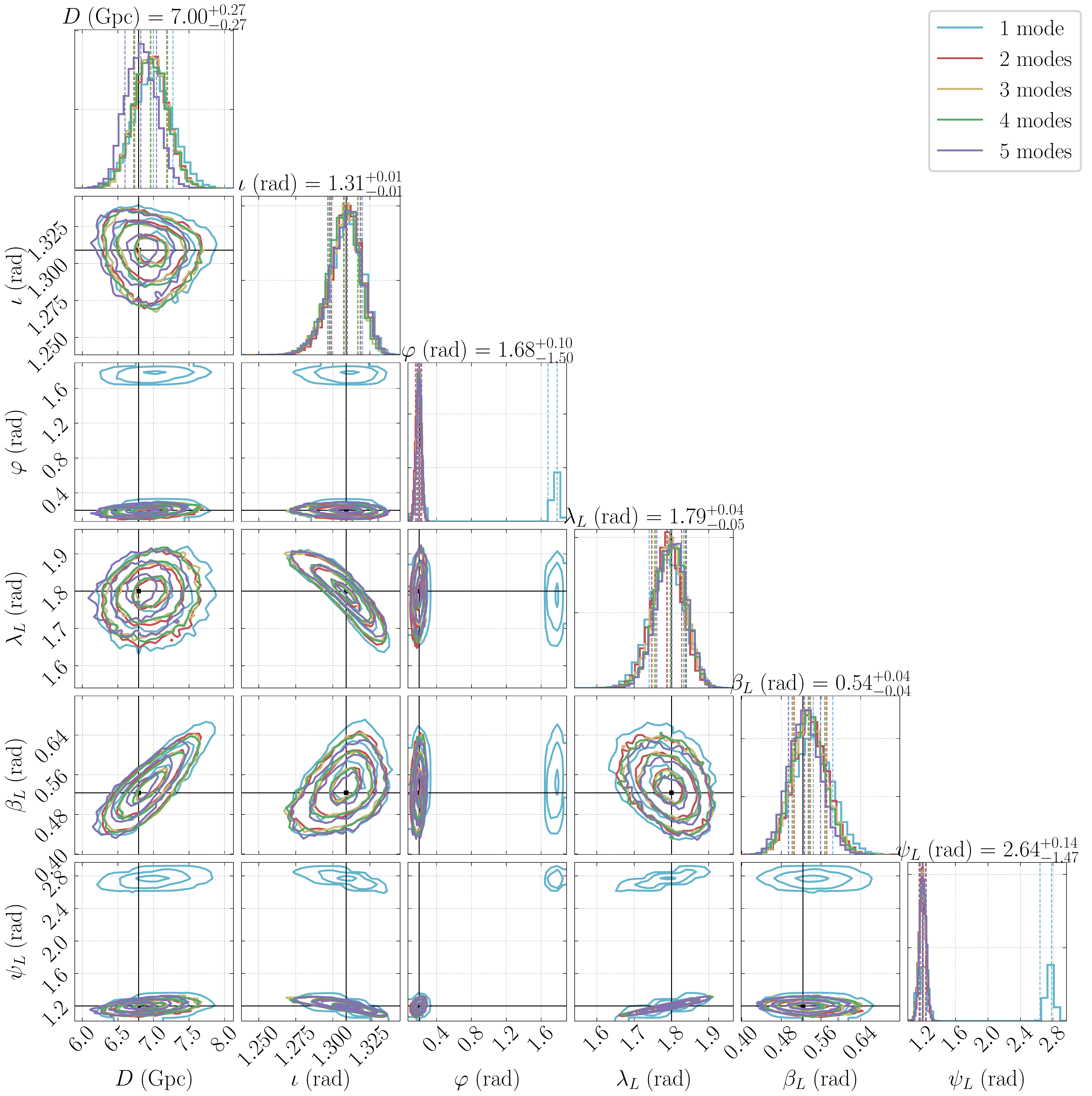} 
\includegraphics[width=0.48\textwidth]{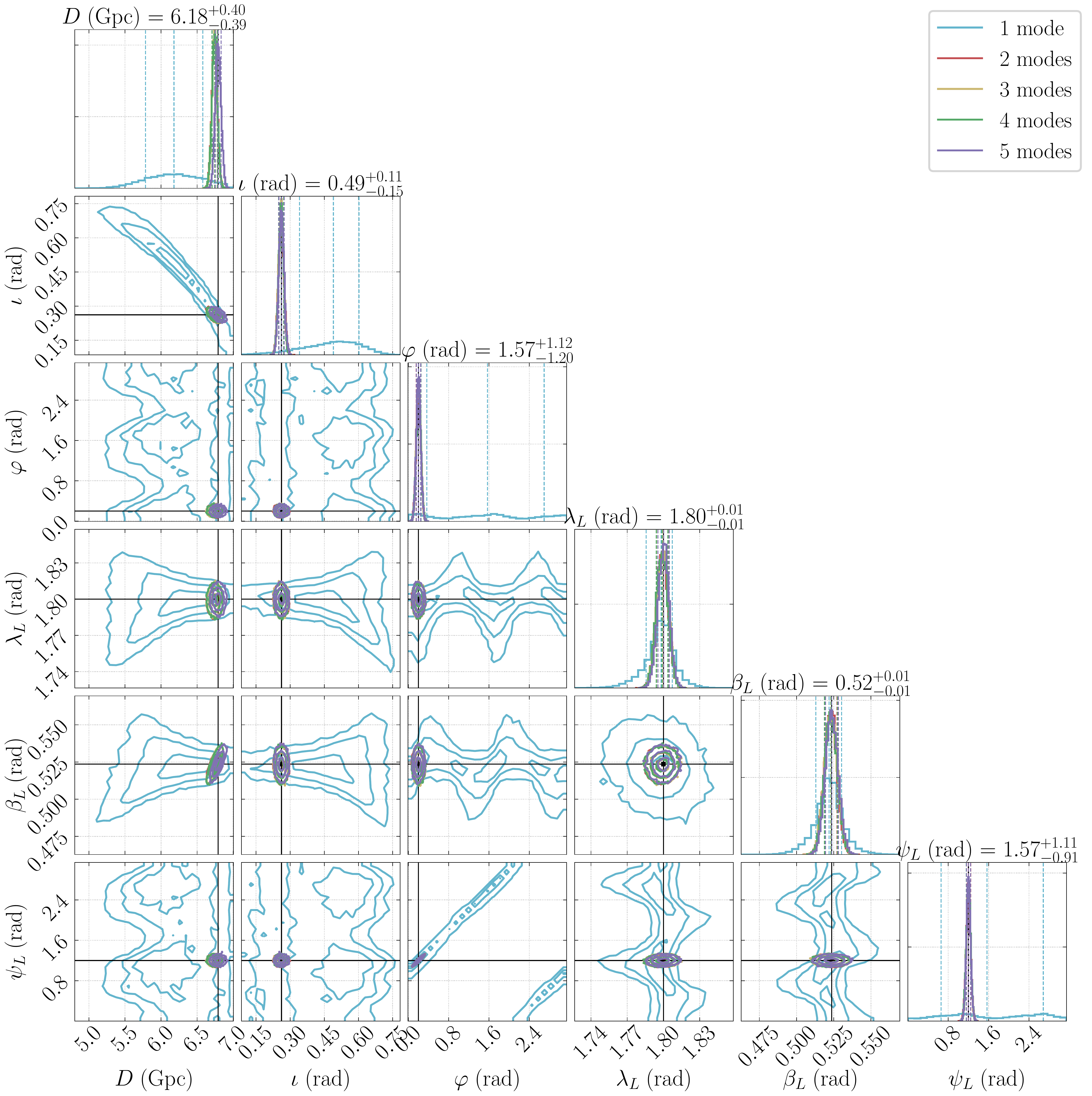} 
\includegraphics[width=0.48\textwidth]{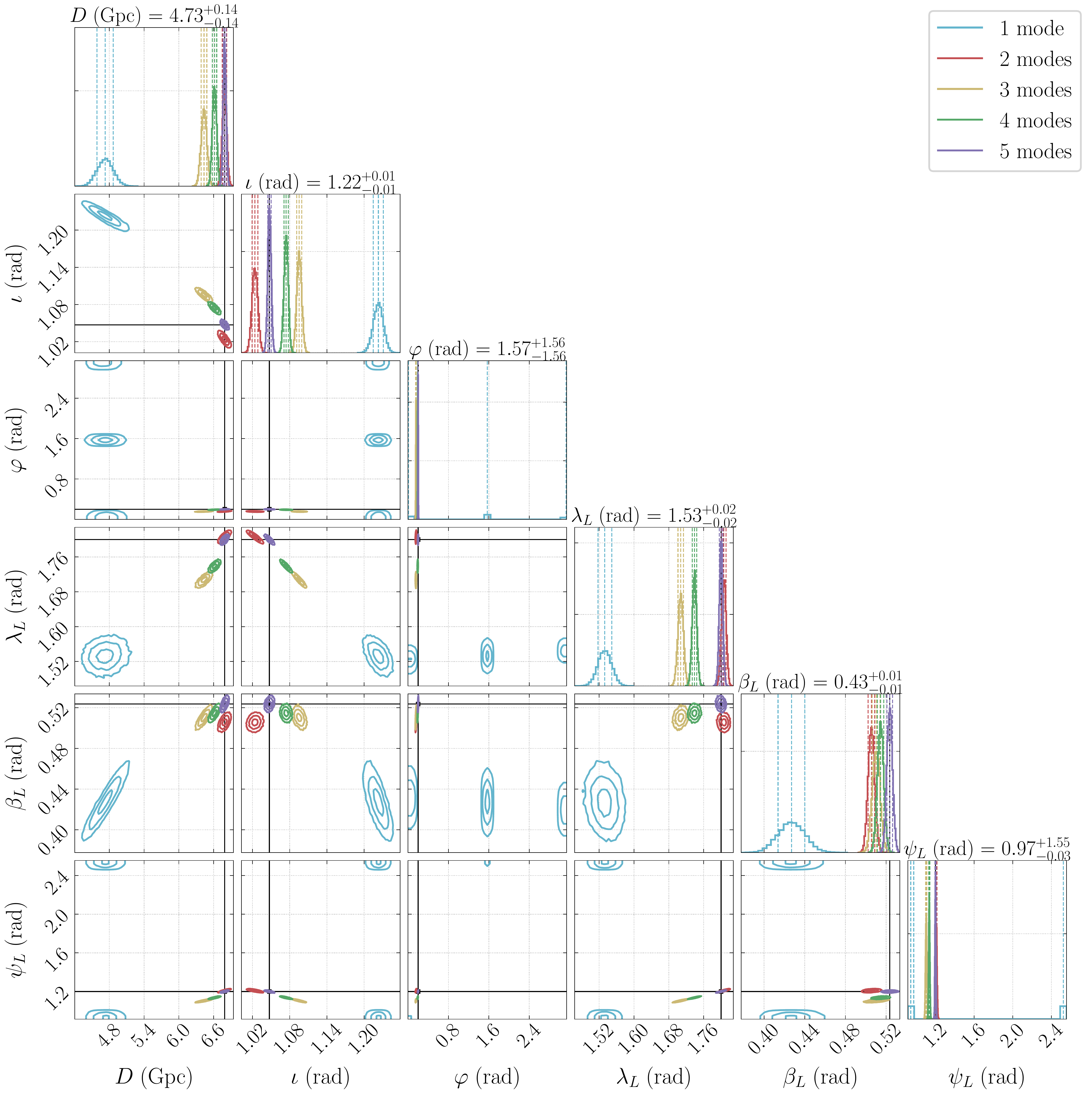} 
\includegraphics[width=0.48\textwidth]{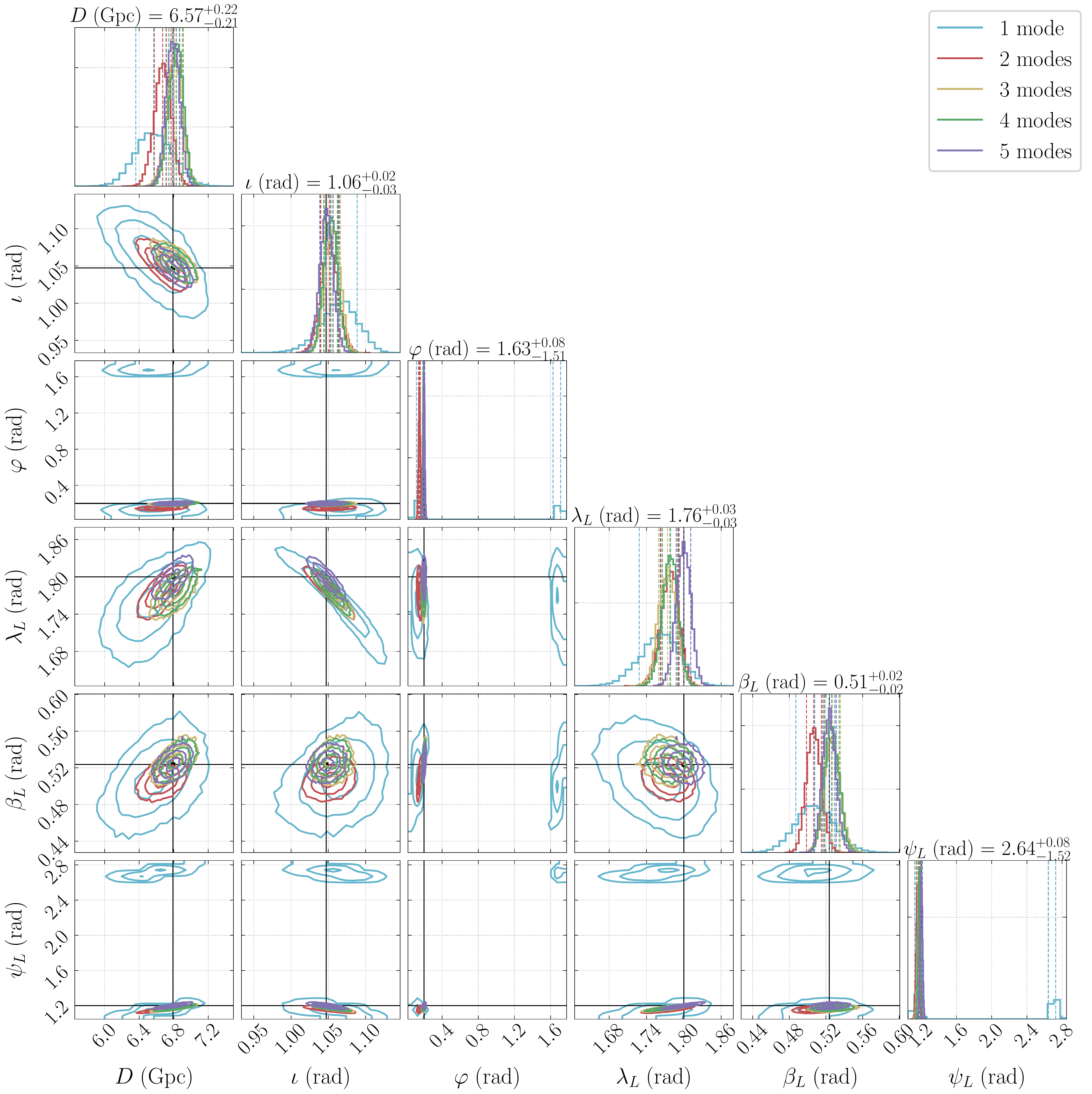} 
\caption{Extrinsic parameter recovery for selected events. Top left: $M=10^5\, M_\odot, \iota=\pi/2-\pi/12, q=4$.
Top right: $M=10^5\, M_\odot, \iota=\pi/12, q=4$.
Bottom left: $M=10^6\, M_\odot, \iota=\pi/3, q=8$.
Bottom right: $M=3\times10^5\, M_\odot, \iota=\pi/3, q=8$. 
}
\label{fig:extrinsic}
\end{figure*}

\begin{enumerate}
    \item The extrinsic parameters for all the $10^5\, M_\odot$ events with either $\iota=\pi/3$ or $\iota\approx\pi/2$ are recovered with very little bias for all mode configurations, and only degenerate in $\varphi$ and $\psi_L$ for the quadrupole-only templates, as expected. An example is given in the top left panel of Fig.~\ref{fig:extrinsic} ($10^5\, M_\odot,q=4,\iota\approx\pi/2$). The recovery with only the (2,~2) mode is noticeably worse for the nearly face-on events with this total mass (compare top left and right panels of Fig.~\ref{fig:extrinsic}). We have observed that, for a nearly face-on, equal-mass system with this lower total mass, higher-order modes become barely distinguishable, so that adding them to the waveform template does not change the PE significantly (but we do not show the corresponding plot for brevity).
    \item As for the intrinsic parameters, the biases in extrinsic parameter posteriors become severe for the $10^6\, M_\odot$ binaries (example shown in bottom left panel of Fig.~\ref{fig:extrinsic}), with injected values generally only being correctly recovered when all five modes are included in the template. Notably, however, adding just the (3,~3) mode still breaks the degeneracies in $\varphi$ and $\psi_L$ that exist in parameter recovery with just the (2,~2) mode (see point 4).
    \item As expected, the results for the $3\times10^5\, M_\odot$ events are somewhere in between cases 1 and 2, with moderate biases in extrinsic parameter recovery (see bottom right panel of Fig.~\ref{fig:extrinsic}). 
    \item While the degeneracy along constant lines of $\varphi\,+\,\psi_L$ and $\varphi\,-\,\psi_L$ noted in Ref.~\cite{Marsat:2020rtl} is evident in nearly all posteriors obtained with (2,~2)-mode only templates, this degeneracy is consistently lifted with the inclusion of even just one subdominant mode. This can be seen in any of the four panels of Fig.~\ref{fig:extrinsic}.
    \item The constraints on the source localization, i.e., the LISA-frame ecliptic longitude and latitude ($\lambda_L,\beta_L$) (among other parameters), improve drastically with the inclusion of one additional mode beyond the (2,~2). This behavior is also noted in Ref.~\cite{Marsat:2020rtl}. Note that not only is the bias significantly reduced when one mode is added, but the size of the statistical error (i.e., width of the contours) also decreases considerably. The precision of sky localization recovery does not change nearly as significantly between runs with multiple higher-order modes. This could be important information for rapid inference of an MBHB's sky position for, e.g., attempted electromagnetic follow-ups. We discuss this point further in Appendix~\ref{app:sky_loc}.
    \item Similarly, the distance-inclination degeneracy commonly observed with quadrupole-only templates is broken by the presence of even one additional mode. This is expected, given the different dependence of the different angular harmonics on the inclination angle. As with the sky position angles, the statistical error on $D_L$ also shrinks considerably with the addition of at least one mode beyond the (2,~2).%
\end{enumerate}

\section{Understanding systematic biases without full parameter estimation}\label{sec:CV_results}

We now turn our attention to using the CV and direct likelihood optimization approaches introduced in Secs.~\ref{sec:CV_intro} and~\ref{sec:nelder-mead} to quickly and cheaply estimate the sorts of biases observed in Secs.~\ref{sec:results} and~\ref{sec:results_extrinsic}. In Sec.~\ref{sec:compare_cheap_and_pe}, we discuss how well the two methods predict the biases seen in PE. In Sec.~\ref{sec:heatmaps}, we estimate the boundaries on parameter space where unbiased PE is possible using the direct likelihood optimization technique. %

\subsection{Comparison of cheap methods with parameter estimation}\label{sec:compare_cheap_and_pe}
We begin by showing how well the CV and direct likelihood optimization approaches are able to recover the systematic biases observed in PE. 

\begin{figure*}[t]
\centering
\includegraphics[width=0.95\textwidth]{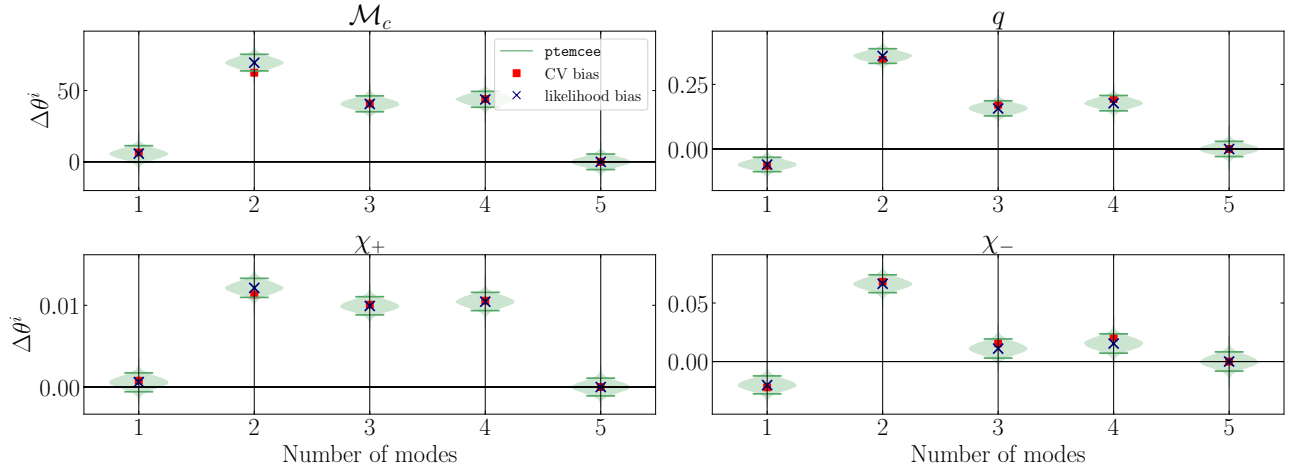}
\caption{Systematic bias on intrinsic parameters estimated within the linear signal approximation (``CV bias'') and with the direct likelihood optimization method (``likelihood bias'') compared to biases recovered in full PE (``\texttt{ptemcee}''). Results are shown for the $M=10^6\, M_\odot, q=8, \iota=\pi/12$ event. For this event, both ``cheaper'' methods are able to recover the biases in PE for intrinsic parameters, with direct likelihood optimization performing slightly better. }
\label{fig:CV-and-NM-int}
\end{figure*}

\begin{figure*}[t]
\centering
\includegraphics[width=0.95\textwidth]{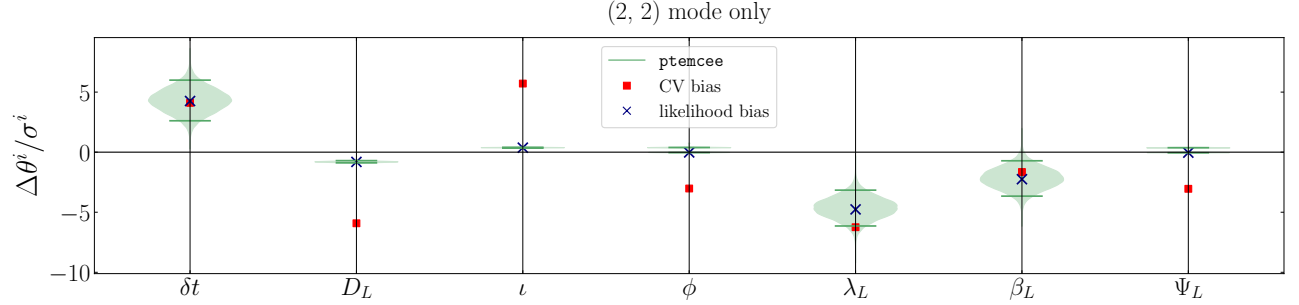}
\caption{Biases on extrinsic parameters for the same $M=10^6\, M_\odot, q=8, \iota=\pi/12$ event shown in Fig.~\ref{fig:CV-and-NM-int}, as recovered with the (2,~2)-mode-only template. Note that for visualization purposes, we plot $\Delta \theta^i/\sigma^i$ %
in this figure, as opposed to $\Delta \theta^i$ as in Fig.~\ref{fig:CV-and-NM-int}.} %
\label{fig:CV-and-NM-ext-22}
\end{figure*}

In Fig.~\ref{fig:CV-and-NM-int}, we plot the biases on intrinsic MBHB parameters as estimated in the CV approximation (red squares), with the direct likelihood optimization method (navy crosses), and as recovered in our PE runs (green violin plots, with 5\% and 95\% quantiles marked by the green lines). Each panel shows results for a single MBHB parameter, with five different plots per panel corresponding to different numbers of modes in the waveform template as indicated on the $x$-axis. In this figure, we show results for one of the events with significant systematic biases ($M=10^6\, M_\odot,q=8,\iota=\pi/12$). 

Both the CV and direct likelihood optimization methods recover the biases in PE fairly well in Fig.~\ref{fig:CV-and-NM-int}, with direct likelihood optimization sometimes recovering the center of the posteriors from PE with slightly better accuracy. Both methods likewise perform fairly well in recovering biases on the extrinsic parameters of this event (again, with direct likelihood optimization performing slightly better), except in the (2,~2)-mode-only case. We show this case in Fig.~\ref{fig:CV-and-NM-ext-22}, where we plainly see that direct likelihood optimization performs better than the CV approach. The rest of the extrinsic parameter results for this event are given in Appendix~\ref{app:CV_cont.}. %

We find this superior performance of the direct likelihood optimization method in cases of severe bias to generally be true, particularly for nearly equal-mass events. \begin{figure*}[t]
\centering
\includegraphics[width=0.9\textwidth]{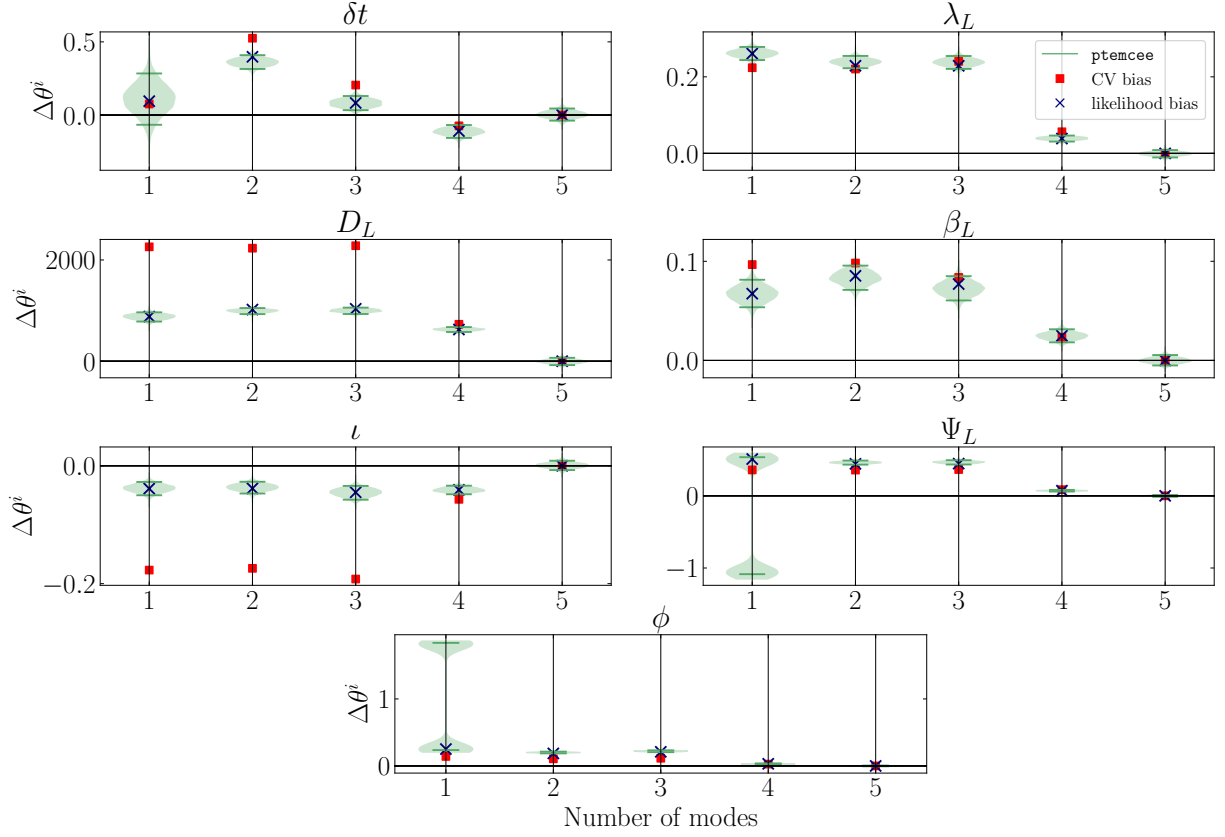}
\caption{Biases on extrinsic parameters for the $M=10^6\, M_\odot, \iota=\pi/3, q=1.1$ event. Direct likelihood optimization performs significantly better than CV in this case.} %
\label{fig:CV-and-NM-ext-2}
\end{figure*}
To illustrate this further, in Fig.~\ref{fig:CV-and-NM-ext-2} we plot the biases on extrinsic parameters for the $M=10^6\, M_\odot, \iota=\pi/3, q=1.1$ event. Here, it is evident that the biases are recovered much more accurately when we directly maximize the likelihood than when we use the linear signal approximation. This is also true for the intrinsic parameters (not shown), although the CV performance is not quite as bad. 

Notably, calculating the bias with direct likelihood optimization for the events shown in Fig.~\ref{fig:CV-and-NM-int} and Fig.~\ref{fig:CV-and-NM-ext-2} took 11 seconds and 34 seconds, respectively, when running in a \texttt{Jupyter Notebook} on a single core. The results we show in these figures were obtained starting with an initial simplex with parameters set 20$\sigma$ away from their injected values, and when iterating the procedure 3 and 15 times, respectively. For all but one of the $M=10^6\, M_\odot$ events, we found that with an initial simplex of this size (20$\sigma$), 15 iterations were generally more than enough to converge to similarly accurate estimations of the biases as the ones shown in Figs.~\ref{fig:CV-and-NM-int}--\ref{fig:CV-and-NM-ext-2}. For some events (e.g., the $q=4, \iota=\pi/3$ and the $q=8,\iota=\pi/3$ events), no iteration of the Nelder-Mead optimization algorithm was necessary; that is, running the algorithm only once was sufficient to find a maximum likelihood that agreed with PE. In such cases, the estimated bias calculations take as little as 9 seconds.

As mentioned previously, there is one case in which neither the CV approach nor the direct likelihood optimization is able to predict the systematic biases very well, and that is for the event with $M=10^6 \,M_\odot, q=1.1, \iota=\pi/12$. We note that this is the case for which the quadrupole is so dominant that interpreting the effect of adding higher-order modes is not straightforward. 

Before concluding this section, we note that the CV approach requires us to be able to take derivatives of the waveform with respect to the parameters: see Eqs.~\eqref{eqn:FIM} and~\eqref{eqn:CV_full}. As mentioned before, there are sometimes stability issues in differentiating the (3,~2) mode in the \textsc{IMRPhenomXHM} family, so including this mode in CV estimates would require careful monitoring of the differentiation at each step.
For this reason, the direct likelihood optimization via Nelder-Mead (which performs gradient-free minimization) has another advantage compared to the CV approach. 

\subsection{Using direct likelihood optimization to set boundaries on unbiased parameter space}\label{sec:heatmaps}

Having determined that directly optimizing the likelihood is the better method of cheaply estimating systematic biases due to neglecting higher-order modes, we proceeded to make the heatmap shown in the introduction (Fig.~\ref{fig:heatmap-inc3}) using this method. For each point of fixed total mass and mass ratio, we compute the systematic biases, $\Delta_{\rm{sys}}\theta^i$, using the direct likelihood optimization method. %
To account for the fact that the optimization algorithm occasionally fails, we draw a random redshift value from $z_{\rm{opt}}\in[1,10]$ until a redshift is found at which 15 steps of the optimization algorithm can be successfully completed. (Note: 15 steps were generally completed with the very first random redshift that was drawn, consistent with the good performance of optimization via the Nelder-Mead algorithm discussed in Sec.~\ref{sec:nelder-mead}). We stress that there is nothing physically significant about randomly drawing redshift values here; it is merely a convenient way to initialize the optimization process multiple times when necessary, that is, in case it occasionally fails. We find that there are some regions (in particular, around $4\leq q\leq7$) in which the optimization algorithm seems more sensitive to initial conditions. To mitigate this effect, we continue looking for values of $z_{\rm{opt}}$ at which the optimization is successfully completed until we find multiple initializations (values of $z_{\rm{opt}}$) which return the same values for the systematic biases. We then take such a systematic bias estimation to be ``good.''

We next compute the minimum redshift at which our PE is unbiased, i.e., where
\begin{eqnarray}
    \Delta_{\rm{sys}} \theta^i\, \leq 2\sigma^i\;,
    \label{eqn:unbiased_criterion}
\end{eqnarray}
with the statistical error on the $i^{\rm{th}}$ parameter,  $2\sigma^i$, computed via the Fisher information matrix. To find the redshift at which this occurs, we leverage the simple scaling of errors computed via Fisher analysis, i.e., 
\begin{eqnarray}
    \frac{\sigma^i(z_{\rm{opt}})}{\sigma^i(z)} = \frac{D_L(z_{\rm{opt}})}{D_L(z)} \,.
    \label{eqn:Fisher_z_scaling}
\end{eqnarray}
This relation holds for statistical errors on all parameters except $\theta^i = D_L$, in which case the right-hand side of Eq.~\eqref{eqn:Fisher_z_scaling} is squared. (On the other hand, the systematic bias on $D_L$ also scales with the distance, unlike the systematic biases on all other parameters. The final expression for minimum distance in Eq.~\eqref{eqn:unbiased_criterion_complete} therefore also ultimately holds for the bias on luminosity distance as well.) The luminosity distance at which we have unbiased parameter inference is then 
\begin{eqnarray}
    D_L(z) = \frac{\Delta_{\rm{sys}} \theta^i \,D_L(z_{\rm{opt}})}{2\sigma^i(z_{\rm{opt}})} \,,
    \label{eqn:unbiased_criterion_complete}
\end{eqnarray}
from which one can get the corresponding redshift. (The minimum redshift we consider is $z=0.1$; if the redshift determined via Eq.~\eqref{eqn:unbiased_criterion_complete} is less than this value, we simply set it to $z=0.1$.) For the sky localization, we look for the redshift at which

\begin{eqnarray}
    \begin{pmatrix}
        \Delta_{\rm{sys}} \lambda_L\\ 
        \Delta_{\rm{sys}} \beta_L
    \end{pmatrix}^{\rm{T}} 
    \Sigma^{-1}
    \begin{pmatrix}
        \Delta_{\rm{sys}} \lambda_L\\ 
        \Delta_{\rm{sys}} \beta_L
    \end{pmatrix}
    \leq \chi^2_{\rm{ppf}}(0.95,\rm{df}=2)\,,
    \label{eqn:sky_loc_uncertainty}
\end{eqnarray}
with $\Sigma$ the $2\times2$ covariance matrix %
between variables $\lambda_L$ and $\beta_L$, and $\chi^2_{\rm{ppf}}(0.95,\rm{df}=2)$ the percent-point function for a chi-squared distribution with two degrees of freedom at the 95\% confidence level. 

The ``minimum redshift'' calculated in this manner for unbiased PE of the intrinsic parameters is indicated by the color in each bin of Fig.~\ref{fig:heatmap-inc3}. The more blue the square, the less biased the PE is, and the nearer a system can be without finding biases due to excluding the (3,~2) mode. %

\begin{figure}[t]
\centering
\includegraphics[width=0.48\textwidth]{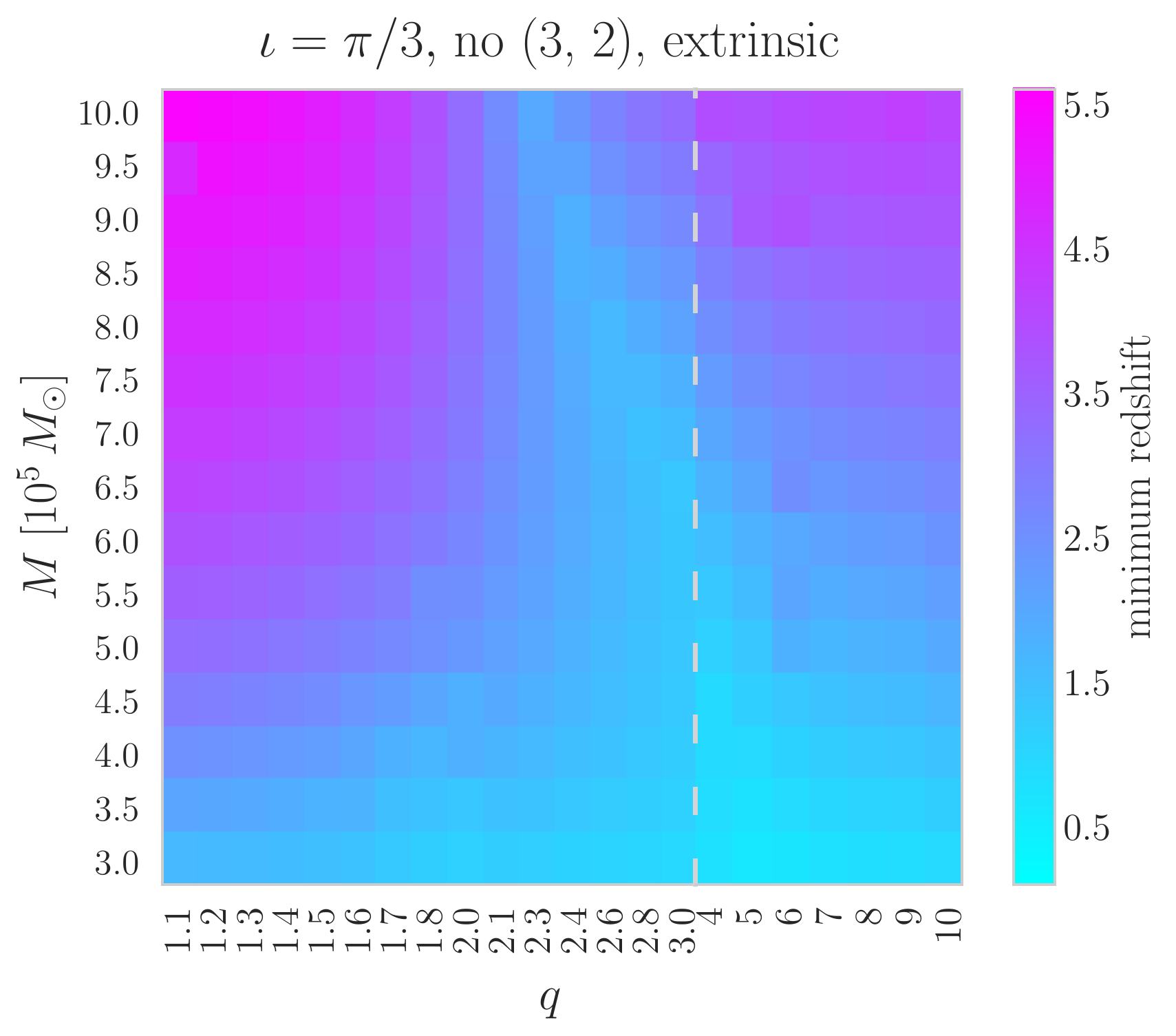}
\caption{Minimum redshift at which parameter estimation on the luminosity distance and sky position with $\iota= \pi/3$ is unbiased with the (3,~2) mode excluded from the template for recovery (e.g., parameter estimation is biased for $z\lesssim3$ for $M=6\times10^5\, M_\odot$ and $q=2$).}
\label{fig:heatmap-inc3-ext}
\end{figure}

In Fig.~\ref{fig:heatmap-inc3-ext}, we show the minimum redshift at which the inference of the distance and sky localization is unbiased when neglecting the (3,~2) mode, again for the inclination $\iota=\pi/3$, as in Fig.~\ref{fig:heatmap-inc3}. The presence of two separate regions of larger bias for the sky localization (more pink) can be explained on the one hand by larger signal strength (enhanced for lower $q$ at a fixed total mass, i.e., higher $\mathcal{M}_c$) and, on the other hand, longer duration of the signal (enhanced for lower $\mathcal{M}_c$).

\begin{figure*}[t]
\centering
\includegraphics[width=0.95\textwidth]{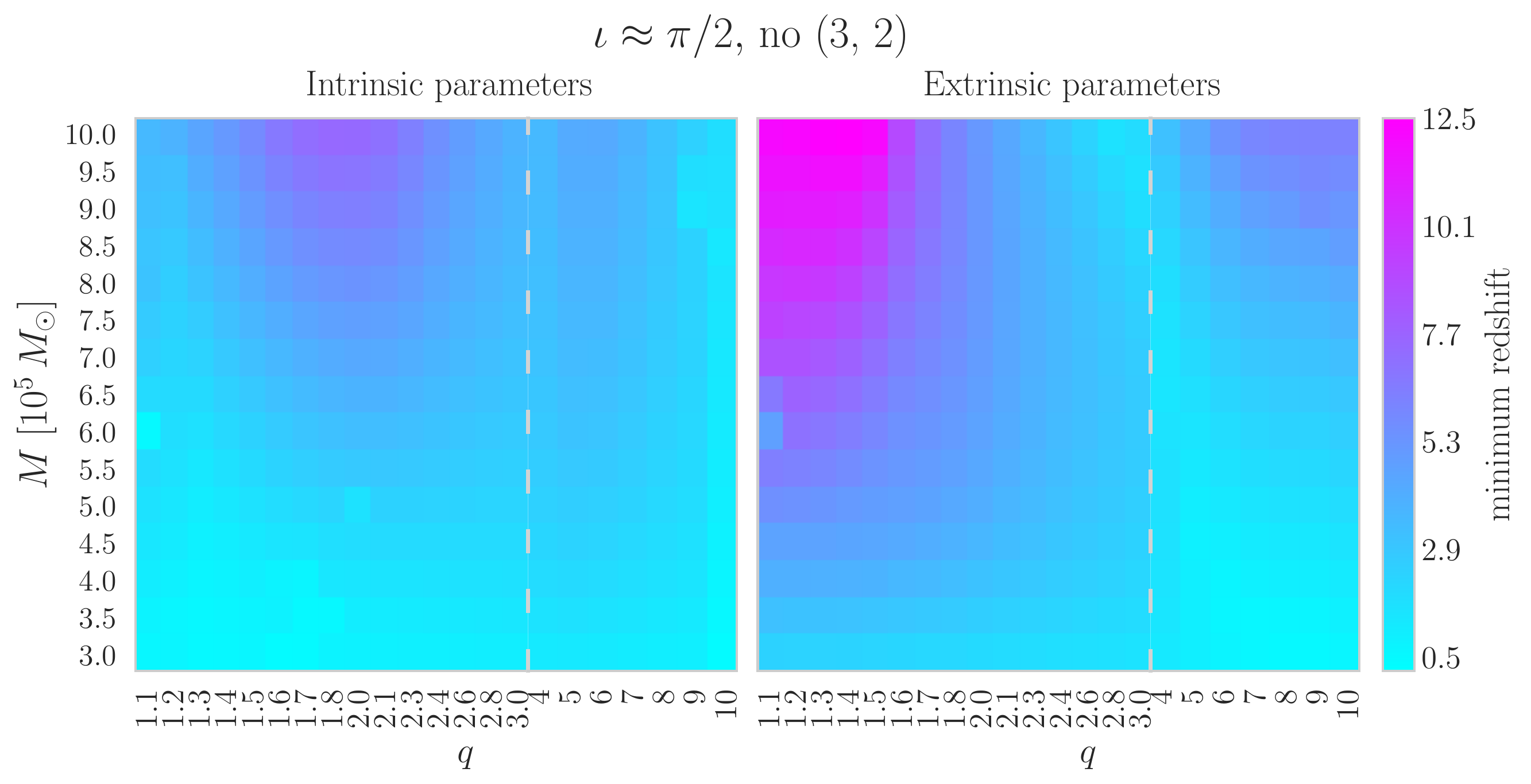}
\caption{Same as Fig.~\ref{fig:heatmap-inc3-ext} but for nearly-edge-on systems. We plot the minimum redshift for unbiased parameter recovery without the (3,~2) mode. Plots are shown for parameters [$\mathcal{M}_c,q,\chi_+,\chi_-$] on the left, and for the sky localization and distance on the right. }%
\label{fig:heatmap-inc2-both}
\end{figure*}

In Fig.~\ref{fig:heatmap-inc2-both}, we show the corresponding plots for when the (3,~2) mode is excluded in PE on nearly edge-on systems ($\iota\approx\pi/2$). While the morphology is about the same as in the plots with $\iota=\pi/3$ for both the intrinsic and extrinsic parameters, we note that the minimum redshift range is noticeably higher for edge-on systems. This is consistent with the observation that higher modes generally become more significant as the inclination approaches $\pi/2$. 

The statistical errors in Figs.~\ref{fig:heatmap-inc3},~\ref{fig:heatmap-inc3-ext}, and~\ref{fig:heatmap-inc2-both} also vary over the parameter space, in addition to the systematic errors. As a consequence, these plots effectively convey the point at which biases become the limiting factor in accurate parameter estimation, rather than showing the absolute size of the biases. Note also that the choice of a $2\sigma$ threshold here is somewhat arbitrary; if we instead use a threshold of $1\sigma$, the upper bound on the minimum redshift range increases to about 10.5 and 19.5 for the inclination angles $\iota=\pi/3$ and $\iota \approx\pi/2$, respectively, while the shapes of the plots look the same.

Finally, we comment on the fact that we do not include plots for the nearly face-on events ($\iota=\pi/12$). On the one hand, the quadrupole is quite dominant at this inclination angle, so we already expect higher modes to have a less significant impact. In addition, recalling Fig.~\ref{fig:mode-ordering}, we find that at this inclination, it is either the (4,~4) or the (2,~1) that is the quietest mode throughout the parameter space we consider here, rather than the (3,~2). To examine the impact of removing the quietest mode as done in Figs.~\ref{fig:heatmap-inc3},~\ref{fig:heatmap-inc3-ext}, and~\ref{fig:heatmap-inc2-both}, we would therefore perform the likelihood optimization excluding one of these modes. However, we found that the likelihood is difficult to optimize when the (3,~2) mode is included, with different initializations of the Nelder-Mead algorithm often giving different final results for the systematic bias. %
We suspect that mode-mixing for the (3,~2) mode contributes to the instability we find when trying to optimize the likelihood. We note that one can still use the formalism we introduce here to study the effect of removing any other modes, as long as the (3,~2) is also excluded to avoid this stability issue. For example, one could reproduce Fig.~\ref{fig:heatmap-inc2-both} with ``injected’’ modes [(2,~2), (3,~3), (2,~1), (4,~4)] and ``recover’’ with modes [(2,~2), (3,~3), (2,~1)] to study the effect of removing the (4,~4) mode.

\section{Extrapolating the mode content: approximating \texorpdfstring{$\ell\geq4$}{l greater than 4} waveforms}\label{sec:approx_HM}

In the above results, we see clearly that parameter recovery for a GW signal with radiation modeled in five modes can be significantly biased even when neglecting only one higher-order mode. Given this result, we cannot assume that the biases cease here; that is, we are not guaranteed that biases will be negligible when injecting a signal with six modes and recovering with five or less, injecting with seven modes and recovering with six or less, etc., particularly when we look at potentially even higher-mass events (i.e., the ``golden'' MBHB LISA sources with total mass $\sim10^7\,M_\odot$ or higher~\cite{LISA:2024hlh}),  
which will have both a larger overall SNR and more merger/ringdown-dominated signals, where higher-order modes become more important. %
While a thorough investigation of such effects is only possible with waveform templates that accurately model gravitational radiation in higher angular harmonics for a range of BBH configurations, here we show that with rough approximations to such waveforms motivated by PN theory, we can already begin to see the effects of neglecting the modes that are currently absent in waveform approximants. %

\subsection{Approximate higher-order mode waveforms from PN theory}

\subsubsection{Approximate amplitudes}\label{sec:approx_amps}

To construct rough approximations to higher-order mode amplitudes, we follow the procedure used in Ref.~\cite{Wadekar:2023kym} to construct higher-order mode templates for searches of the LVK O3 data. This is essentially the \emph{quadrupole mapping} procedure used to construct the \textsc{IMRPhenomHM} waveform family~\cite{London:2017bcn}. In this procedure, higher-order mode amplitudes are constructed by rescaling the (2,~2) amplitude by the lowest-order PN amplitude terms for each higher multipole. These terms are taken from Sec.~3 of Ref.~\cite{Mishra:2016whh} (with updated versions for $\ell\leq4$ terms in Appendix~E of Ref.~\cite{Garcia-Quiros:2020qpx}). Explicitly, the expressions for the first five ($\ell=m$) terms are 
\begin{eqnarray}
    H_{22}&\equiv&-1+\left(\frac{323}{224}-\frac{451\eta}{168}\right)V_2^2+\mathcal{O}(V_2^3)\,,\nonumber\\
    H_{33}&\equiv&-\frac{3}{4}\sqrt{\frac{5}{7}}\left(\frac{1-q}{1+q}\right)V_3+\mathcal{O}(V_3^3)\,,\nonumber\\
    H_{44}&\equiv&-\frac{4}{9}\sqrt{\frac{10}{7}}\left(1-3\eta\right)V_4^2+\mathcal{O}(V_4^4)\,,\\
    H_{55}&\equiv&-\frac{125}{96}\sqrt{\frac{5}{33}}\left(\frac{1-q}{1+q}\right)(1-2\eta)V_5^3+\mathcal{O}(V_5^5)\,,\nonumber\\
    H_{66}&\equiv&-\frac{18}{5}\sqrt{\frac{3}{143}}\left(1-5\eta+5\eta^2\right)V_6^4+\mathcal{O}(V_6^5)\,,\nonumber
    \label{eqn:Hlms}
\end{eqnarray}
where $\eta$ is the symmetric mass ratio, $\eta=m_1m_2/M^2$ (with $M=m_1+m_2$), and the $V_k$ are frequency-domain PN parameters given by $V_k(f)=[2\pi M f/k]^{1/3}$.
These $H_{\ell m}$'s in PN theory are related to the gravitational waveform by the following equations (see Eqs.~(9)--(11) of Ref.~\cite{Mishra:2016whh}):
\begin{eqnarray}
    h(\theta,\phi)&=&\sum^{+\infty}_{\ell=2} \sum^{\ell}_{m=-\ell} h_{\ell m} \,_{-2}Y_{\ell m}(\theta,\phi)\,,\\
    \tilde{h}_{\ell m}&=& \frac{M^2}{D_L} \pi \sqrt{\frac{2\eta}{3}}V_m^{-7/2}e^{-i\left(m\Psi_{\mathrm{SPA}}+\pi/4\right)}H_{\ell m}\,,
    \label{eqn:PN_hlms}
\end{eqnarray}
where $\Psi_{\mathrm{SPA}}$ is the phase in the stationary phase approximation, and we have placed the tilde over $h_{\ell m}$ in the second line to emphasize that these are Fourier-domain expressions for the modes. Both $\Psi_{\mathrm{SPA}}$ and $H_{\ell m}$ in the second line are functions of $V_m$.

Given this information, the authors of Ref.~\cite{Wadekar:2023kym} then construct effective amplitude equations for higher-order modes, given the full waveform for the quadrupole, by dividing both the angular dependence and the $H_{\ell m}$'s of each higher-order mode by that of the quadrupole, i.e., 
\begin{eqnarray}
    \label{eqn:hlm_over_h22}
    \left|\frac{h_{33}(3f)}{h_{22}(2f)}\right|&\simeq&\frac{3\sqrt{3}}{4\sqrt{2}}\left(\frac{1-q}{1+q}\right) (2\pi M f)^{1/3} \sin(\iota) \,,\nonumber\\
    \left|\frac{h_{44}(4f)}{h_{22}(2f)}\right|&\simeq&\frac{2\sqrt{2}}{3} (1-3\eta) (2\pi M f)^{2/3} \sin^2(\iota)\,,\\
    \left|\frac{h_{55}(5f)}{h_{22}(2f)}\right|&\simeq&\frac{125}{192}\sqrt{\frac{5}{2}} \left(\frac{1-q}{1+q}\right)(1-2\eta) (2\pi M f) \sin^3(\iota)\,,\nonumber \\
    \left|\frac{h_{66}(6f)}{h_{22}(2f)}\right|&\simeq&\frac{27\sqrt{3}}{40} (1-5\eta+5\eta^2) (2\pi M f)^{4/3} \sin^4(\iota)\nonumber\,.
\end{eqnarray}
Note that compared to Eq.~(11) of Ref.~\cite{Wadekar:2023kym}, we have the additional expressions for $\ell=m=5$ and $\ell=m=6$. 

In this work, we are interested in reconstructing modes with $\ell \geq5$, at variance with Ref.~\cite{Wadekar:2023kym}, which focused on building only the (3,~3) and (4,~4) modes from the quadrupole. In this case, it is interesting to ask whether more accurate higher-order mode waveforms can be built by approximating $\ell \geq5$ modes from the (3,~3) or (4,~4) modes, rather than the quadrupole. To investigate this, we additionally compute the following terms:
\begin{eqnarray}
    \label{eqn:hlm_from_44}
    \left|\frac{h_{44}(4f)}{h_{33}(3f)}\right|&\simeq& \frac{16}{9\sqrt{3}}\left(\frac{1+q}{1-q}\right)(1-3\eta)  (2\pi M f)^{1/3} \sin(\iota),\\
    \left|\frac{h_{55}(5f)}{h_{44}(4f)}\right|&\simeq&\frac{125\sqrt{5}}{256} \left(\frac{1-q}{1+q}\right)\frac{(1-2\eta)}{(1-3\eta)} (2\pi M f)^{1/3} \sin(\iota)\,,\nonumber \\
    \left|\frac{h_{66}(6f)}{h_{44}(4f)}\right|&\simeq&\frac{81}{80}\sqrt{\frac{3}{2}}\frac{(1-5\eta+5\eta^2)}{1-3\eta} (2\pi M f)^{2/3} \sin^2(\iota)\nonumber\,.
\end{eqnarray}

\begin{figure}[t]
\centering
\includegraphics[width=0.48\textwidth]{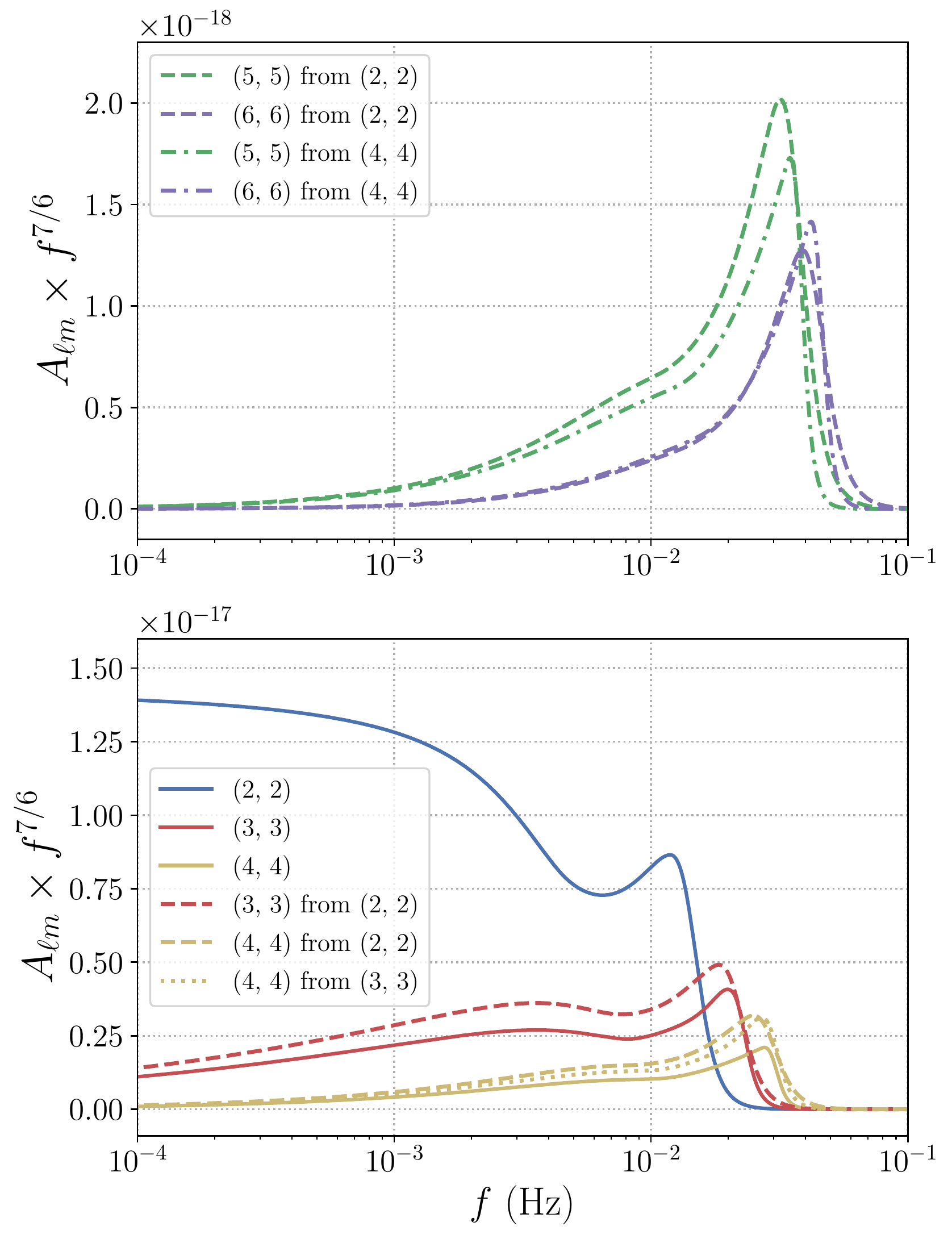} 
\caption{
Neglecting higher-order modes ($\ell\geq5$) which are not currently modeled within the \textsc{IMRPhenomXHM} approximant could introduce biases in PE. We therefore approximately model these modes as shown here, and then test them in a PE run in Fig.~\ref{fig:HM_PE}. Top: we model the (5,~5) and (6,~6) modes by rescaling the \textsc{IMRPhenomXHM} modes for (2,~2) and (4,~4) using the PN-motivated relations in Eqs.~\eqref{eqn:hlm_over_h22} and~\eqref{eqn:hlm_from_44}. Bottom: as a test of our rescaling method, we show that modeling (3,~3) and (4,~4) using just the (2,~2) \textsc{IMRPhenomXHM} waveform gives fairly accurate results. All mode amplitudes are plotted for an event with $M=10^6\,M_\odot, q=8$. Dashed, dotted and dot-dashed lines correspond to approximate amplitudes obtained by rescaling the ``exact'' amplitudes for the (2,~2), (3,~3), and (4,~4) modes, respectively. }
\label{fig:approx_HM_PN}
\end{figure}

In the top panel of Fig.~\ref{fig:approx_HM_PN}, we plot the approximate amplitudes we obtained for $h_{55}$ and $h_{66}$, rescaling from both the (2,~2) (dashed lines) and (4,~4) (dot-dashed lines) waveforms. The system plotted in Fig.~\ref{fig:approx_HM_PN} has total mass and mass ratio $M=10^6\,M_\odot$ and $q=8$. For comparison, we show in the bottom panel how the approximate (3,~3) and (4,~4) amplitude expressions rescaled from the quadrupole (dashed lines) compare to the ``exact'' \textsc{IMRPhenomXHM} amplitudes (solid lines). This bottom panel is essentially a reproduction of Fig.~6 of Ref.~\cite{Wadekar:2023kym}, but for a much heavier BBH system.
Compared to the top panel of Fig.~6 of Ref.~\cite{Wadekar:2023kym}, we also show an approximate (4,~4) waveform that is rescaled from the (3,~3) \textsc{IMRPhenomXHM} waveform (dotted yellow line). 

\subsubsection{Approximate phases}\label{sec:approx_phases}

\begin{figure}[t]
\centering
\includegraphics[width=0.48\textwidth]{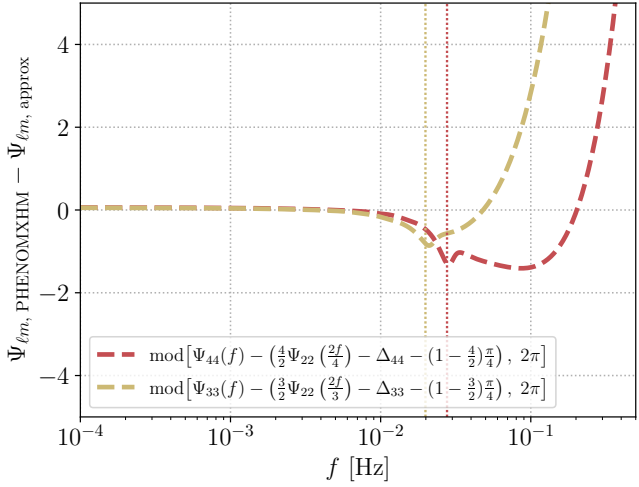} 
\caption{Similar to the lower panel of Fig.~\ref{fig:approx_HM_PN}, but comparing our approximate PN-based model for the phases of higher modes (see Eq.~\eqref{eqn:PN_phases}) instead of amplitudes. 
We see that our model is fairly accurate at low frequencies. The accuracy begins to degrade slightly as we approach the merger frequencies, shown with dotted lines for each mode. %
The difference between our approximate phases and the ``exact'' phases from the \textsc{IMRPhenomXHM} approximant is shown for the same event with amplitudes given in Fig.~\ref{fig:approx_HM_PN}.
}
\label{fig:approx_HM_PN_phases}
\end{figure}

To approximate the phases of higher-order modes, we begin by taking %
\begin{eqnarray}
    \Psi_{\ell m, \; \rm{approx}} (f) \approx \frac{m}{2} \Psi_{22}\left(\frac{2f}{m}\right)&-&\Delta_{\ell m}%
    \nonumber\\ &-& \left(1-\frac{m}{2}\right)\frac{\pi}{4}\,,
    \label{eqn:PN_phases}
\end{eqnarray}
with $\Delta_{\ell m}$ given by 
\begin{eqnarray}
    \label{eqn:delta_lm}
    \Delta_{\ell m} = \mathrm{mod}\left(\frac{m \pi}{2}-\pi,2\pi\right)
\end{eqnarray}
[see, e.g., Ref.~\cite{Cotesta:2020qhw}, although note the sign error in their Eq.~(3.5)].
In Fig.~\ref{fig:approx_HM_PN_phases}, we plot the difference between the approximate phases computed via Eq.~\eqref{eqn:PN_phases} and the ``exact'' phases from the \textsc{IMRPhenomXHM} waveform for the same $M=10^6\,M_\odot, \,q=8$ system with amplitudes shown in Fig.~\ref{fig:approx_HM_PN}. The approximate phases perform quite well at low frequencies and begin to veer off around the merger. We show the location of the merger frequency (i.e., the point at which the \textsc{IMRPhenomXHM} (3,~3) and (4,~4) amplitudes are at their maximum) with dotted lines. 

As done for the amplitudes, we additionally observed whether higher-order mode phases might be better approximated from the (3,~3) or (4,~4) modes than the (2,~2), i.e., if we replace $m/2\times\Psi_{22}\left(2f/m\right)$ in Eq.~\eqref{eqn:PN_phases} with $m/3\times\Psi_{33}\left(3f/m\right)$ or $m/4\times\Psi_{44}\left(4f/m\right)$. While we sometimes found that approximating from the (3,~3) resulted in a smaller difference from the ``exact'' phase and/or a more stable phase through merger, we only examined these phenomena for a handful of systems. For the present study, we build the phases simply following Eq.~\eqref{eqn:PN_phases}, and we leave a more thorough investigation of optimal ways to approximate higher-order modes for future study. 

\subsection{Biases with approximate waveforms}

\begin{figure}[ht!]
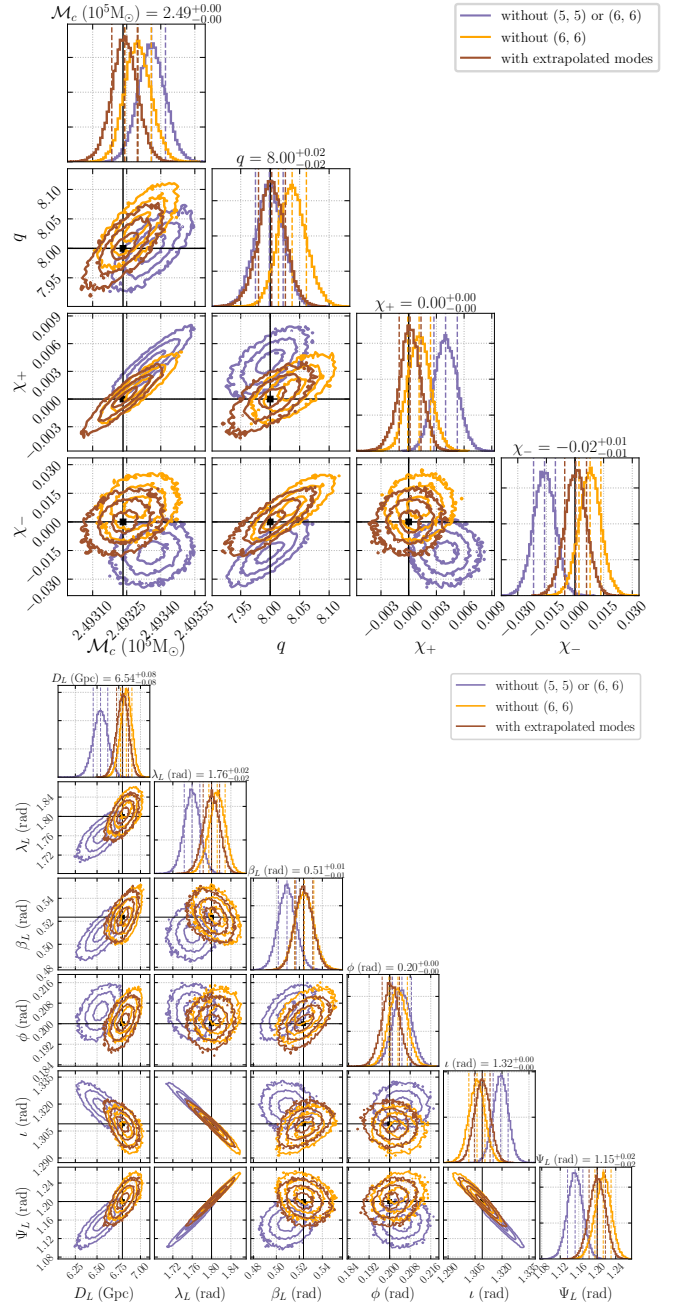

\centering
\includegraphics[width=0.49\textwidth]{extrapolated_modes_PE_intrinsic_inc2_corrected.pdf} \includegraphics[width=0.47\textwidth]{extrapolated_modes_PE_extrinsic_inc2_corrected.pdf} 
\caption{
Parameter recovery with and without the extrapolated (5,~5) and (6,~6) modes (see Figs.~\ref{fig:approx_HM_PN} and~\ref{fig:approx_HM_PN_phases}).
For reference, the extrapolated modes have SNR$_{55}$=76.5 and SNR$_{66}$=34.4, compared to SNR$_{22}$=780.1. There are moderate biases when excluding the extrapolated modes from the template for recovery. While the biases are noticeable at this inclination angle near $\pi/2$, they are indiscernible for nearly face-on binaries, where the spin-weighted spherical harmonics for $(\ell,|m|) = (5,5)$ and $(6,6)$ are zero or very close to zero.}
\label{fig:HM_PE}
\end{figure}

We now observe whether we can see noticeable systematic biases of the same sort seen in Secs.~\ref{sec:results} and~\ref{sec:results_extrinsic}, when we inject a signal with the (5,~5) and (6,~6) modes as modeled with the expressions in Eqs.~\eqref{eqn:hlm_over_h22}--\eqref{eqn:delta_lm} and recover with these $\ell\geq5$ modes removed. The results of such an injection and recovery for the $M = 10^6\,M_\odot, \, q=8,\, \iota \approx \pi/2$ event are shown in Fig.~\ref{fig:HM_PE}.

When injecting the signal with the extrapolated (5,~5) and (6,~6) modes and recovering without the (6,~6) (yellow contours and histograms), we observe very slight biases in the recovery of both the intrinsic and extrinsic parameters. When removing the (5,~5) mode as well (purple contours), the biases on all parameters generally become somewhat worse. %

We note that the relative importance of the (5,~5) and (6,~6) modes is heavily dependent upon the value of the inclination angle. In particular, at this inclination near $\pi/2$, the spin-weighted spherical harmonics $_{-2}Y_{55}(\iota,\phi=0.2)$ and $_{-2}Y_{66}(\iota,\phi=0.2)$ are larger than the harmonics for any other modes in our templates [i.e., for $(\ell,|m|)=$(2,~2), (3,~3), (2,~1), (4,~4), or (3,~2)]. The single-mode SNRs of the extrapolated modes at $\iota=\pi/2$ are SNR$_{55}$=76.5 and SNR$_{66}$=34.4, which are fairly comparable to the SNR of the (3,~2) mode at this inclination (SNR$_{32}$=60.8); the total SNR of this system is 912.9. In contrast, at $\iota=\pi/12$ with all other parameters the same, we have SNR$_{55}$=3.4 and SNR$_{66}$=0.4 (compared to the total SNR=1769.4). The PE in this case would be virtually indistinguishable whether we exclude just the (6,~6) mode, both the (5,~5) and (6,~6), or neither extrapolated mode in the template for recovery. 

In short, we find that depending on the inclination angle, waveforms containing only modes with $\ell\leq4$ may not be sufficiently accurate to perform PE on MBHBs with total mass up to $10^6M_\odot$. Within the scope of this paper, we have only checked to see whether the (5,~5) and (6,~6) modes might be necessary within the parameter space under consideration. Eventually, we will need to determine how many modes are ``enough''; will we also need $\ell\geq7$ modes? How about $\ell>4$ modes with $\ell\neq |m|$? We plan to explore these questions in future work. Note also that the observable volume is different for systems with different inclinations (e.g., face-on systems have a larger observable volume compared to edge-on systems). In an upcoming study, we plan to marginalize over inclination including the observable volume effect. This will help us more accurately characterize the relative impact of higher-order modes for a system with given $M,q$ (for LISA and also for current and future ground-based detectors).%
Finally, the biases could be more pronounced for systems with spinning progenitor BHs, a consideration we do not address here.

\section{Conclusions}
\label{sec:conclusions}

We have assessed how the exclusion of higher-order modes would bias our inference of MBHB signals detected with LISA. We find that for events with masses starting at a few times $10^5\, M_\odot$, excluding even just the (3,~2) mode can result in significantly biased inference of both intrinsic and extrinsic parameters. We demonstrate how such bias depends on the total mass, mass ratio, and inclination angle of the event by running PE on a coarse grid spanning these parameters. We then fill in this grid with substantially more data points using the direct likelihood optimization method, which we found produces good estimates of the systematic biases due to excluding higher-order modes. The result is an approximate boundary around a region of unbiased PE in the parameter space of total mass, mass ratio, and redshift for a few different inclination angles. We then began to explore how the same effects might persist for the exclusion of higher modes that are not yet modeled with current waveform approximants, using our own crude waveforms motivated by PN theory to begin to answer this question. 

There are many avenues for future work. In the first place, it will eventually be important to understand how these biases would appear for binaries with spinning progenitors, including precessing systems. In fact, we expect higher-order modes to be more important for systems with high aligned spins (see, e.g.,~\cite{Varma:2016dnf}). %
Similarly, we will eventually have to account for eccentric systems, systems with even higher mass ratios than what we have explored thus far ($q>10$), etc. 

To carry out some of the above work, it could be useful to further develop some of the tools implemented here. For instance, there is certainly room to improve on the direct likelihood optimization method. One could explore any number of alternative optimization algorithms and see whether they perform better than the Nelder-Mead algorithm for these particular high-dimensional GW problems. Even keeping the Nelder-Mead algorithm, one could perform a more extensive study of how the results depend on the number of iterations, size of the initial simplex, etc. 

Moreover, having found potential methods to estimate systematic bias, one could apply the above formalism to investigate systematic biases for BBH sources of ground-based detectors (both current and future). The direct likelihood optimization method could also be useful in understanding systematic biases due to other waveform inaccuracies, besides the exclusion of higher-order modes (e.g., failing to model precession, eccentricity, etc.). 

Finally, much work remains to be done in going beyond the technique we have used in Sec.~\ref{sec:approx_HM} to develop accurate waveforms with higher-order mode content. One could begin by examining in more detail whether rescaling from higher-order modes than the (2,~2) is beneficial across the parameter space. Ultimately, it will be important to answer the question of how many modes are enough to perform unbiased PE on the MBHB sources that will be observed by LISA. 
 
\begin{acknowledgments}
We thank Luca Reali, Veome Kapil, Rohit Chandramouli, Nicolás Yunes, Will Farr, and Leo Stein for helpful discussions. S.Y. is supported by the NSF Graduate Research Fellowship Program under Grant No. DGE2139757. S. Y., F. I., D. W, and E. B. are supported by NSF Grants No.~AST-2307146, No.~PHY-2513337, No.~PHY-090003, and No.~PHY-20043, by NASA Grant No.~21-ATP21-0010, by John Templeton Foundation Grant No.~62840, by the Simons Foundation [MPS-SIP-00001698, E.B.], by the Simons Foundation International [SFI-MPS-BH-00012593-02], and by Italian Ministry of Foreign Affairs and International Cooperation Grant No.~PGR01167. The work of F.I. is supported by a Miller Postdoctoral Fellowship. S.M. acknowledges support form the French space agency CNES in the framework of LISA. This work was carried out at the Advanced Research Computing at Hopkins (ARCH) core facility (\url{rockfish.jhu.edu}), which is supported by the NSF Grant No.~OAC-1920103. 
\end{acknowledgments}

\appendix 

\section{{\fontfamily{cmr}\textsc{IMRPhenomXHM} vs. \textsc{IMRPhenomHM}}}
\label{app:HMvsXHM}
In this Appendix, we show how the use of different approximants (namely, \textsc{IMRPhenomHM}~\cite{London:2017bcn} and \textsc{IMRPhenomXHM}~\cite{Garcia-Quiros:2020qpx}) %
affects our PE results. Despite belonging to the same waveform family, these two approximants feature very different procedures to approximate the higher-order mode content of the signal, as we shall see below. %
Before we proceed, we note that to truly have a direct comparison between results with the different approximants, one would need to carefully align the waveforms in phase, which we do not do here. Our purpose is merely to show qualitatively the effect of using different methods to approximate higher-order modes.

\begin{figure*}[t]
\centering
\includegraphics[width=0.48\textwidth]{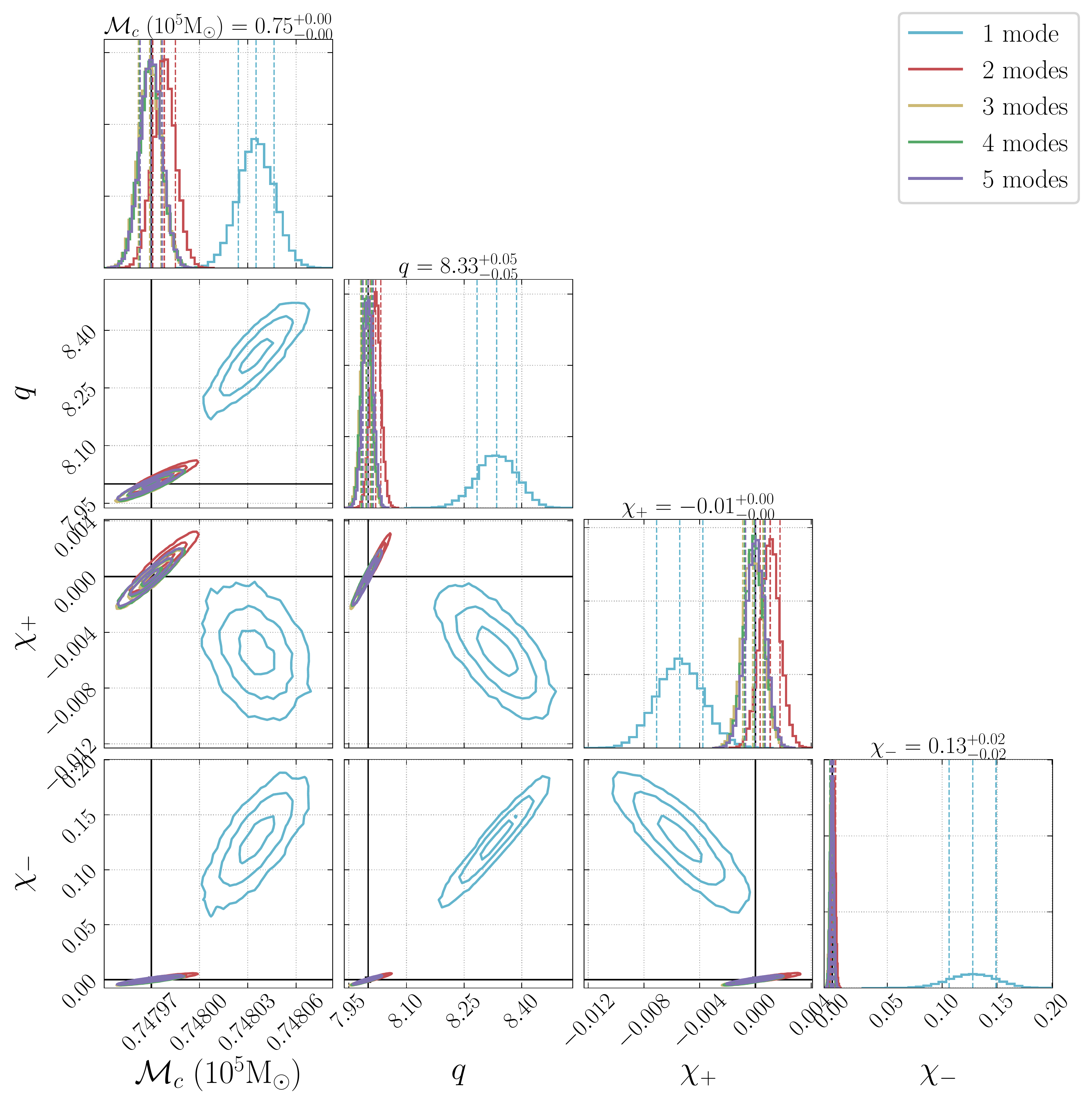}
\includegraphics[width=0.48\textwidth]{physM3e5s0q8inc3DS.pdf} 
\includegraphics[width=0.48\textwidth]{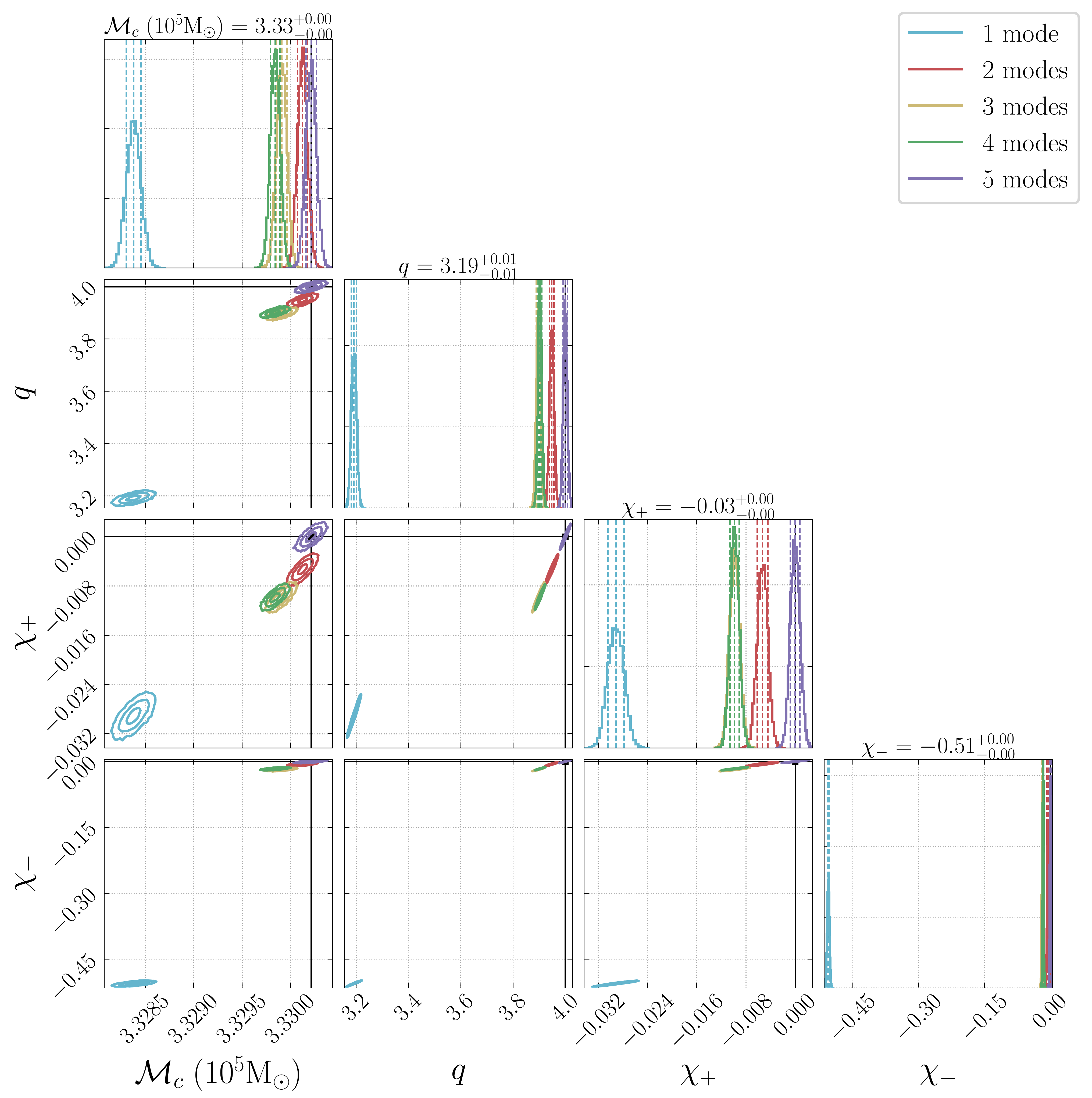}
\includegraphics[width=0.48\textwidth]{physM1e6s0q4inc2DS.pdf} 
\caption{PE results for a few example systems using the \textsc{IMRPhenomHM} (left) and \textsc{IMRPhenomXHM} (right) approximants. Top row: $M=3\times10^5\, M_\odot, \iota=\pi/3, q=8$. Bottom row: $M=10^6\, M_\odot, \iota=\pi/2, q=4$. }
\label{fig:XHM_vs_HM_intrinsic}
\end{figure*}

In Fig.~\ref{fig:XHM_vs_HM_intrinsic}, we plot posteriors on intrinsic parameters as obtained in PE runs with \textsc{IMRPhenomHM} (left column) and \textsc{IMRPhenomXHM} (right column). For brevity, in this section we will drop the common ``\textsc{IMRPhenom}'' portion of each approximant name and simply use the labels ``XHM'' or ``HM.'' We show the comparisons for both a moderate-bias case ($M=3\times10^5\, M_\odot$, top row) and a case with more severe biases ($M=10^6\, M_\odot$, bottom row). 

In the moderate-bias case, the injected values of the intrinsic parameters are recovered fairly accurately with comparatively fewer modes in the template with HM. At first glance, this may be unexpected, since XHM is the newer approximant family (with all modes calibrated to numerical relativity simulations). However, we note that this is not necessarily indicative of HM performing better; on the contrary, it could be that the effect of higher-order modes is underestimated with this approximant, such that including more modes does not accurately capture the full extent of changes in the physical signal. This possibility is supported by the fact that XHM is much better than HM at recovering signals injected with NR surrogate waveforms (see, e.g., Fig. 14 of Ref.~\cite{Garcia-Quiros:2020qpx}). It is therefore likely that compared to XHM models, the older HM models do not adequately reflect the full dependence of the total waveform on higher-order mode content. A potential consequence of this would be that the HM models are therefore not as useful in understanding the full extent of biases due to neglecting higher-order modes. 

In the case of severe bias, we see much of the same phenomenon. The mass ratio $q$ and $\chi_-$ are recovered considerably worse with HM than XHM when the 1-mode template is used. The recovery of chirp mass and $\chi_+$ with this template is comparable between approximants, but the recovered values of these parameters are again closer to injections with HM than with XHM when using 2, 3, and 4 mode templates. We stress again that this could really be a result of HM underestimating the importance of higher modes for a given system. Extrinsic parameter recovery (not shown) is roughly comparable with either waveform approximant in the cases of both moderate and severe biases.

It is interesting to note that the SNR ordering is generally not the same when determined with the different waveform families. In particular, HM tends to put comparatively more weight in the (4,~4) harmonic, so that the SNR in this harmonic is generally larger than the SNR in the (2,~1) harmonic, whereas the opposite is true for XHM. 
We note that HM has an additional available mode (the (4,~3) mode) compared to XHM. To make the comparison more straightforward, we compared PE runs between the approximants using only the five modes available with either family. 

To summarize, we see that the systematic bias due to insufficient mode content does not appear in the same way when using different waveform approximants (with the former caveat that we are not looking at exactly the same systems, due to the phases not being exactly aligned). Importantly, in the HM family, the higher modes are more simply rescaled from the (2,~2) mode, indeed in much the same manner as we employ in Sec.~\ref{sec:approx_HM} of the main text. In contrast, each mode in the XHM family is calibrated to numerical relativity simulations, such that the modes are in a sense more truly independent from one another. Adding or removing one higher mode can therefore be more consequential with XHM than with HM (i.e., it can alter the waveform more significantly). %
This sort of phenomenon is important to keep in mind as we move toward more accurate waveform models containing higher-order modes. %

\section{Cutler-Vallisneri formalism}\label{app:CV_formalism}

In this Appendix, we outline for completeness the linear signal approximation introduced by Cutler and Vallisneri~\cite{Cutler:2007mi}, which we use as a cheap method to approximate systematic biases due to insufficient mode content in waveform templates.

We write the total signal at a GW detector $s(t)$ as
\begin{equation}
    s(t) = h_{\rm{GW}}(t)+n(t)\,,
\end{equation}
where $h_{\rm{GW}}$ is the GW signal and $n(t)$ the detector noise, which we take to be stationary and Gaussian. The log likelihood is then given by 
\begin{equation}
    \mathrm{ln} \,p(s|\bm{\theta}) \propto -\frac{1}{2} \left( s-h_m | s-h_m \right) \,,
    \label{eqn:ln_likelihood}
\end{equation}
where $\bm{\theta}$ are the waveform parameters and the approximate waveform template, $h_m$, will always differ from the true waveform $h_{\rm{GW}}$ by some amount
\begin{equation}
    \delta h(\bm{\theta})=h_{\rm{GW}}(\bm{\theta})-h_m(\bm{\theta}) \,.
\label{eqn:delta-h}
\end{equation}
The inner product in Eq.~\eqref{eqn:ln_likelihood} is given by
\begin{equation}
    \left(a|b\right)=4 \mathrm{Re} \int_0^\infty \frac{\tilde{a}(f)\tilde{b}^*(f)}{S_n(f)}df \,,
    \label{eqn:inner_prod}
\end{equation}
where the asterisk denotes complex conjugation,  $\tilde{a}(f)$ and $\tilde{b}(f)$ are the Fourier transforms of time-domain signals $a(t)$ and $b(t)$, respectively, and $S_n(f)$ is the noise PSD.

The maximum of the likelihood function occurs at some best-fit parameters, $\bm{\theta}_{\rm{bf}}$, such that
\begin{equation}
    \left( \partial_i h_m(\bm{\theta}_{\rm{bf}}) | s- h_m(\bm{\theta}_{\rm{bf}})\right)=0\,,
    \label{eqn:max_likelihood}
\end{equation}
where $\partial_i=\partial/\partial\theta^i$ denotes the partial derivative with respect to parameter $\theta^i$. (The expression in Eq.~\eqref{eqn:max_likelihood} can be obtained by demanding that the gradient of the likelihood with respect to $\theta^i$ vanishes at $\bm{\theta}=\bm{\theta}_{\rm{bf}}$.) The linear signal approximation now enters, where for a small perturbation $\bm{\theta}_{\rm{tr}}=\bm{\theta}_{\rm{bf}}+\Delta\bm{\theta}$, with $\bm{\theta}_{\rm{tr}}$ the true GW parameters, we can expand the waveform near $\bm{\theta}_{\rm{bf}}$ as
\begin{equation}
    h_m(\bm{\theta}_{\rm{tr}})-h_m(\bm{\theta}_{\rm{bf}}) \approx - \Delta \theta^i \partial_i h_m(\bm{\theta}_{\rm{bf}}) \,.
    \label{eqn:small_bias}
\end{equation}
Then we can write the quantity $s-h_m(\bm{\theta}_{\rm{bf}})$ as 
\begin{eqnarray}
    s-h_m(\bm{\theta}_{\rm{bf}})&=& n+h_{\rm{GW}}(\bm{\theta}_{\rm{tr}}) \nonumber\\
    &-&h_m(\bm{\theta}_{\rm{tr}})+h_m(\bm{\theta}_{\rm{tr}})-h_m(\bm{\theta}_{\rm{bf}})\nonumber\\
    &\approx& n +\delta h(\bm{\theta}_{\rm{tr}}) - \Delta \theta^i \partial_i h_m(\bm{\theta}_{\rm{bf}}) \, .
    \label{eqn:s-h_approx}
\end{eqnarray}
Equation~\eqref{eqn:max_likelihood} then becomes
\begin{eqnarray}
     \left( \partial_i h_m(\bm{\theta}_{\rm{bf}}) | s- h_m(\bm{\theta}_{\rm{bf}})\right) &\approx& \left( \partial_i h_m(\bm{\theta}_{\rm{bf}}) | n \right) \\
     \;\;\;&+& \left( \partial_i h_m(\bm{\theta}_{\rm{bf}}) | \delta h(\bm{\theta}_{\rm{tr}}) \right) \nonumber\\
     \;\;\;&-&\Delta \theta^j\left( \partial_i h_m(\bm{\theta}_{\rm{bf}} )| \partial_j h_m(\bm{\theta}_{\rm{bf}}) \right)\, . \nonumber
    \label{eqn:max_likelihood_LSA}
\end{eqnarray}
Identifying the Fisher matrix $\Gamma_{ij}$ as
\begin{equation}
     \Gamma_{ij}\equiv \left( \partial_i h_m(\bm{\theta}_{\rm{bf}}) | \partial_j h_m(\bm{\theta}_{\rm{bf}}) \right) \,,
    \label{eqn:FIM}
\end{equation}
we can rearrange Eq.~\eqref{eqn:max_likelihood_LSA} to obtain 
\begin{eqnarray}
     \Delta \theta^i &=& \left(\Gamma^{-1}(\bm{\theta}_{\rm{bf}})\right)^{ij} \left( \partial_i h_m(\bm{\theta}_{\rm{bf}}) | n \right) \nonumber\\
     &+& \left(\Gamma^{-1}(\bm{\theta}_{\rm{bf}})\right)^{ij} \left( \partial_i h_m(\bm{\theta}_{\rm{bf}}) | \delta h(\bm{\theta}_{\rm{tr}}) \right) \,.
    \label{eqn:CV_full}
\end{eqnarray}
The first piece on the right-hand side, $\left(\Gamma^{-1}(\bm{\theta}_{\rm{bf}})\right)^{ij} \left( \partial_i h_m(\bm{\theta}_{\rm{bf}}) | n \right)$, describes errors due to noise and is not the focus of our study here. The second piece, which we will denote $\Delta_{\rm{sys}} \theta^i =\left(\Gamma^{-1}(\bm{\theta}_{\rm{bf}})\right)^{ij} \left( \partial_i h_m(\bm{\theta}_{\rm{bf}}) | \delta h(\bm{\theta}_{\rm{tr}}) \right)$, is the systematic bias due to imperfect approximation of the true waveform $h_{\rm{GW}}$ by $h_m$ (due to, in this case, too few harmonics in the template). We calculate $\Delta_{\rm{sys}} \theta^i$ for our example events and examine how well this quantity predicts the biases observed in the PE results.

\begin{figure*}[htbp]
\centering
\includegraphics[width=0.95\textwidth]{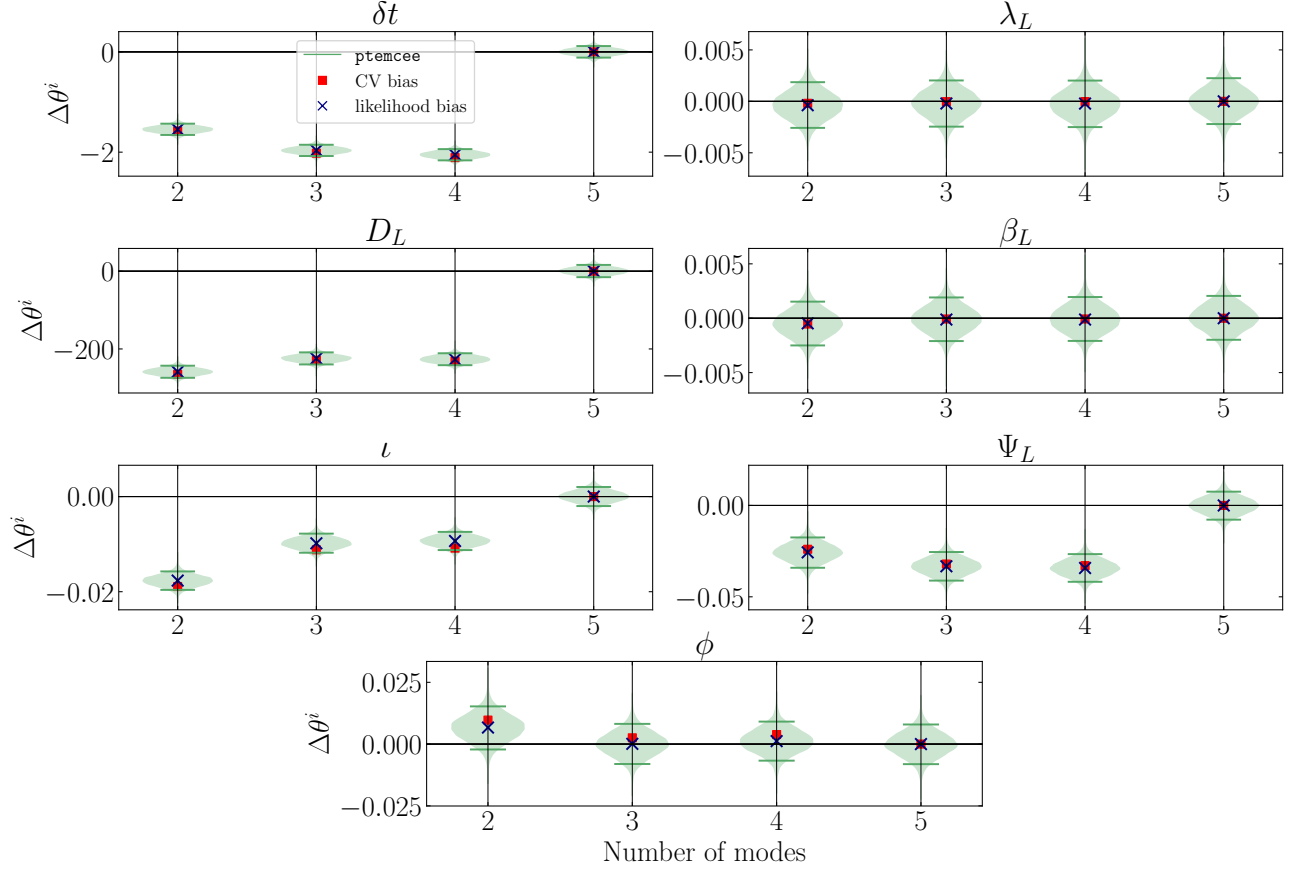}
\caption{Systematic bias on extrinsic parameters estimated within the linear signal approximation (``CV bias'') and with the direct likelihood optimization method (``likelihood bias'') compared to biases recovered in full PE (``\texttt{ptemcee}''). Results are shown for the $M=10^6 M_\odot,\, q=8,\, \iota=\pi/12$ event also discussed in Sec.~\ref{sec:CV_results}. }
\label{fig:CV-and-NM-ext}
\end{figure*}

\section{Additional results for the CV and direct likelihood optimization estimates of systematic biases}\label{app:CV_cont.}

In Fig.~\ref{fig:CV-and-NM-ext}, we plot biases as determined by the CV and direct likelihood optimization methods, as well as PE, for the remainder of the extrinsic parameters of the $M=10^6 M_\odot, \,q=8,\, \iota=\pi/12$ event discussed in the main text (Figs.~\ref{fig:CV-and-NM-int} and~\ref{fig:CV-and-NM-ext-22}). Both the CV and direct likelihood optimization methods recover the biases in PE fairly well for the more moderate biases observed with at least two modes in the waveform template, compared to the quadrupole-only case shown in Fig.~\ref{fig:CV-and-NM-ext-22}.

\section{Effect of higher harmonics on the precision of sky localization estimates}\label{app:sky_loc}

 In this Appendix, we expand on the fifth point of Sec.~\ref{sec:results_extrinsic} regarding the improved precision of the inference of sky position angles when using more than the (2,~2) mode. While the focus of the main text is on how parameter estimation on MBHBs becomes biased when higher harmonics are not adequately modeled in waveform templates, it is also true that the precision (in addition to the accuracy) of parameter estimation is increased considerably with the addition of higher-order modes. 

\begin{figure*}[t]
\centering
\includegraphics[width=0.98\textwidth]{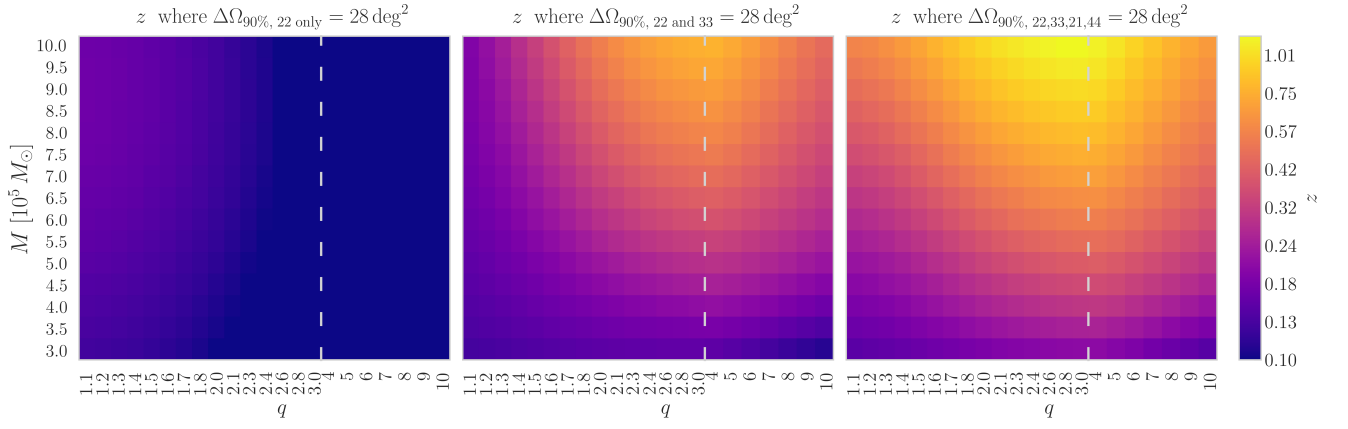} 
\caption{Redshift at which MBHBs with the given total mass and mass ratio can be localized within 28 deg$^2$ when using the (2,~2) mode only (left), the (2,~2) and (3,~3) modes (center), and the (2,~2), (3,~3), (2,~1) and (4,~4) modes (right). It is clear that sky localization improves considerably with just one additional mode beyond the (2,~2), while adding further modes beyond that does not introduce much further improvement at the masses we consider.}
\label{fig:skyloc}
\end{figure*}

To illustrate this point, in Fig.~\ref{fig:skyloc} we plot the maximum redshift at which an MBHB with the given total mass and mass ratio can be localized within $28\;\rm{deg}^2$ (i.e., the area in which GW170817 was localized) when using a quadrupole-only template (left panel), and then when adding one or more modes to the template (center and right panels, respectively). In these panels, the sky area $\Delta \Omega_{\rm{X}\%}$ at a confidence level $\rm{X}$ is computed as in Eq. (28) of Ref.~\cite{Iacovelli:2022bbs}:
\begin{eqnarray}\label{eqn:sky_area} 
\Delta\Omega_{\rm{X}\%} &=& -2\pi | \cos\beta_L|\ln\left(1 - \frac{\rm{X}}{100}\right) \\
&\times& \sqrt{(\Gamma^{-1})_{\beta_L\beta_L}(\Gamma^{-1})_{\lambda_L\lambda_L} - (\Gamma^{-1})^2_{\beta_L\lambda_L}}\nonumber \,,
\end{eqnarray}
where $\Gamma$ is the Fisher information matrix and $(\Gamma^{-1})_{\theta^i\theta^j}$ denotes the component of the inverse Fisher (i.e., covariance) matrix for parameters $\theta^i$ and $\theta^j$. 
We stress that because we use Fisher matrices to compute the errors here, we are not accounting for the loss of precision due to the multimodal nature of the posteriors on $\lambda_L$ and $\beta_L$. While the posteriors on these parameters are for the most part unimodal in the parameter space considered within the scope of this paper (see Fig.~\ref{fig:extrinsic}), degeneracies in the sky position start to become more severe for more massive events, which last for a shorter amount of time. 
Moreover, in the (2,~2)-only case in particular, one can find apparent bias in the sky position posteriors that comes from projecting the multimodal posterior into the subspace of extrinsic parameters (see Eq.~(79) of Ref.~\cite{Marsat:2020rtl}). We therefore caution the reader against taking the plots in Fig.~\ref{fig:skyloc} as perfectly confident estimates of the precision with which we will be able to localize MBHB events. Nevertheless, it is interesting to notice how dramatically the sky localization precision improves with the addition of even one mode beyond the (2,~2), whereas adding further modes beyond that does not make much of a difference at the masses we consider here. We point out that the (3,~2) mode is not included in the estimate of statistical error in the right panel of Fig.~\ref{fig:skyloc} because, as mentioned several times in the main text, this mode cannot be easily incorporated in the calculation of Fisher matrices due to differentiability issues. 

Finally, we note that the sky localization depends heavily on the inclination angle. Results are shown in Fig.~\ref{fig:skyloc} for MBHB systems with a median inclination angle of $\iota=\pi/3$, but localization can improve significantly for more face-on binaries (and likewise worsen for more edge-on systems). The same plot made for $\iota=\pi/12$ results in an ability to localize systems within 28 deg$^2$ out to redshift $z\sim15$ when using four modes in the waveform.

\bibliography{lisaHM}

\end{document}